\documentclass[5p,twoside,final,authoryear]{elsarticle_edit}
\usepackage{graphicx,setspace,multicol,fixltx2e,float}
\usepackage{amsfonts,amssymb}
\usepackage[fleqn]{amsmath}
\usepackage{fancyhdr}
\usepackage{hyperref}
\usepackage{times}

\hypersetup{colorlinks,linkcolor=blue,filecolor=blue,urlcolor=blue,citecolor=blue}

\newcommand{\mnras}{\textsc{MNRAS}}
\newcommand{\apj}{ApJ}
\newcommand{\apjl}{ApJL}
\newcommand{\apjs}{ApJS}
\newcommand{\aap}{A\&A}
\newcommand{\aaps}{A\&AS}
\newcommand{\pasp}{\textsc{PASP}}

\long\def\symbolfootnote[#1]#2{\begingroup%
\def\thefootnote{\fnsymbol{footnote}}\footnote[#1]{#2}\endgroup} 
\def\blfootnote{\xdef\@thefnmark{}\@footnotetext}

\newcommand{\sfrac}[2]{\genfrac{}{}{}{1}{#1}{#2}}

\newcommand{\q}{\theta}
\newcommand{\uq}{u_{\q}}
\newcommand{\bgrad}{\mbox{\boldmath $\nabla$}}
\newcommand{\bdot}{\mbox{\boldmath $\cdot$}}
\newcommand{\bF}{\boldsymbol{F}}

\newcommand{\bi}{\boldsymbol{i}}
\newcommand{\bI}{\boldsymbol{I}}
\newcommand{\bq}{\boldsymbol{q}}
\newcommand{\br}{\boldsymbol{r}}
\newcommand{\bpsi}{\boldsymbol{\psi}}
\newcommand{\bS}{\boldsymbol{S}}
\newcommand{\bsigma}{\boldsymbol{\sigma}}
\newcommand{\bu}{\boldsymbol{u}}
\newcommand{\bw}{\boldsymbol{w}}
\newcommand{\bx}{\boldsymbol{x}}

\newcommand{\divm}{\bgrad \bdot \,}
\newcommand{\DF}{\Delta F}
\newcommand{\Dt}{\Delta t}
\newcommand{\Dx}{\Delta x}
\newcommand{\p}{\partial}
\newcommand{\pt}{\partial t}
\newcommand{\px}{\partial x}
\newcommand{\pz}{\partial z}
\newcommand{\pr}{\partial r}
\newcommand{\pq}{\partial \theta}
\newcommand{\prho}{\partial \rho}

\begin{document}

\pagestyle{mypagestyle}
\thispagestyle{empty}
\mathindent=0pt
\journal{New Astronomy}
\twocolumn[
\begin{@twocolumnfalse}
\vspace*{0.cm} 
\parbox{0.85\textwidth}{\Large\textbf{RAPID: A fast, high resolution, flux-conservative algorithm designed 
for planet-disk interactions}}

\vspace{1.5cm}

\large\textbf{L. R. Mudryk$^{*}$, N. W. Murray}

\normalsize
\vspace{12pt}

\footnotesize\textit{Canadian Institute for Theoretical Astrophysics, Toronto, ON, CANADA, M5S 3H8}

\vspace{12pt}

\hrulefill

\vspace{8pt}

\begin{tabular}{@{}l @{}l @{}l}
\textbf A R T I C L E \hspace{1em} I N F O  & \hspace{0.05\textwidth} &\textbf
A B S T R A C T \\
\hrulefill & & \hrulefill \\
\begin{minipage}{0.35\textwidth}
Published in \textit{New Astronomy}
\hspace{0em} \\
\hspace{0em} 
95.75.Pq \\ 
97.10.Gz \\
97.82.Jw \\

Hydrodynamics \\
Methods: numerical \\
Planetary systems: formation \\
\end{minipage} & &
\begin{minipage}{0.6\textwidth}
\small We describe a newly developed hydrodynamic code for studying accretion
disk processes. The numerical method uses a finite volume, nonlinear,
Total Variation Diminishing (TVD) scheme to capture shocks and control
spurious oscillations. It is second-order accurate in time and space
and makes use of a FARGO-type algorithm to alleviate
\textit{Courant-Friedrichs-Lewy} time step restrictions imposed by the
rapidly rotating inner disk region.  OpenMP directives are implemented 
enabling faster computations on shared-memory, multi-processor
machines. The resulting code is simple, fast and memory efficient.  
We discuss the relevant details of the numerical method and provide
results of the code's performance on standard test problems. We also
include a detailed examination of the code's performance on planetary
disk-planet interactions.  We show that the results produced on the
standard problem setup are consistent with a wide variety of other
codes. \vspace{5pt}
\end{minipage} \\
\hrulefill & & \hrulefill 
\end{tabular}

\vspace{5em}

\end{@twocolumnfalse}]
\symbolfootnote[0]{\hspace{-1em}$*$ \hspace{0.15em} \scriptsize{Corresponding author.}}
\symbolfootnote[0]{\scriptsize{\textit{Email addresses:} \href{mailto:mudryk@cita.utoronto.ca}{mudryk@cita.utoronto.ca}
 (L.R. Mudryk), \href{mailto:murray@cita.utoronto.ca}{murray@cita.utoronto.ca} (N.W. Murray).} \\ \\
doi:10.1016/j.newast.2008.05.002}

\section{\textbf{Introduction}}

The study of almost all astronomical objects relies on an
understanding of their hydrodynamics.  Indeed, for many such objects
the involved hydrodynamics are complex enough to require numerical
modeling.  The process of numerical modeling usually proceeds by
writing partial differential equations describing the behavior of a
continuous medium as an equivalent set of algebraic equations for a
finite set of discretized elements.  This discretization can generally
be performed in two different ways.  In an Eulerian approach, one
discretizes the spatial domain into volumes termed grid cells. The
fluid is considered to move through this fixed 
background grid.  By contrast, in a Lagrangian approach the fluid is
discretized into fluid elements (or `particles') which can then move
freely according to their initial velocities, and only their
interactions need to be modeled.  Lagrangian methods work well in
situations with large background flows where Eulerian methods would
spend the bulk of their time advecting the (uninteresting) balanced
flow, accumulating numerical errors with the numerous iterations
required. Lagrangian methods have a large dynamic range in length but
not in mass, achieving good spatial resolution in high-density regions
but performing poorly in low-density regions.  In addition, the usual 
implementations of Lagrangian methods, based on Smoothed Particle
Hydrodynamics (SPH), do not easily allow the higher spatial accuracy
that grid methods can employ nor do they capture shocks as accurately
as grid methods. By contrast, Eulerian methods provide a large dynamic
range in mass but not in length. In general they are also
computationally faster by several orders of magnitude, easier to
implement, and easier to parallelize.

The RAPID code (Rapid Algorithm for Planets In Disks), which we
present here, uses an Eulerian approach, adapted for a cylindrical
grid.  While we focus on planet-disk interactions in this paper, the
code is intended for the general study of accretion disks 
containing a dominant central mass.  In such systems the gas disk
surrounding the central object is of a small enough mass that
its self-gravity may be ignored.  Such disks will have a roughly
Keplerian velocity profile resulting from the mass of the central
object.  In order to obtain higher algorithm efficiency in the
presence of this Keplerian flow, we make use of a FARGO-type algorithm
\citep{masset00}.  The algorithm's underlying strategy is to subtract
off the bulk flow, which can be considered simply a translation of
grid quantities, leaving the dynamically important residual velocity.
RAPID is second-order accurate in space and time. Advection is
accomplished through a nonlinear Total Variation Diminishing (TVD)
scheme, which helps to control spurious oscillations.  Time-stepping
is accomplished through a standard Runge-Kutta scheme.  Operator
splitting is used to account for multiple dimensions and source terms
such as those due to gravitational potentials and viscosity. 

In Section \ref{sec:eulerian} we outline the fluid equations to be
solved and discuss considerations of angular momentum important for
accuracy on cylindrical grids.  In Section \ref{sec:numericalmethod}
we discuss the details of the RAPID algorithm.  We provide results of
basic hydrodynamical tests in Section \ref{sec:basictests} and
demonstrate the code's performance on typical planetary disk setups in
Section \ref{sec:planettests}.  Conclusions are presented in Section
\ref{sec:conclusions}.

\section{Eulerian hydrodynamics} \label{sec:eulerian}

The Navier-Stokes equations may be written as
\begin{align}
&    \frac{\prho}{ \pt} + \bgrad \bdot \rho \bu = 0 \label{eq:densityflux}  \\
&    \frac{\prho \bu}{\pt} + \bgrad \bdot [\rho \bu \bu + p\bI ] 
       = -\rho \bgrad \phi + \bgrad \bdot \bsigma \label{eq:momentumflux}\\ 
&    \frac{\p \rho e}{\pt} + \divm [(\rho e + p)\bu] 
       = -\rho \bu \bdot (\bgrad \phi) + \divm [ \bu \bdot \bsigma ] - \divm \bpsi, \label{eq:energyflux}
  \end{align}
for the mass density $\rho$, momentum density $\rho\bu = \rho
(u_1,u_2,u_3)$, and total energy density $\rho e = \rho\varepsilon +
\frac{1}{2}\rho \bu^2$ of a fluid volume.  The symbols $\varepsilon$
and $p$ represent the internal energy per mass and the pressure of the
fluid, $\phi$ represents the potential due to a body force (such as
that from an external gravitational field), $\bsigma$ represents the
non-isotropic component of the stress-strain tensor for the fluid, and
$\bpsi$ represents any heat flux.  We use the symbol $\bI$ for the
identity matrix and note that the combination $\rho \bu \, \bu \equiv
\rho u_i u_j$ is a direct product yielding a matrix for the momentum
fluxes.

These equations express the transfer of mass, linear momentum, and
energy within the fluid volume written out in an arbitrary coordinate
system.  In terms of a general solution vector $\bq = (\rho, \rho u_1, \rho
u_2, \rho u_3, \rho e)$, flux tensor $\bF(\bq)$, and
source vector $\bS$, these equations all have the same formally
simple form

\begin{equation}
\frac{\p \bq }{ \pt} + \divm \bF = \bS. 
 \label{eq:fluxform}
\end{equation}
In the case where $\bS =0$ the equations reduce to the conservation
form of the Euler equations, expressing non-dissipative advection of
fluid quantities. 

In Cartesian coordinates and for the solution vector $\bq = (\rho$, $\rho
u_x$, $\rho u_y, \rho u_z, \rho e)$, the Euler equations describe the
evolution of five conserved scalar quantities.  However, written in
cylindrical coordinates, where $\bx = (r,\q,z)$, and for the natural
choice of solution vector $\bq = (\rho, \rho u_r, \rho u_\q, \rho u_z, \rho
e)$, only three components of this vector are conserved scalar 
quantities.  The quantities $\rho u_r$ and $\rho u_\q$ are not
conserved. We thus choose to solve for the solution vector $\bq =
(\rho, \rho u_r, {\cal H}, \rho u_z, \rho e)$, where the quantity
${\cal H} = \rho r(u_\q + r\Omega)$ is the fluid's angular momentum in
the inertial frame (we will refer to this quantity as the \emph{inertial
angular momentum}).  It includes contributions from the fluid's angular
velocity $\omega = u_\q/r$, as well as the reference frame's angular
velocity $\Omega$, assumed to be oriented along the z-axis. Because
inertial angular momentum is conserved in a rotating system, it is a
more natural physical variable to use and doing so improves the
accuracy of the results (see \S \ref{subsec:influences}).  In the
appendix we write out modified versions of equations
(\ref{eq:densityflux})--(\ref{eq:energyflux}) for this choice of
solution vector, expressed in  cylindrical coordinates.

\section{Numerical method} \label{sec:numericalmethod}

The solution method we describe is based on the relaxing Total
Variation Diminishing (TVD) method by \citet{jin95}.  \hspace{0.2pc} This method has 
been successfully applied to solve the Euler equations on Cartesian grids in 
astrophysical simulations by \citet{pen98} and \citet{trac03,trac04}.

\subsection{Relaxation system} \label{subsec:relaxation}

The relaxing TVD method solves the Euler equations by assuming the
equations may be split into components corresponding to leftward and
rightward travelling waves.  In place of the Euler equations
(eq.\ [\ref{eq:fluxform}] with $\bS =0$), the following coupled system
is solved along a single grid direction for the solution vector
$\bq$: 
\begin{align}
& \frac{\p \bq}{\pt} + \frac{\p}{\px} (c\bw) = 0 \label{eq:coupled1} \\
& \frac{\p \bw}{\pt} + \frac{\p}{\px} (c\bq) = 0.\label{eq:coupled2}
\end{align}
The relations $\bq = \bq^R + \bq^L$, and $\bw = \bF/c = \bq^R -
\bq^L$, define the solution variables in terms of the leftward and
rightward-travelling waves.  Equation (\ref{eq:coupled2}) represents a
separate equation for the evolution of the normalized flux vector
$\bw$.  The variable $c$ is a positive-definite function which has
the interpretation of a speed associated with a particular grid cell.
The solution is stable in the sense that its total variation (see
\S \ref{subsec:tvdschemes}) decreases as long as all values of $c$ are greater than or
equal to the largest eigenvalue of the flux Jacobian $\p \bF(\bq)/\p \bq$
\citep{jin95}.  Because the the waves are split into separate
rightward and leftward components, the maximum eigenvalue of the
Jacobian is limited for both components by the value $c_i = |u_i| +
c_s$ where $c_s$ is the sound speed for the cell.  Substituting these
definitions into equations (\ref{eq:coupled1}) and (\ref{eq:coupled2})
decouples the system and yields
\begin{equation}
\frac{\p \bq}{\pt} + \frac{\p \bF^R}{\px\,\,\,} - \frac{\p \bF^L}{\px\,\,\,} = 0, \label{eq:coupled}
\end{equation}
where $\bF^L = c\bq^L$, and $\bF^R = c\bq^R$.  The original coupled
system, equations (\ref{eq:coupled1}) and (\ref{eq:coupled2}), is then
equivalent to the solutions of the two separate leftward- and
rightward-moving waves given in equation (\ref{eq:coupled}).  It is
now possible to separately solve for each of the travelling waves and
add the results to determine the full solution along a single
direction.

\subsection{Solution of wave-split, one-dimensional Euler equations} \label{subsec:wavesplitsolution}

In order to solve for the advection of the separately travelling
waves, we implement a second-order Runge-Kutta scheme which uses a TVD
flux-interpolation scheme to control spurious oscillations. Consider
the integral form of the classical Euler equations (eq.
[\ref{eq:fluxform}] with $\bS=0$) in one dimension and for a single
conserved fluid quantity so that $\bq(\bx,t) \equiv q(x,t)$.  We can
write this form as
\begin{equation}
\frac{\p }{ \pt} \int_{x_A}^{x_B} q(x,t) dx + \frac{\p}{\px}\int_{x_A}^{x_B} F(x,t)dx = 0. \label{eq:strongform}
\end{equation}
We discretize the $N$-dimensional spatial domain into a uniform\-ly
spaced grid of points, $\bx_{\bi}$, defined to be the centers of
uniform\-ly packed rectangular $N$-volumes (cells).  Using this simple
dis\-cre\-tization, fluid quantities defined at cell-centered grid
points may be interpreted as the cell-averaged value of the solution
vector for that grid point.  For a single one-dimensional cell at
location $x_i$ and with boundaries at $x_A = x_{i-1/2}$ and $x_B =
x_{i+1/2}$, the integrals $\int_{x_A}^{x_B}qdx$ and
$\int_{x_A}^{x_B}F(x,t)dx$ in equation (\ref{eq:strongform}) represent
the cell-averaged fluid quantities $q_i$ and $F_i$.  Discretizing
equation (\ref{eq:strongform}) in terms of these cell-averaged
quantities yields
\begin{equation}
\frac{q^{t+\Dt}_i - q^t_i}{\Dt} + \frac{F^t_{i+1/2} - F^t_{i-1/2}}{\Dx} = 0, \label{eq:discretized}
\end{equation}
where superscripts reference the specific time step and subscripts
reference the spatial cell.  Computing $q^{t+\Dt}_i$ for any grid
cell thus requires interpolating a value for that cell's boundary fluxes
$F_{i\pm1/2}$, based on the cell-centered fluxes of neighboring cells.  

The interpolation process introduces diffusive and dispersive errors.
Diffusive errors result from excessive clipping and averaging
occurring during the reconstruction process, resulting in the smearing
of an initially sharp profile. Such errors are unavoidable in
computational codes, but can be minimized.  Dispersive errors result
from spurious over- and undershoots occurring during the
reconstruction process.  They result in ringing-type oscillations
occurring near sharp discontinuities.  We control these latter errors
by using a method whose total variation decreases at each successive
time step.  In order to update $q_i^{t+\Delta t}$ in equation
(\ref{eq:discretized}), we use a second-order Runge-Kutta method: for
the half-timestep we interpolate the fluxes at the boundaries using 
the first-order upwind method; for the the full time step, we use a
second-order flux-limited correction to the upwind method.  These
details are described below. 

\subsection{TVD schemes} \label{subsec:tvdschemes}

\citet{harten83} introduced the Total Variation Diminishing (TVD) 
condition as a nonlinear stability condition to ensure mo\-notonicity
preservation.  The \textit{total variation} of an ordered series of n
points, ${q_i}$ at a given time step, may be defined as
\begin{equation}
TV(q^t) = |q_n^t-q_1^t| + 2(\sum q_{max}^t - \sum  q_{min}^t). \label{eq:tvddef2}
\end{equation}
The variables $q_{max}^t$ and $q_{min}^t$ refer to local maxima and
minima in the set.  Spurious oscillations increase the number of
extrema and thus increase the total variation.  A solution is said to
be TVD stable if its total variation at a given time step is less than
or equal to that at the previous time step.  A flux-assignment scheme
to interpolate the fluxes at the boundaries of cells may be similarly
designated as TVD if the total variation of the assigned fluxes
decreases with each successive time step.

The only linear flux-assignment schemes that are TVD are upwind methods
\citep{godunov}.  They assign flux based on the direction in 
which fluid is advecting by assuming that all of the flux at a given
cell boundary comes from the cell upwind of the boundary location.
Considering a one-dimensional flow where flow to the right (larger
indices) is positive, a simple upwind scheme can be described by 
\begin{alignat}{2}
     F_{i+1/2}^U & = F_i,     & \hspace{1pc} u_i     & > 0 \label{eq:upwindright} \\ 
     F_{i+1/2}^U & = F_{i+1},  & \hspace{1pc} u_{i+1} & < 0. \label{eq:upwindleft} 
 \end{alignat}
Equation (\ref{eq:upwindright}) stipulates that for flow to the right,
the flux at the rightward boundary of a given cell is assigned to be
the value at the cell's center, that is, the flux as defined just
upwind of the boundary location.  Equation (\ref{eq:upwindleft})
stipulates that for flow to the left, the flux at the leftward
boundary of a given cell is assigned to be the value at the cell's
center, again, the flux upwind of the boundary location.

It is possible to improve upon the above first-order upwind scheme by 
considering second-order corrections to the assigned fluxes. For flow
to the right there are two neighboring second-order corrections to
equation (\ref{eq:upwindright}):
\begin{equation}
  \begin{aligned}
     \DF_{i+1/2}^L & = \frac{F_{i} - F_{i-1}}{2} \\ 
     \DF_{i+1/2}^R & = \frac{F_{i+1}-F_{i}}{2}.
   \end{aligned}  
\end{equation}
These two corrections consider the influence of flux from the cells
further to the left and to the right of the $i$th (upwind) cell.  In a
similar manner, for flow to the left there are two neighboring
second-order corrections to equation (\ref{eq:upwindleft}) that 
consider the influence of flux from cells to the left and right of the
($i$+1)-th (upwind) cell: 
\begin{equation}
 \begin{aligned}
   \DF_{i+1/2}^L & = \frac{F_{i+1} - F_{i}}{2} \\ 
   \DF_{i+1/2}^R & = \frac{F_{i+2}-F_{i+1}}{2}.
 \end{aligned}   
\end{equation}

To determine the actual value of the correction to the pure\-ly upwind
flux, we apply a \textit{flux limiter} $\phi(\DF^L,\DF^R)$, which 
specifies the relative weight of the two corrections.  For a given
limiter, the second-order boundary flux used for the full Runge-Kutta
time step can be written as $F_{i+1/2}^{U} + \DF_{i+1/2}$, where
$\DF_{i+1/2} = \phi(\DF_{i+1/2}^L,\DF_{i+1/2}^R)$.  

We consider four established limiters: Minmod, Van Leer,
Monotonized Central-difference (MC), and Superbee limiters, all designed to
satisfy the TVD condition (eq. [\ref{eq:tvddef2}]), as well as a new
scheme.  These limiters are defined in terms of the leftward and
rightward corrections in numerous sources; see \citet{leveque}, for
example.   

\begin{figure}
\begin{center}
\resizebox{0.45\textwidth}{0.45\textwidth}{\includegraphics{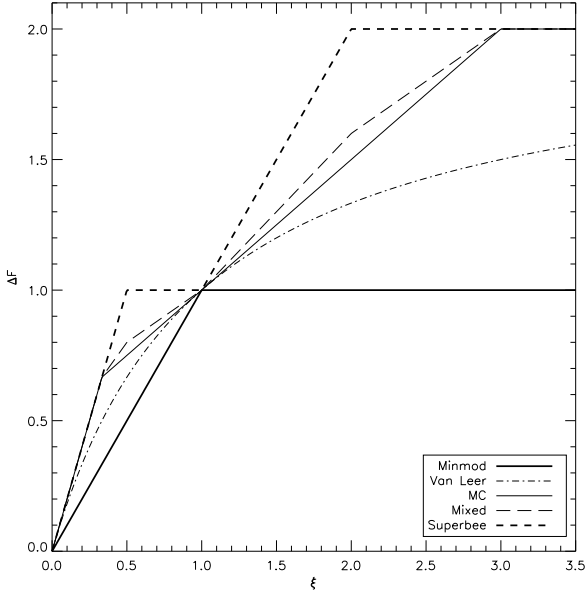}}
\end{center}
\caption{ \label{fig:limiters}
  Flux corrections for various limiters: Minmod (heavy,
  solid), Van Leer (dot-dashed), MC (fine, solid), mixed (fine, dashed),
  and Superbee (heavy, dashed).}
\end{figure}

In Figure \ref{fig:limiters} we portray these limiters graphically as
a function of flux ratio $\xi=\DF^L/\DF^R$ for the corresponding
magnitude of the rightward flux correction $\DF = \DF^R$.  Note that
all the above limiters are zero when the flux corrections are of
opposite sign (when $\xi$ is negative) as occurs near an
extremum. This feature prevents growth of the extremum and ensures the
interpolation remains TVD. They are also all symmetric under exchange
of $\DF^L$ and $\DF^R$ (in which case, 
$\xi\rightarrow \DF^R/\DF^L$ and $\DF \rightarrow \DF^L$).

The region defined by $\DF < \max(0,\min(2,2\xi))$ satisfies
non-linear stability conditions determined to be in the general
class of TVD-stable limiters \citep{Sweby}.  Limiters that satisfy
the above condition and that are also second-order accurate are
found within the area bounded by the Superbee and Minmod limiters.

The Minmod limiter is the most diffusive because it always takes
the minimum value of the possible second-order corrections; thus,
any flux not assigned by the second-order reconstruction ends up
being smeared out over more than one grid cell.  The Van Leer, 
MC, and mixed limiters (see below) are progressively less
diffusive, as they assign more and more of the possible flux
correction to a definite cell.  Superbee is the least diffusive
second-order limiter possible but at the cost of increased
instability. These tradeoffs are discussed further in Section
\ref{sec:basictests}.

Also shown on the graph is the mixed limiter we designed, which
sometimes exhibits a better compromise between stability and
higher-order accuracy.  The mixed limiter is a normalized linear
combination of the MC and Superbee schemes.  In practice we usually
weight the scheme as 80\% MC and 20\% Superbee as drawn in Figure
\ref{fig:limiters}.

\subsection{Operator splitting} \label{subsec:operatorsplitting}

The above description suffices to solve the Euler equations
in one dimension. Multiple dimensions and additional source terms
present in the full Navier-Stokes equations are accounted for by using
the operator splitting technique of \citet{strang}.  A full time step
is performed as a double sweep through an ordered sequence of
operators comprising the full equation, first in forward sequence,
then in reverse.  To illustrate this process for a single double
sweep, we write equation (\ref{eq:fluxform}) using operators as
\begin{equation}
\frac{\p q_j}{\pt} + \displaystyle\sum_i^{N_D} L_i[q_j] - S_{g,\phi}[q_j] - V[q_j] = 0,
  \label{eq:operatorform}
\end{equation}
where $N_D$ is the number of physical dimensions being modeling.  The
operators $L_i[q_j]$ represent the update of $q_j$ in a single
direction due to advective terms $\sum_i^{N_D}u_i \p q_j/\px_i$,
performed by solving the relaxation system with the \hspace{0.1em}Runge-Kutta/TVD 
scheme outlined above.  The operators $S_{g,\phi}$ and $V$ represent
additional routines which differ numerically from the TVD algorithm.
They account for source terms (due to both gravitational potentials
and cylindrical geometry) and for viscosity, respectively.  We have
lumped the source terms due to gravity and geometry into the same
operator as they are both accounted for in the same subroutine.  This
subroutine updates the solution vector due to the source forcing using
a second-order Runge-Kutta routine.  The update of the solution vector
due to the viscosity is performed as a direct second-order accurate
difference of the stress tensor (see \S
\ref{subsec:viscosityimplementation} below). Updates of the solution
vector accounting for each of the above operators are performed in the
sequence  
\begin{equation}
q^{t+\Dt}_j = V S_{g,\phi} L_3 L_2 L_1[q_j].
\end{equation}
A second sweep is then performed using the same time step $\Dt$ to
yield the completely updated solution
\begin{equation}
q^{t+2\Dt}_j = L_1 L_2 L_3 S_{g,\phi} V V S_{g,\phi} L_3 L_2 L_1[q_j].
\end{equation}
As discussed in \citet{strang}, this procedure ensures second-order
accuracy.

\subsection{Implementation of viscosity}  \label{subsec:viscosityimplementation}

Viscosity is implemented by updating the fluid quantities due to the 
viscosity operator equation $\p q_j/\pt = V[q_j]$, where $V$ is
written out explicitly in the appendix in terms of the stress-strain
tensor $\sigma_{ij}$.  Considering the update of the radial momentum
component of the solution vector in two dimensions as an example, we
have   
\begin{equation}
\frac{\p q_2}{\pt} = \frac{1}{r}\frac{\p r\sigma_{rr}}{\pr} +
  \frac{1}{r}\frac{\p \sigma_{\q r}}{\p\q} - \frac{\sigma_{\q\q}}{r}, 
\end{equation}
which upon substituting $\sigma_{ij}$ becomes 
\begin{align}
\begin{split}
\frac{\p q_2}{\pt} &= \frac{1}{r}\frac{\p}{\pr} \left [2\rho r \nu 
  \left ( \frac{\p u_r}{\p r} - \frac{1}{3}\left( \frac{1}{r}\frac{\p ru_r}{\pr} +
  \frac{1}{r}\frac{\p \uq}{\pq}\right ) \right ) \right ] \\
  &+ \frac{1}{r}\frac{\p }{\p\q} \left [ \rho \nu \left ( 
    \frac{1}{r}\frac{\p u_r}{\pq} + \frac{\p \uq}{\pr} - \frac{\uq}{r}
    \right )\right] \\
  &- \left [\frac{2\rho}{r} \nu \left (\frac{1}{r}\frac{\uq}{\pq} 
   + \frac{u_r}{r} - \frac{1}{3}\left(\frac{1}{r}\frac{\p ru_r}{\pr} 
   + \frac{1}{r}\frac{\p\uq}{\pq}\right) \right) \right ].  \label{eq:viscosity}
\end{split}
\end{align}
We implement the density and velocity derivatives which result on the
right-hand side in the above equation as second-order accurate finite
differences.  We thereby include all components of the viscous tensor
in calculations with added physical viscosity (where $\nu \ne 0$).   

\subsection{Alterations for cylindrical grids}  \label{subsec:cylindricalgrids}

We can express the equations listed in the Appendix (for zero
viscosity and no gravitational potential) in a form similar
to equation (\ref{eq:fluxform}) as 
\begin{equation}
  \frac{\p \bq}{\pt} + \frac{1}{r}\frac{\p r \bF_r}{\pr} +
  \frac{1}{r}\frac{\p \bF_\q}{\p \q} + \frac{\p \bF_z}{\p z} = \bS,
  \label{eq:formr1}
\end{equation}
where $\bS=(0,\rho u_\q^2/r + p/r,0, 0,0)$.  The discretization of
equation (\ref{eq:formr1}) is computationally equivalent to that of
the Euler equations on a Cartesian grid if we use $(\Dx_1,\Delta
x_2,\Delta x_3) = (r_i \Delta r$, $r_i \Delta \q$,$\Delta z)$ but for the
source term in the $q_2$-equation and the extra r-multiplier in front
of the radial flux.  We account for the source term using a
second-order Runge-Kutta scheme implemented with operator splitting as
discussed in Section \ref{subsec:operatorsplitting}.  The extra
r-multiplier is included when solving in the radial direction;
equivalently stated, while the azimuthal and vertical advection
operators act directly on the solution vector as $L_\q [\bq]$ and $L_z
[\bq]$, the radial advection operator acts on the solution vector
scaled by r as $L_r [r \bq]$. 

\subsection{Time step restrictions} \label{subsec:timesteprestrictions}

\textit{Courant-Friedrichs-Lewy} (CFL) conditions are imposed to
insure that numerical information does not propagate at physically
unrealistic speeds.  In practise one limits the value used for the
time step so that waves travel less than one grid cell per time step. 

The one dimensional Euler equations support three types of waves:
entropy waves which move at the fluid's flow speed $u$, and two types
of acoustic waves which travel ``rightward'' and ``leftward'' at the
speed of sound relative to the flow speed $u+c_s$ and $u-c_s$,
respectively.  In a given cell, the maximum and minimum values of
these three speeds determines the speed at which information can
travel to the right and to the left, respectively.  More generally one
can limit the time step due to the two global speed extrema. These 
two speeds are equivalent to the largest and smallest eigenvalue of
the flux Jacobian $\p \bF(\bq)/\p \bq$ as demonstrated in \citet{laney98}.
The time step may be further limited by other physical restrictions
such as the CFL condition imposed by viscosity.  Implementation of the
fast-advection algorithm (\S \ref{subsec:FARGOalgorithm}) places a
further restriction on the time step in order to ensure that
differential rotation does not cause two adjacent annuli of fluid to
shear past one another by more than a half grid cell.

Given these three restrictions, we determine our time step to be the
global minimum from the entire grid of values, determined for each
cell as $\Dt = (\Dt_f^{-2} + \Dt_{vis}^{-2} + \Dt_{sh}^{-2})^{-1/2}$, 
where $\Dt_f = \min[\Delta r/(|u_r|+c_s)$, $r\Delta \q/(|u_\q|+c_s)$, $\Delta
 z/(|u_z|+c_s)]$, $\Dt_{vis} = \min[\Delta r^2/4\nu$, $(r\Delta
\q)^2/4\nu$, $\Delta z^2/4\nu]$ and $\Dt_{sh} =(1/2)(\p\omega/\pq)^{-1}$.  
In simulations of protoplanetary disks, the limitation imposed by 
acoustic waves is almost always the most restrictive.

\subsection{FARGO algorithm} \label{subsec:FARGOalgorithm}

Accretion disks with dominant central masses have velocity profiles
that are nearly Keplerian.  In particular the azimuthal velocity
is dominated by the Keplerian value, which is azimuthally uniform at a
given radius. We adopt the fast advection algorithm described by
\citet{masset00}, in order to increase algorithm efficiency in the
presence of such nearly azimuthally uniform background flows. The
algorithm's underlying strategy is to subtract off the bulk background
flow, which can be considered simply a translation of grid quantities
in the angular direction, leaving the dynamically important residual
velocity.  

In decomposing the velocity, the cylindrical grid is broken up into a
series of annuli, and an averaged background velocity in the azimuthal
direction $u_{_{\rm AVG}}(r)$ is calculated for each annulus.  The first
velocity component corresponds to the residual amount
$u_{_{\rm RES}}(r,\q)$, by which the total velocity differs from its 
azimuthally averaged value $u_{_{\rm AVG}}(r)$.  The averaged background
velocity is further decomposed as $u_{_{\rm AVG}}(r) = u_{_{\rm SH}}(r) +
u_{_{\rm CR}}(r)$.  The former component is constructed to correspond to
the largest possible whole-number shift of grid cells in the azimuthal
direction and the latter to the remaining partial-cell shift.  Neither
of these components depend on the angular variable.  The whole-number
shift is rounded to the nearest integer so that the partial-cell shift
may be positive or negative, but will always correspond to a shift
magnitude less than or equal to half a grid cell.  The total velocity
at any cell on the grid is then given as
\begin{equation}
  u(r,\q) = u_{_{\rm SH}}(r) + u_{_{\rm CR}}(r) + u_{_{\rm RES}}(r,\q).
\end{equation}

The transport of fluid quantities due to the $u_{_{SH}}$ component is
easily accomplished by numerically shifting the fluid variables by the
appropriate number of cells in the azimuthal direction.  Because the
shift is integral, the process does not introduce any numerical
diffusion, nor does it limit the size of time step permitted by the
CFL conditions.

Transport of a fluid quantity along an annulus due to the partial-cell
velocity component, which is independent of the azimuthal grid cell
number, may be accomplished by interpolating the function and
redetermining the interpolation at a shifted location.  This process
is not unstable; therefore, it does not reduce the CFL-allowed time
step, but it does introduce diffusion as peaks in the interpolated
quantity are shifted by fractions of a grid cell and must then be
redistributed among more than one cell.

Finally, transport of fluid quantities due to the residual a\-zi\-mu\-thal
component is performed using the relaxing TVD algorithm, just as for
the radial velocity sweep.  This transport step contributes to the
numerical viscosity and also lowers the size of the time step allowed
for stability.  However, if the flow is nearly azimuthally uniform,
the residual velocity will be small compared to the average velocity
at a given radius, and the corresponding time step will be much larger
than that otherwise permitted by the full azimuthal velocity.  In
practise, the allowed time step is increased by a factor of $5-10$
times that allowed without removing the background flow. 

\subsection{Boundary treatments} \label{subsec:bc}

Special boundary conditions are only required in the non-azi\-muthal
directions.  The azimuth is treated as periodic by directly mapping
the ($N_{\q}$+1)-th cell to the 1st cell when calculating the fluxes.
   
In non-azimuthal directions, we use $n_b = 2^{N_D}$ ghost cells on the
inner and outer edges of the computational domain.  This number of
additional cells prevents the effects of lower-order flux
interpolations occurring at the first and last cells from propagating
in towards the center of the domain.  After the solution quantities
are updated at the end of each double sweep, the values of these
boundary cells are redetermined, depending on the physical effect
being modeled.  We consider three treatments here.  The first simply
re-initializes the quantities in the ghost cells to the initial
conditions or some known, prescribed solution.  The second re-assigns
the cell values according to
\begin{alignat}{2}
&  q_{(n_b-i)}       = w q_{(n_b+i+1)},   & \hspace{1pc} i &= 0,n_b-1  \nonumber \\ 
&  q_{(N - n_b+i+1)} = w q_{(N - n_b-i)}, & \hspace{1pc} i &= 0,n_b-1,
\end{alignat}
where $w=1$ for all variables except the radial and vertical
velocities for which $w=-1$.  This boundary treatment approximates
reflecting boundary conditions for which all scalar variables are
symmetric around the boundary while vector variables are
anti-symmetric. The third treatment re-assigns the cell values
according to the prescription
\begin{alignat}{2}
& q_{(i+1)}        = q_{(n_b+i+1)},      & \hspace{1pc} i &= 0,n_b-1 \nonumber  \\ 
& q_{(N - n_b+i+1)}= q_{(N - 2n_b+i+1)}, & \hspace{1pc} i &= 0,n_b-1,
\end{alignat}{2}
for all variables.  This treatment approximates a free-streaming
outflow boundary.

In addition to the boundary treatments discussed above, we sometimes
implement wave-damping conditions near the boundaries, but still
inside the solution domain proper.  These wave-damping conditions are
described by
\begin{equation}
  \bq(\br,t) = \bq(\br,0) +
     [\bq(\br,t)-\bq(\br,0)]e^{|\br-\br_b|t/(\tau R_0)} .
\end{equation}
Such a treatment damps any perturbations about the initial equilibrium
solution that are within the distance $\br_b$ of the boundary, on a
spatial scale $R_0$ and on a time scale $\tau$.

\subsection{Parallelization} \label{subsec:parallelization}

The RAPID code may be parallelized for multi-processor 
machines with both shared memory or distributed memory. We have already
implemented OpenMP directives for shared-mem\-ory machines using data
parallelism.  Because all the processors on such shared memory machines
have access to all the variable arrays, parallelization on such
machines is relatively straightforward. OpenMP \textit{parallel do}
directives are used for any loop-intensive subroutines which operate
on the entire solution array at least once per timestep.  At the
beginning of each such subroutine, the task of updating the solution
array is effectively split into several threads, each of which
operates on a portion of the entire data structure. Each thread is
handled by an available processor and each thread receives its own
local copy of the data required to be updated.  Once all the threads
have updated their portion of the entire data structure, the code is
effectively unsplit, individual pieces of the updated data are
collated together, and the code again proceeds sequentially in an
unbranched manner. Because memory is shared amongst all the
processors, much of the detail of the distribution process is handled
by the OpenMP directives, themselves.   

While not presently implemented as such, the RAPID code is also able to be
parallelized on distributed memory machines using Message Passing
Interface (MPI) protocols.  This type of parallelization is somewhat
more involved as more of the distribution details need to be
explicitly specified.   Typical strategies for this process would
involve splitting the disk into different annular regions with
overlapping boundaries.  The entire solution process for each region
of data is then handled by a separate machine with its own memory.
Because the data is permenantly split among different machines, it is
necessary to communicate updated values of neighboring data cells
between different machines.  Furthermore, the timing of tasks performed
on each of the machines must also be controlled to ensure blocks of
code are performed in the correct sequence.  If each machine (as 
distinguished by separate memory storage) has multiple processors, it 
is possible to implement both MPI and OpemMP directives.  In the
future we intend to implement such a combined MPI/OpenMP
parallelization of the code.

\section{Basic hydrodynamic tests} \label{sec:basictests}

We perform a suite of three hydrodynamic tests: a two-di\-mensional
oblique shock at three angles, a Kelvin-Helmholtz (KH)
instability test and a cylindrical bow-shock test.  The first two
tests are performed in Cartesian coordinates and the last test is
performed in cylindrical coordinates. An adiabatic equation of state
with $\gamma=5/3$ is assumed for all three tests.  The results of these tests are
compared to those from the piece-wise parabolic method
\citep[PPM,][]{colella84}, as implemented in
\textit{VH-1}$^*$\symbolfootnote[0]{\hspace{-1em}$*$ \hspace{0.15em} \href{http://wonka.physics.ncsu.edu/pub/VH-1/}{http://wonka.physics.ncsu.edu/pub/VH-1/}}, 
and when possible with analytical solutions. When referring to results
from the RAPID code, the limiter used for the simulation will
be placed in parentheses.  Due to the number and range of parameters
we wish to explore, the tests presented in this paper are only
two-dimensional.  Various three-dimensional tests of the TVD
algorithm have been performed and a three-dimensional test of a
Sedov-Taylor blast wave may be found in \citet{trac03}.

\subsection{Two-dimensional oblique shock} \label{subsec:hydrotests.2}

The two-dimensional oblique shock is a version of the one-dimensional
Sod shock tube \citep[see][for example]{landau87} set-up in a
two-dimensional box at an angle to the box boundaries. In the
one-dimensional version of a Sod shock, a jump discontinuity is
initialized between two regions of fluid with an initial relative
velocity by requiring that the pressure and density of two regions of
fluid initially be disparate (say, separated by a membrane).  In the
two-dimensional oblique case, the fact that the initial density and
pressure discontinuities are set across the grid at an angle causes
the resulting shock front to propagate obliquely across the box.
Results of the TVD algorithm in the one dimensional case are described
in \citet{trac03}. Here we implement the shock in a two-dimensional
setup with initial conditions given by 
\begin{align}
  \rho  & = \left \{ 
            \begin{alignedat}{2}
               &1,      & \hspace{1pc} x < x_0 \\ 
               &0.125,  & \hspace{1pc} x \ge x_0
            \end{alignedat}  
            \right. \\
  p     & = \left \{
            \begin{alignedat}{2}
               &1,      & \hspace{2pc} x < x_0 \\  
               &0.1,    & \hspace{2pc} x \ge x_0
            \end{alignedat} 
            \right.
\end{align}
with no initial relative velocity.  For a shock with an angle of
$45^{\circ}$, we set $x_0=0.25$ along the vertical and horizontal box
boundaries so that $x_0=\sqrt{2}(0.25)$ along the central diagonal.
A resolution of $N_x \times N_y=180\times 180$ was used.

\begin{figure}
\begin{center}
\resizebox{0.47\textwidth}{0.43\textwidth}{\includegraphics{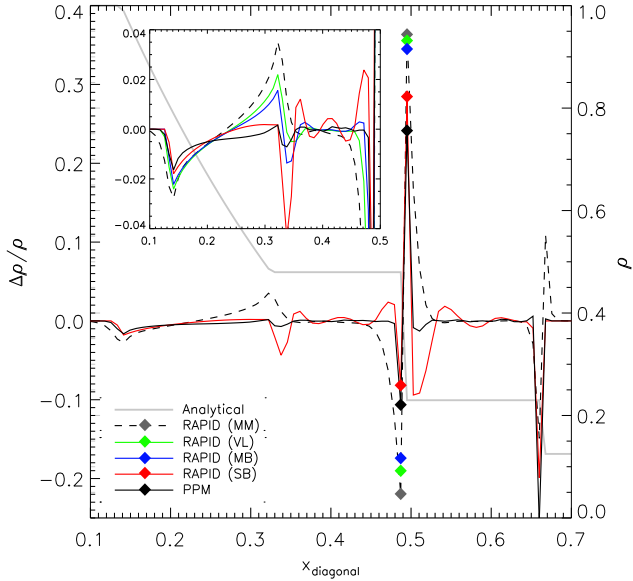}}
\end{center}
\caption{\label{fig:sod.diff} Normalized density differences between
  numerical and analytical solutions for the PPM (solid black), RAPID
  (SB, solid red) and RAPID (MM, dashed black) codes.  Density
  differences (dark lines) are measured by the scale on the left.  The 
  analytical solution is shown in the solid grey line, measured by the
  scale on the right.  Colored diamonds indicate the maximum and
  minimum differences from the analytical solution in the vicinity of
  the contact discontinuity (near $x_{\rm diagonal}=0.5$) for the above
  codes, as well as for the RAPID code with Van Leer and mixed
  limiters. Inset: density differences for all codes over a limited region. }  
\end{figure}

Figure \ref{fig:sod.diff} shows, for the above setup, normalized
\textit{differences} between the computed density and that of the 
analytical solution.  Results are shown for the PPM code as well as for
RAPID with four different choices of limiters. All results are
one-dimen\-sion\-al slices along the central diagonal of the box in the
shock propagation direction.  For clarity, only the Superbee and
Minmod limiters are shown over the entire diagonal extent. Differences
between the computed and analytical pressure show analogous results. 

The PPM code does better than any of the RAPID runs at minimizing the
diffusion near shock fronts and discontinuities.  The diffusion at
these points varies considerably for the RAPID code depending on the
choice of limiter.  Results from the Superbee (SB) limiter display the
least diffusion, and are close to the PPM results except that they
display some spurious oscillations before the shock fronts.  These
oscillations are indicative of instabilities and in practice, if the
simulation has too many shocks, the Superbee limiter can force the
size of the time step allowed by CFL conditions too small to be of
practical use.  Results using the Minmod (MM) limiter display the most
diffusion, but none of the spurious oscillations.  

The level of diffusion (measured roughly by the error incurred at
discontinuities) and dispersion (measured roughly by the amplitude of
the error oscillations) for the remaining limiters fall inbetween
those of the Superbee and Minmod limiters.  The mixed (MB) limiter
scheme (shown in the inset) has only slightly more diffusion than the
Superbee scheme and correspondingly has small\-er pre-shock oscillations
and higher stability.  In practice the mixed scheme has not forced the
time step size to be too small and has proved a good compromise
between stability and lowered diffusion.  The more diffusive MC
limiter (not shown) has almost identical results to the Van Leer (VL)
limiter. 

The PPM code does not exhibit oscillations before shock fronts because
it places further conditions on the dynamics that tend to flatten
gradients both before and after a discontinuity.  It should be noted that
while these extra conditions may increase the stability of the PPM
code, they are not necessarily physical constraints. 

\begin{figure*}
\begin{center}
\resizebox{0.9\textwidth}{0.23\textwidth}{
  \includegraphics{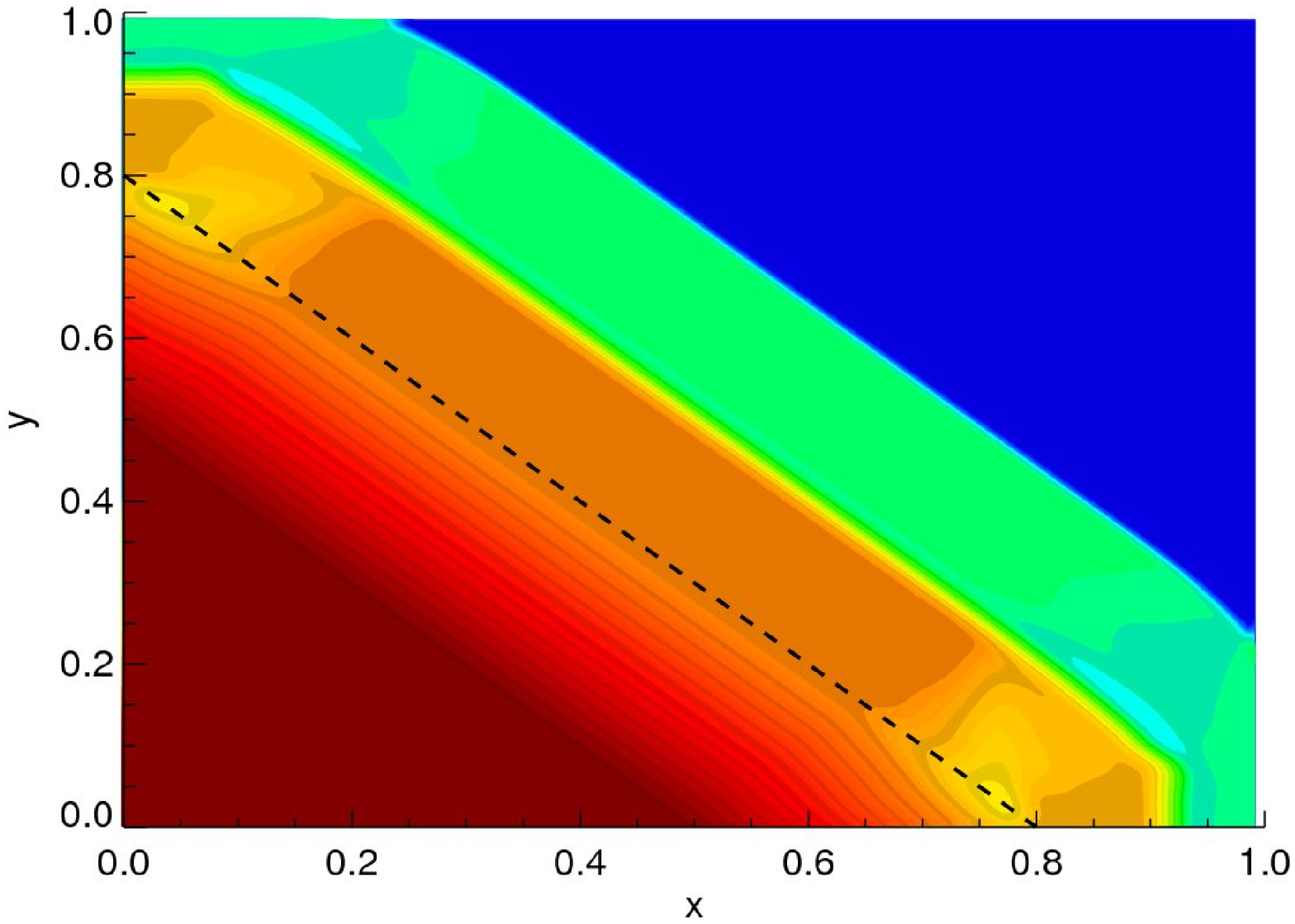}
  \includegraphics{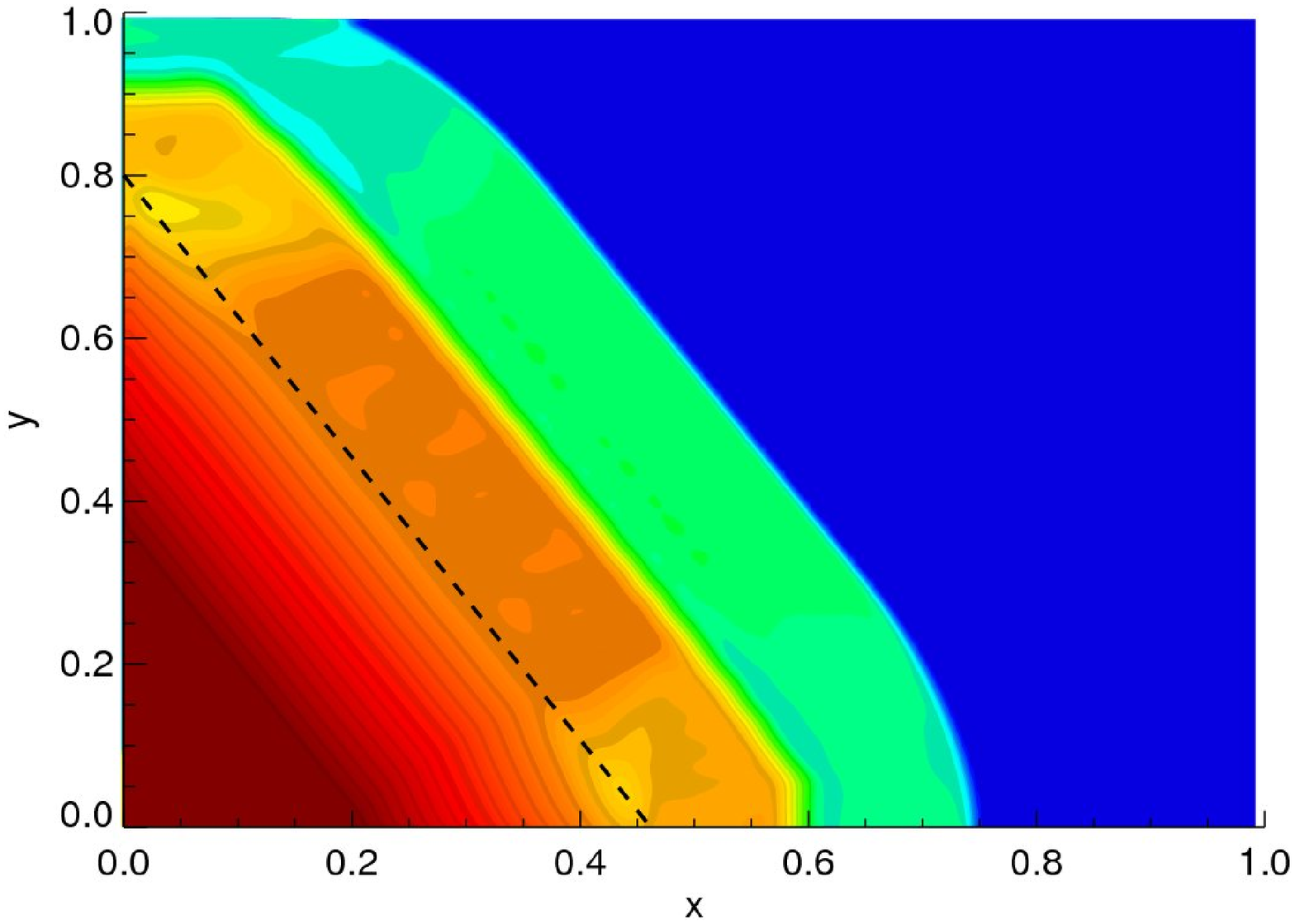}
  \includegraphics{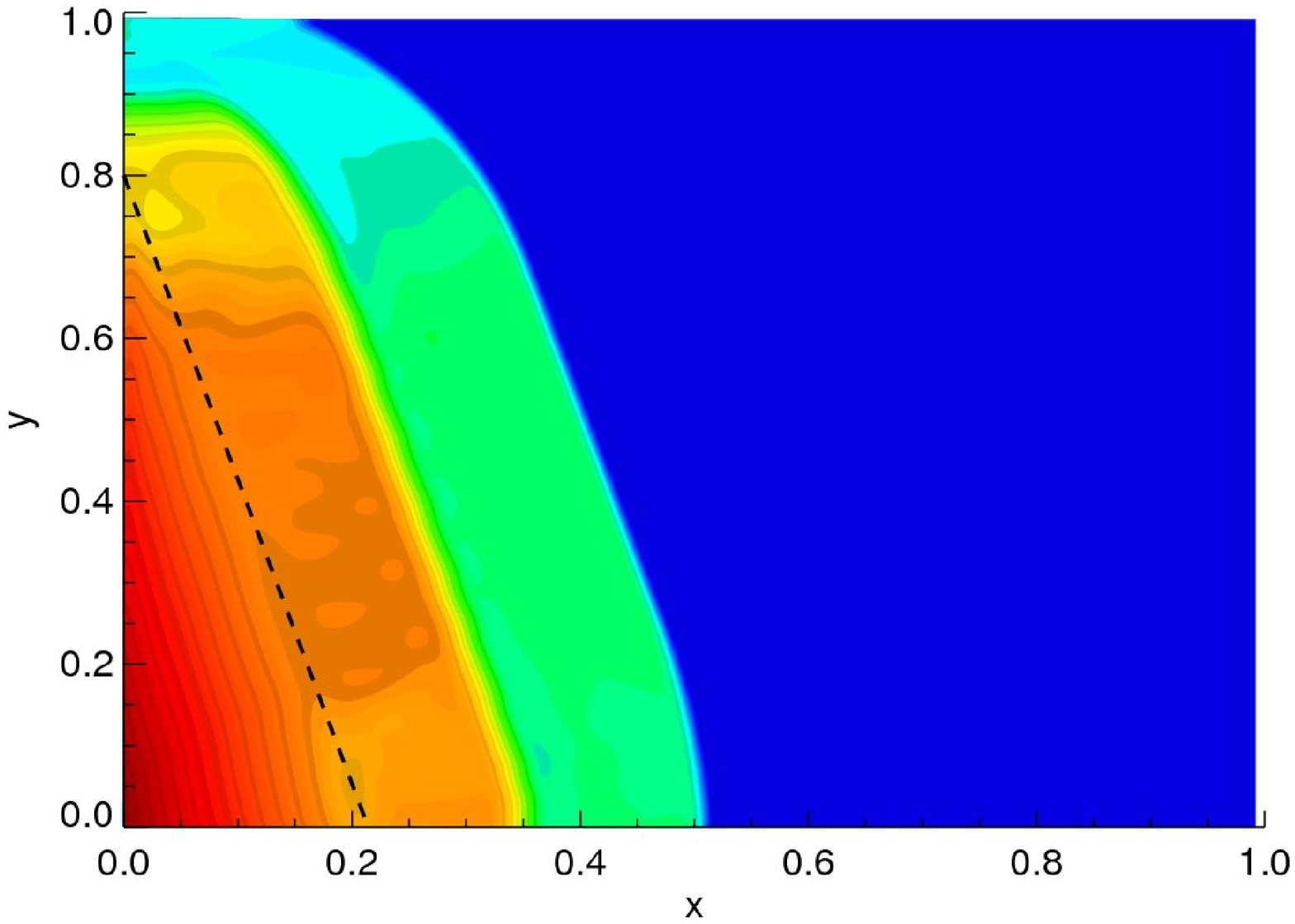}}
\resizebox{0.9\textwidth}{0.23\textwidth}{
  \includegraphics{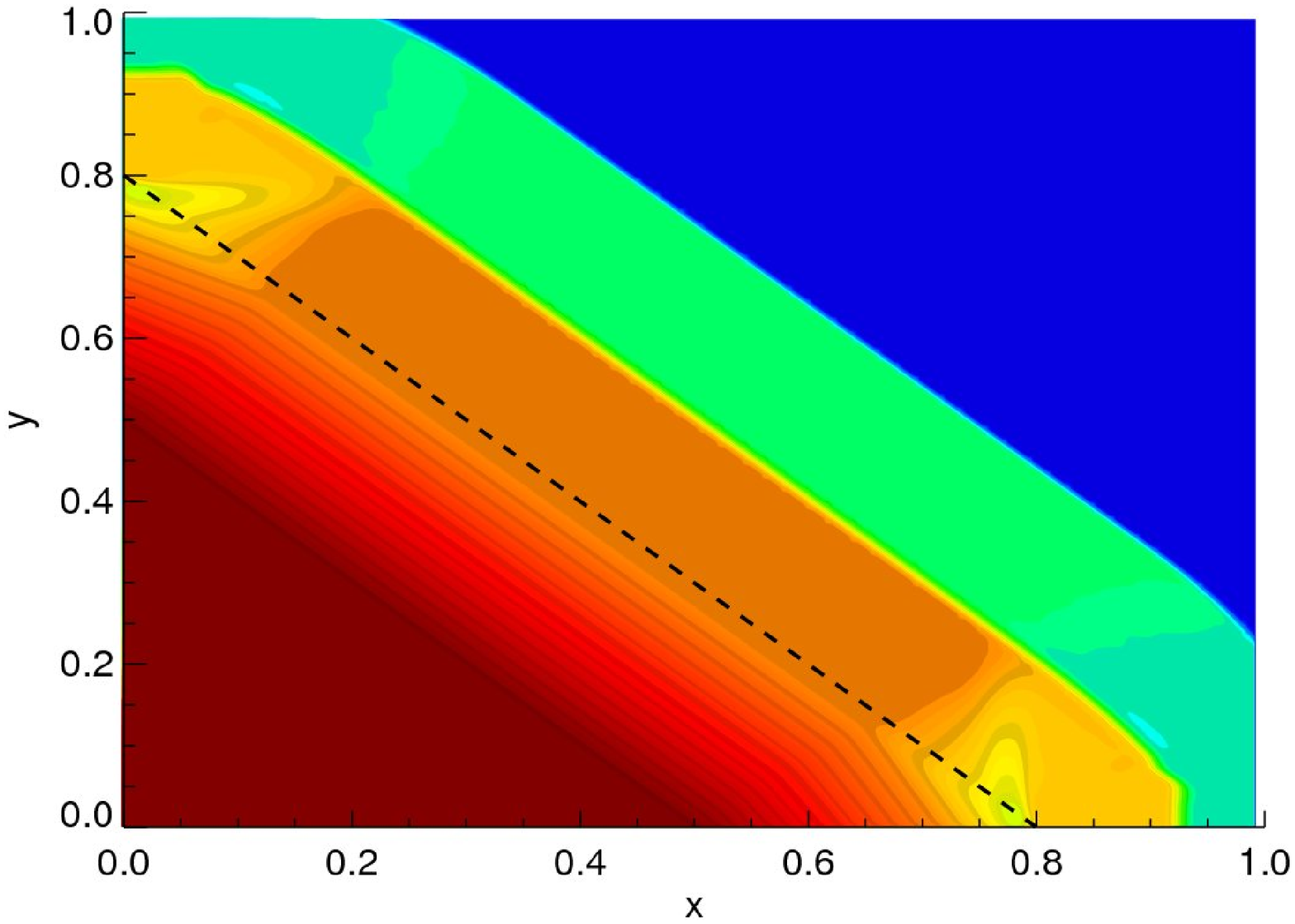}
  \includegraphics{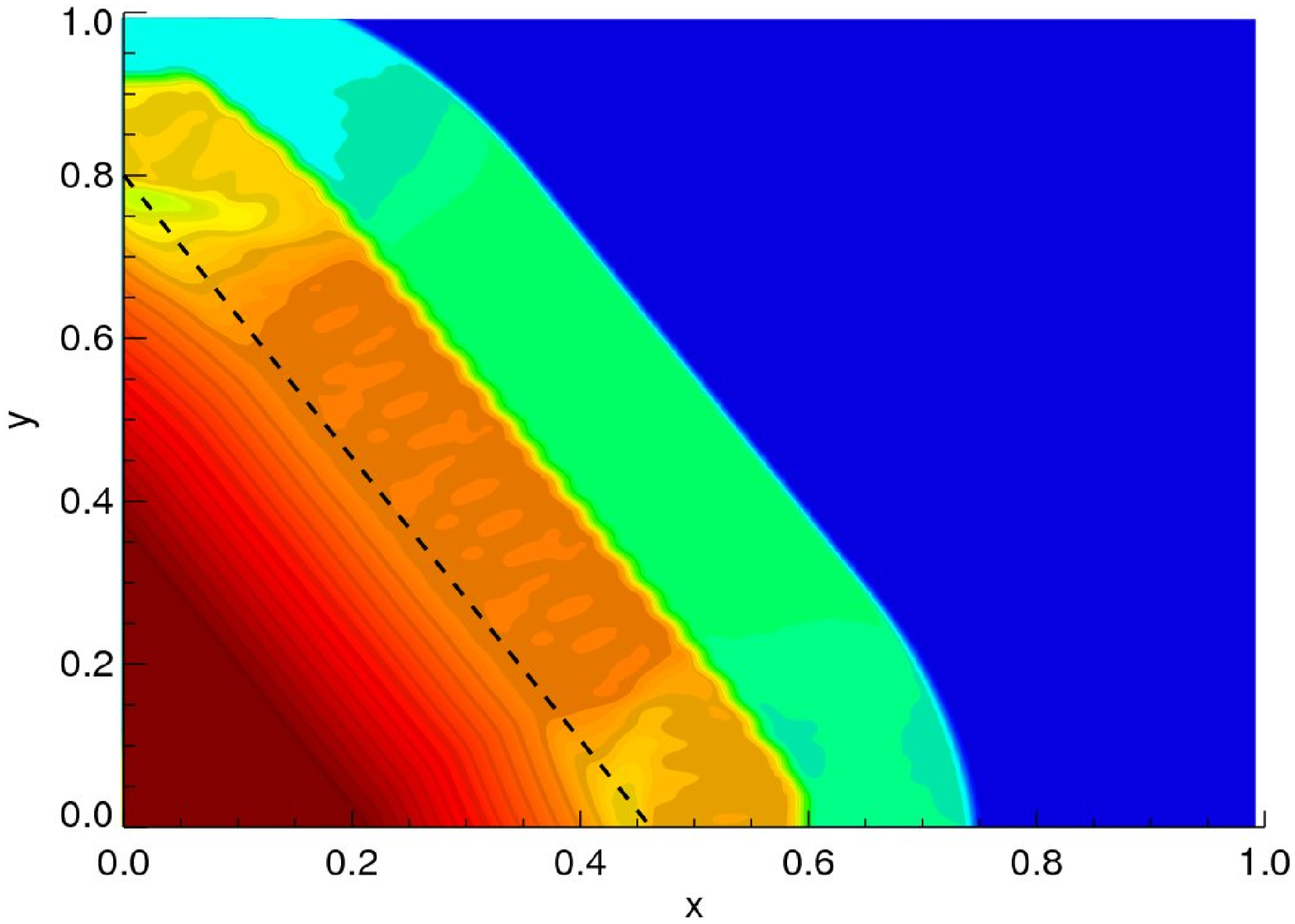}
  \includegraphics{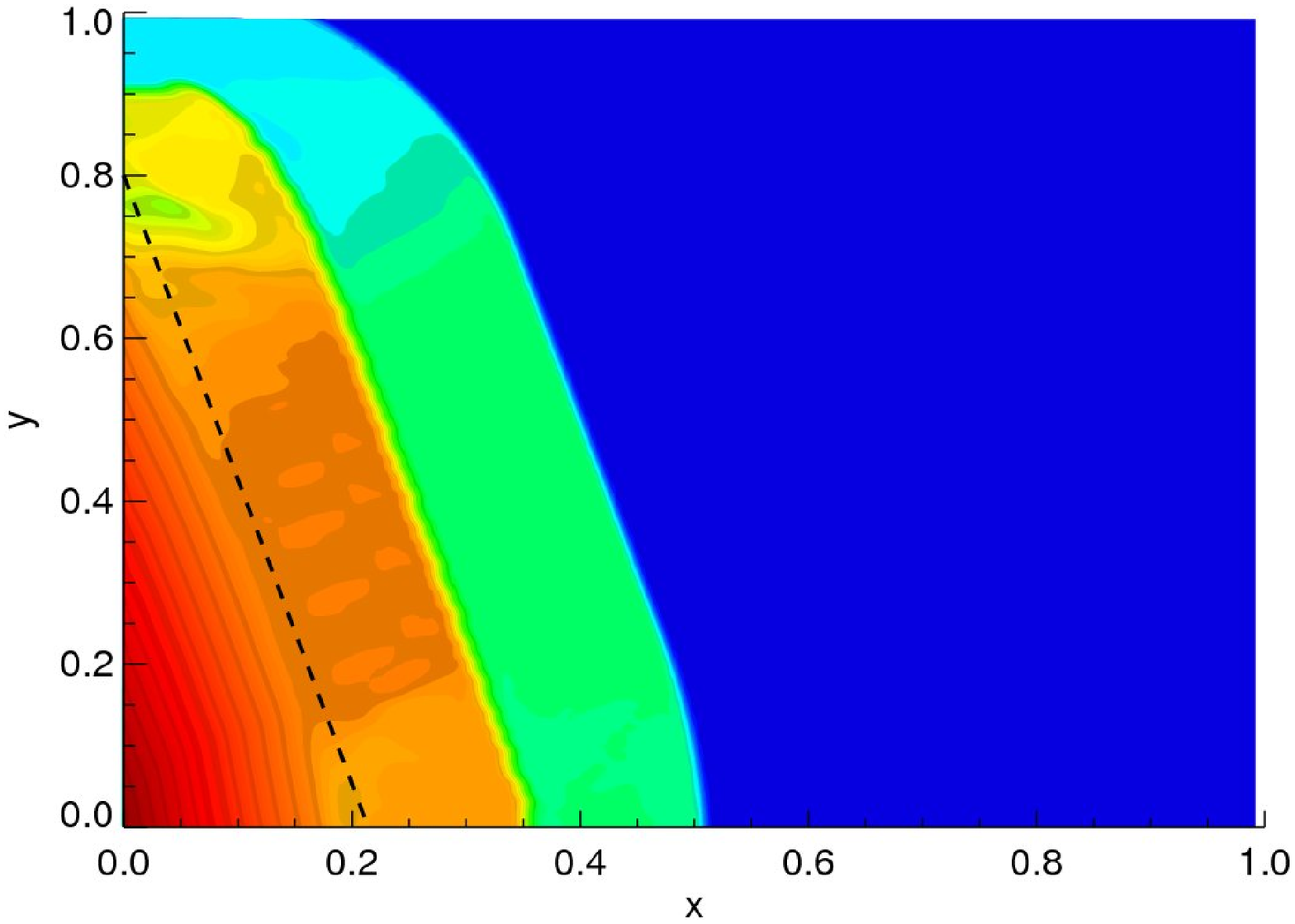}}
\includegraphics{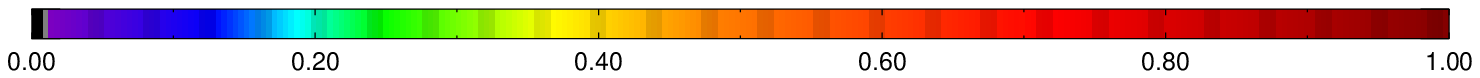}
\end{center}
\caption{\label{fig:sod.contour} Density contour plots illustrating
  propagation of an oblique shock at angles $\q=45^\circ$,
  $\q=30^\circ$, and $\q=15^\circ$ from left to right, respectively.
  Figures in the top row show RAPID(MC) results, figures in the
  bottom row, the PPM results.  The dashed lines show the initial
  location of the discontinuity.}
\end{figure*}

Because the Sod Shock Tube is propagating at an angle, it also proves
a useful examination of any differences caused by dimensional
splitting in the code.  Figure \ref{fig:sod.contour} shows
contour plots of the density from the RAPID(MC) and PPM codes run as
above except with $y_0 = 0.8$, and $x_0$ chosen as appropriate for the
desired shock angle.  For most of the length of the jump discontinuity,
the shock front is straight and propagates at the same speed along the
original diagonal.  This observation indicates that the dimensional
splitting is not leading to asymmetries.  As one gets close to the
ends of the shock front, the fluid is able to ``bleed'' away towards
the open sides (and eventually out of the box once it reaches 
the boundaries).  As a result the shock front begins to diffract at
the edges; this bending increases as the front evolves in time.  Note
that for the $30^\circ$ and $15^\circ$ cases, both codes exhibit some
type of pressure/density waves in the region trailing the contact
discontinuity.   

\subsection{Kelvin-Helmholtz instability} \label{subsec:hydrotests.3}

\begin{figure*}[!th]
\begin{center}
\resizebox{0.9\textwidth}{0.215\textwidth}{
  \includegraphics{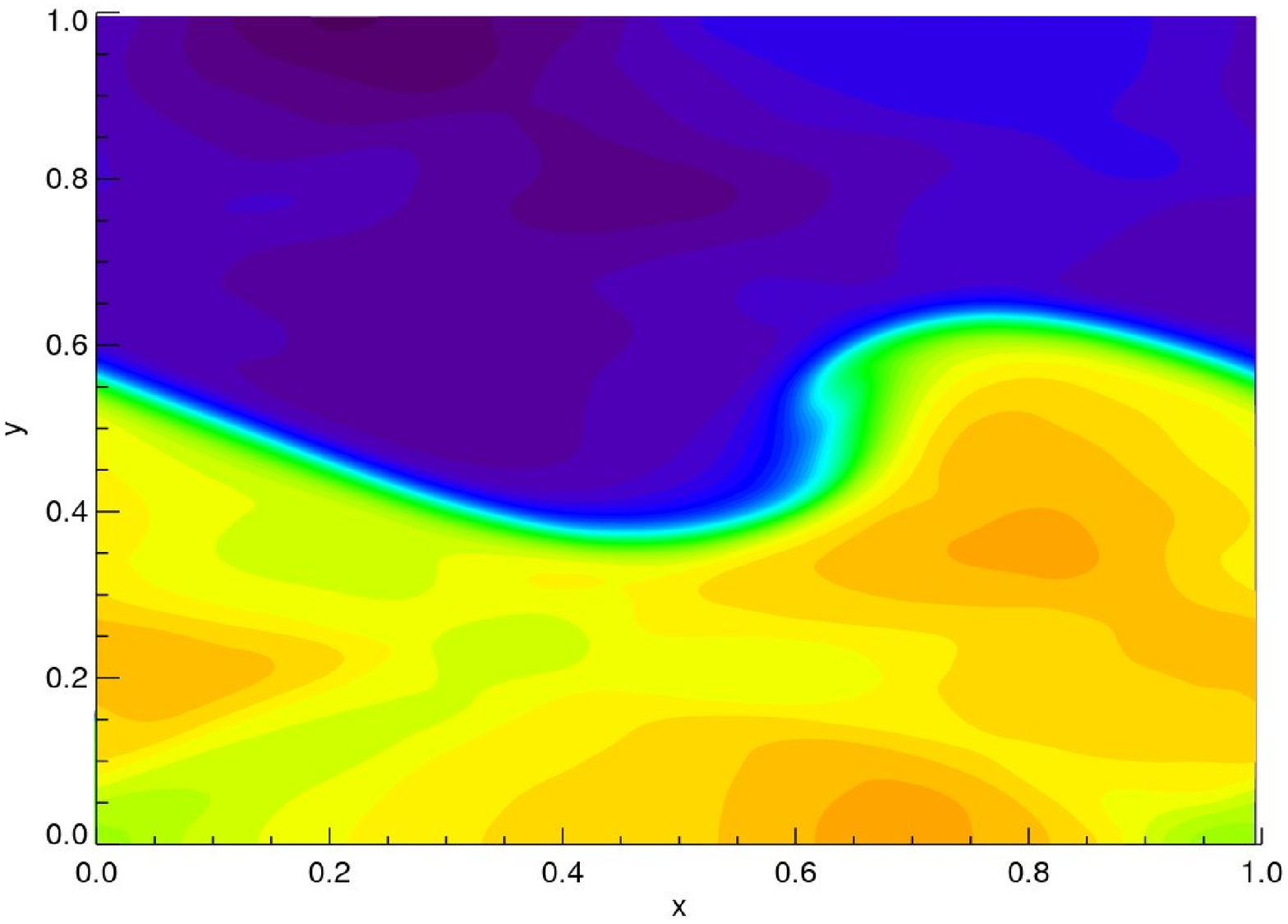}
  \includegraphics{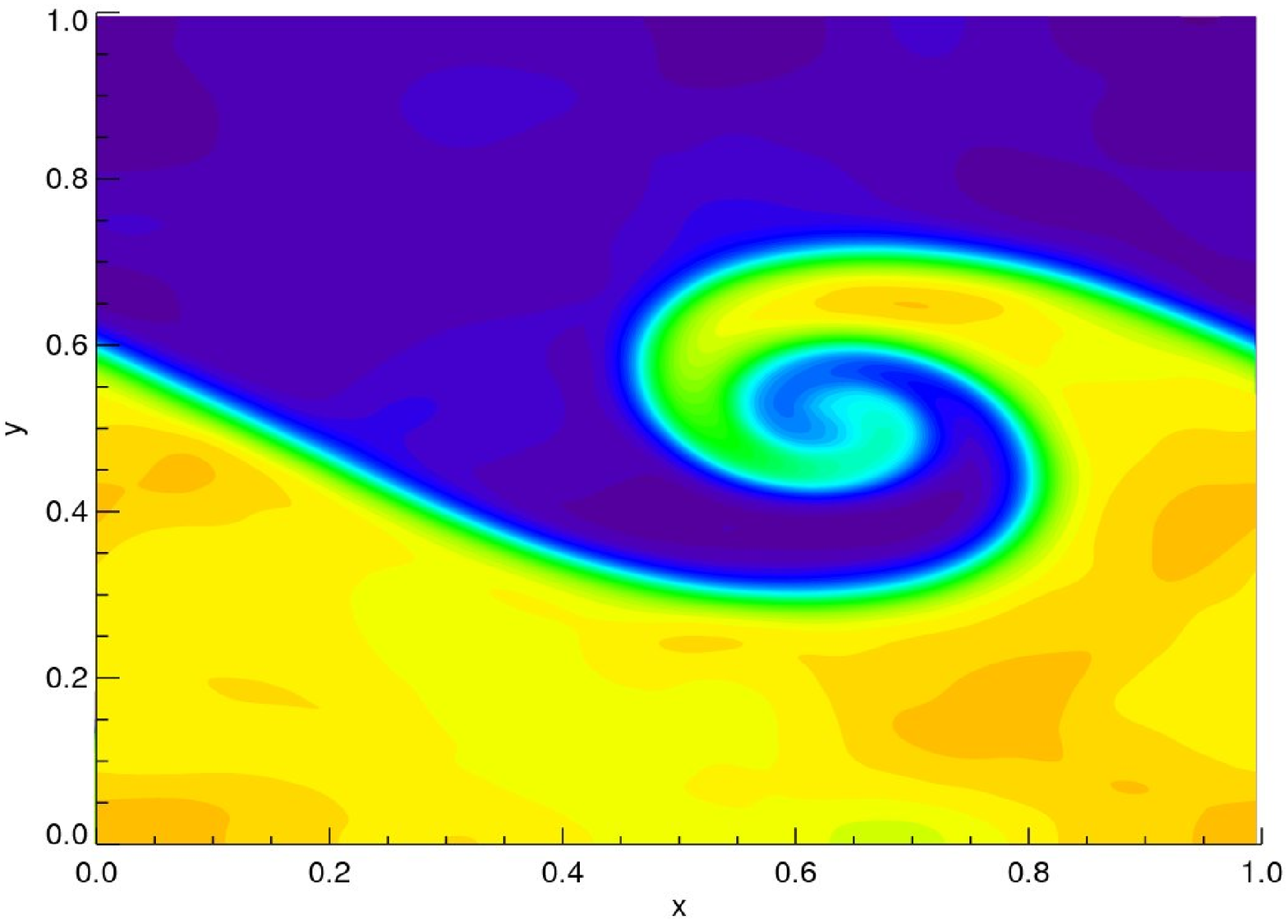}
  \includegraphics{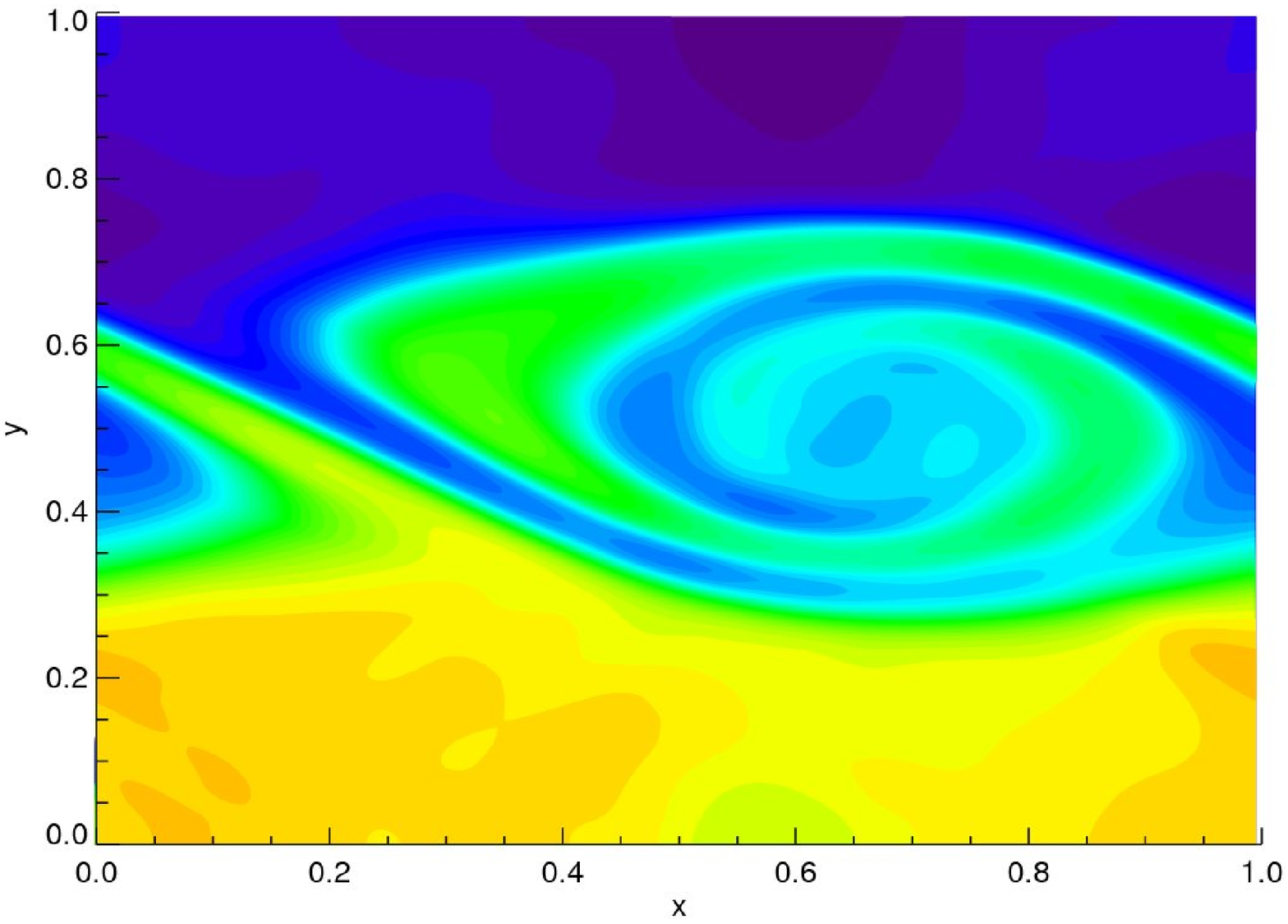}}
\resizebox{0.9\textwidth}{0.215\textwidth}{
  \includegraphics{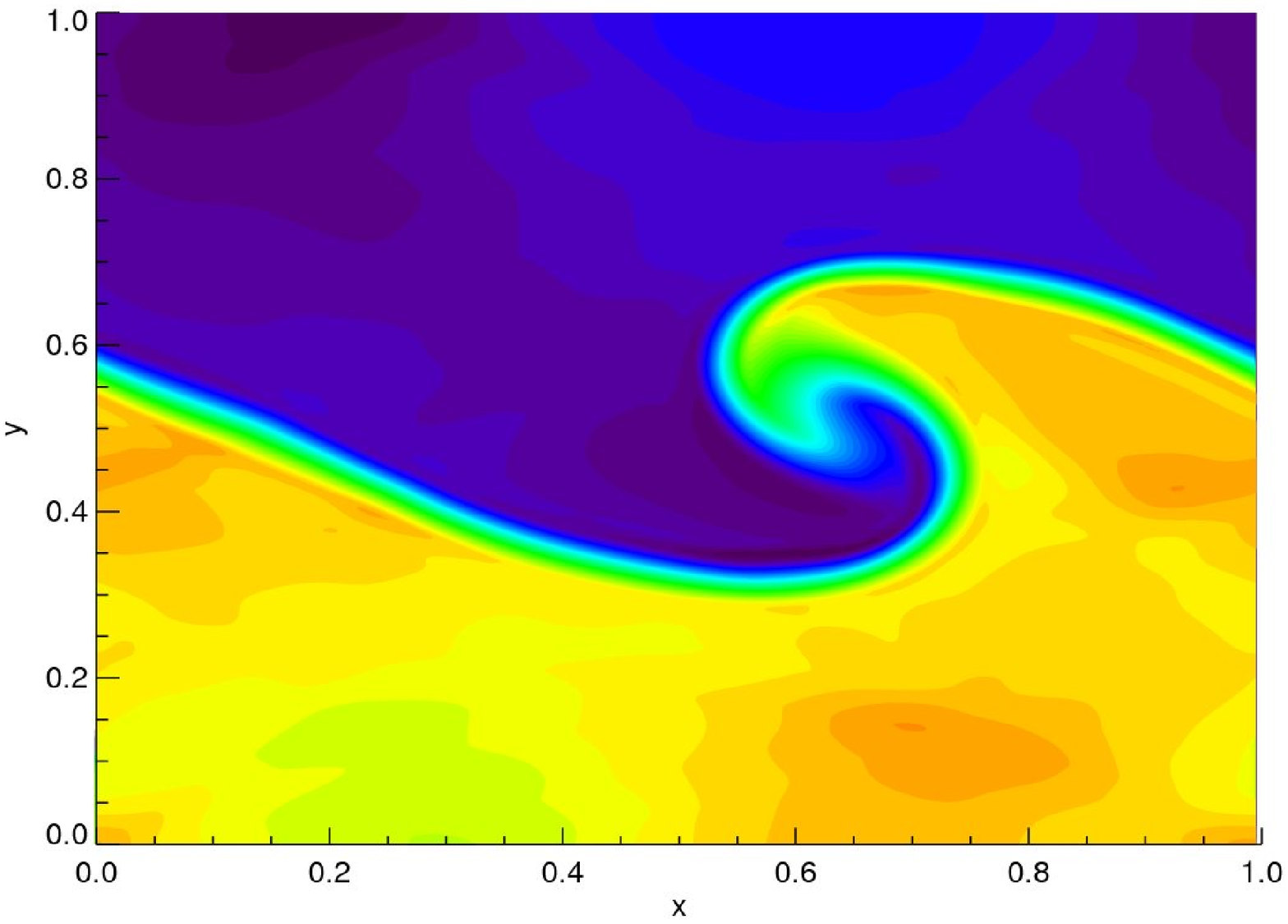}
  \includegraphics{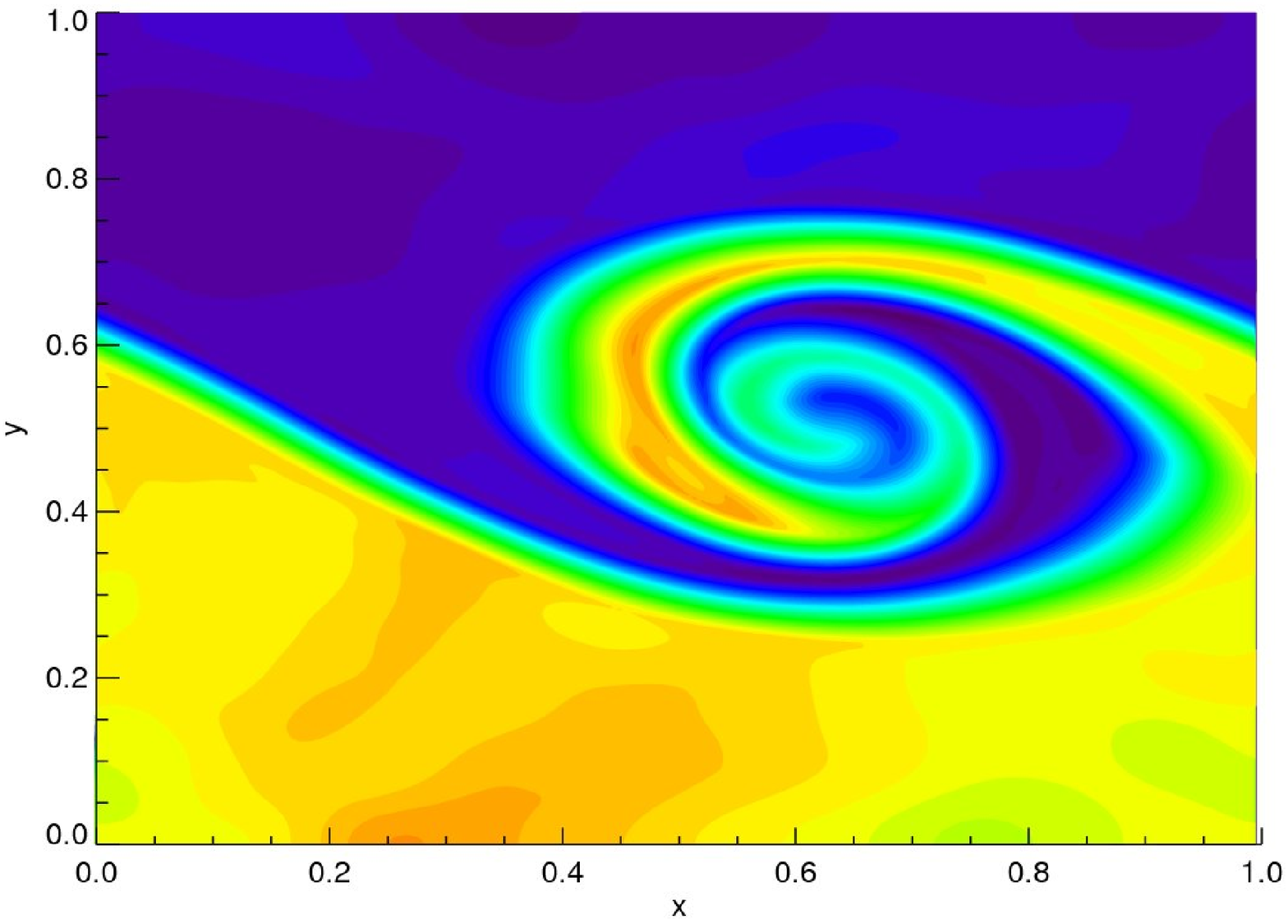}
  \includegraphics{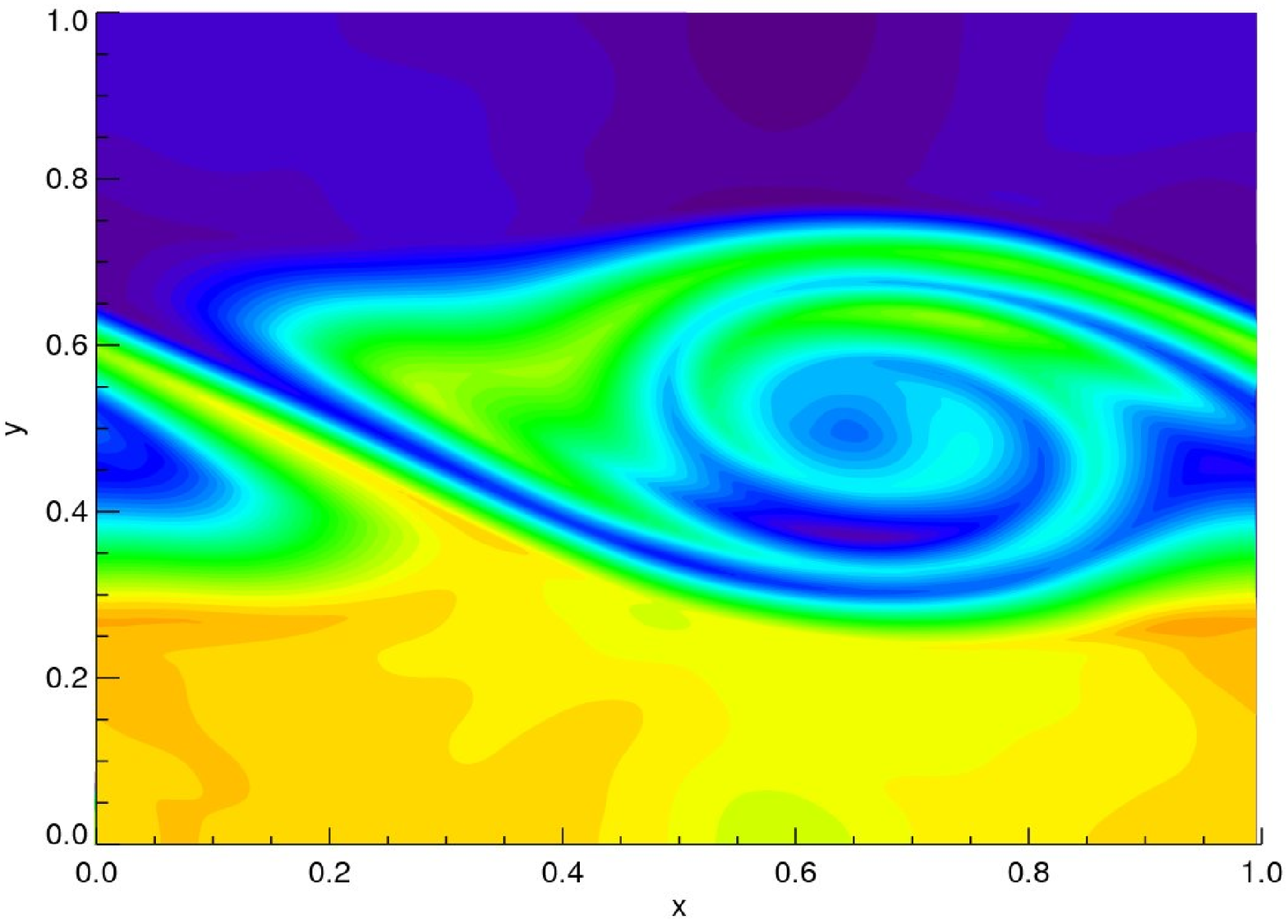}}
\resizebox{0.9\textwidth}{0.215\textwidth}{
  \includegraphics{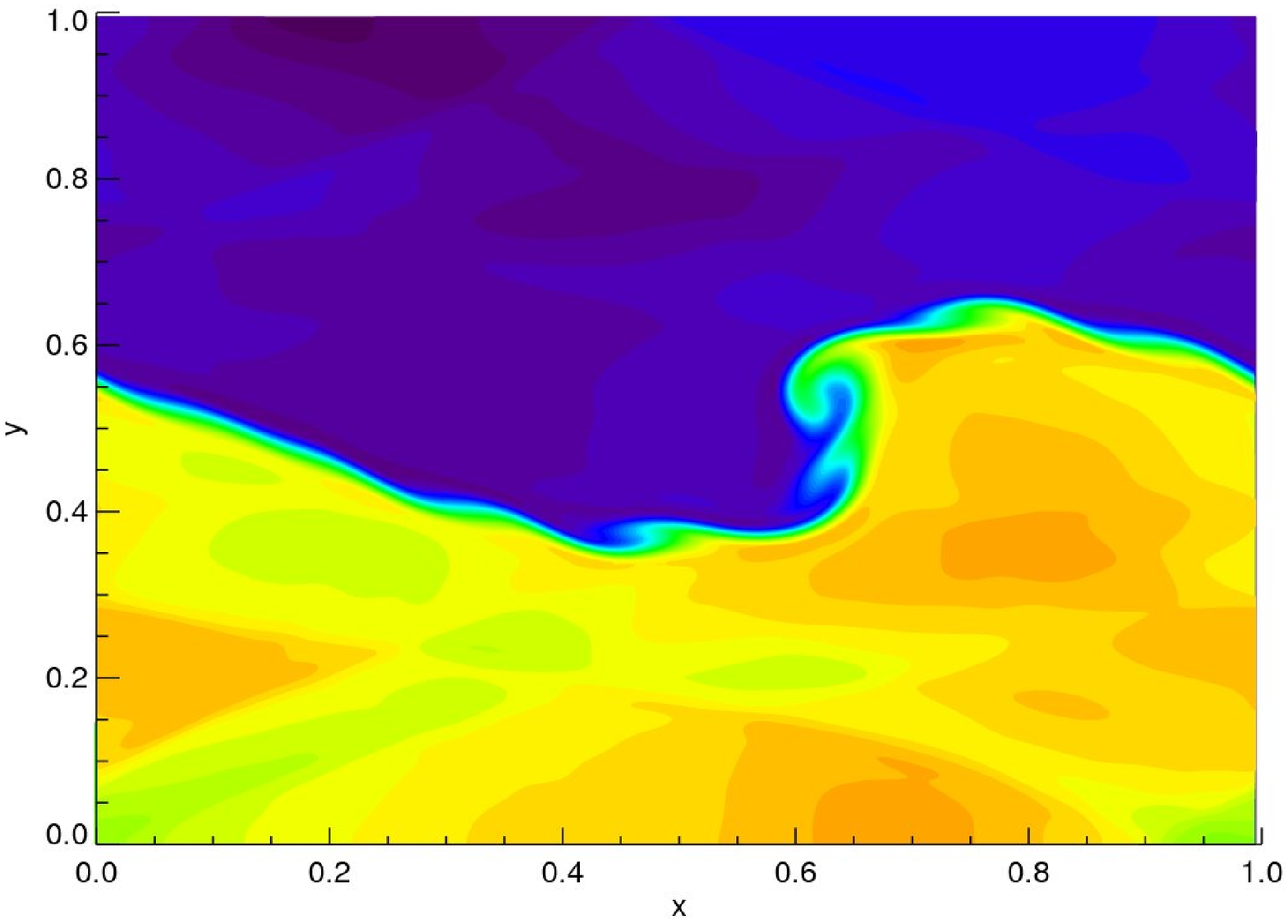}
  \includegraphics{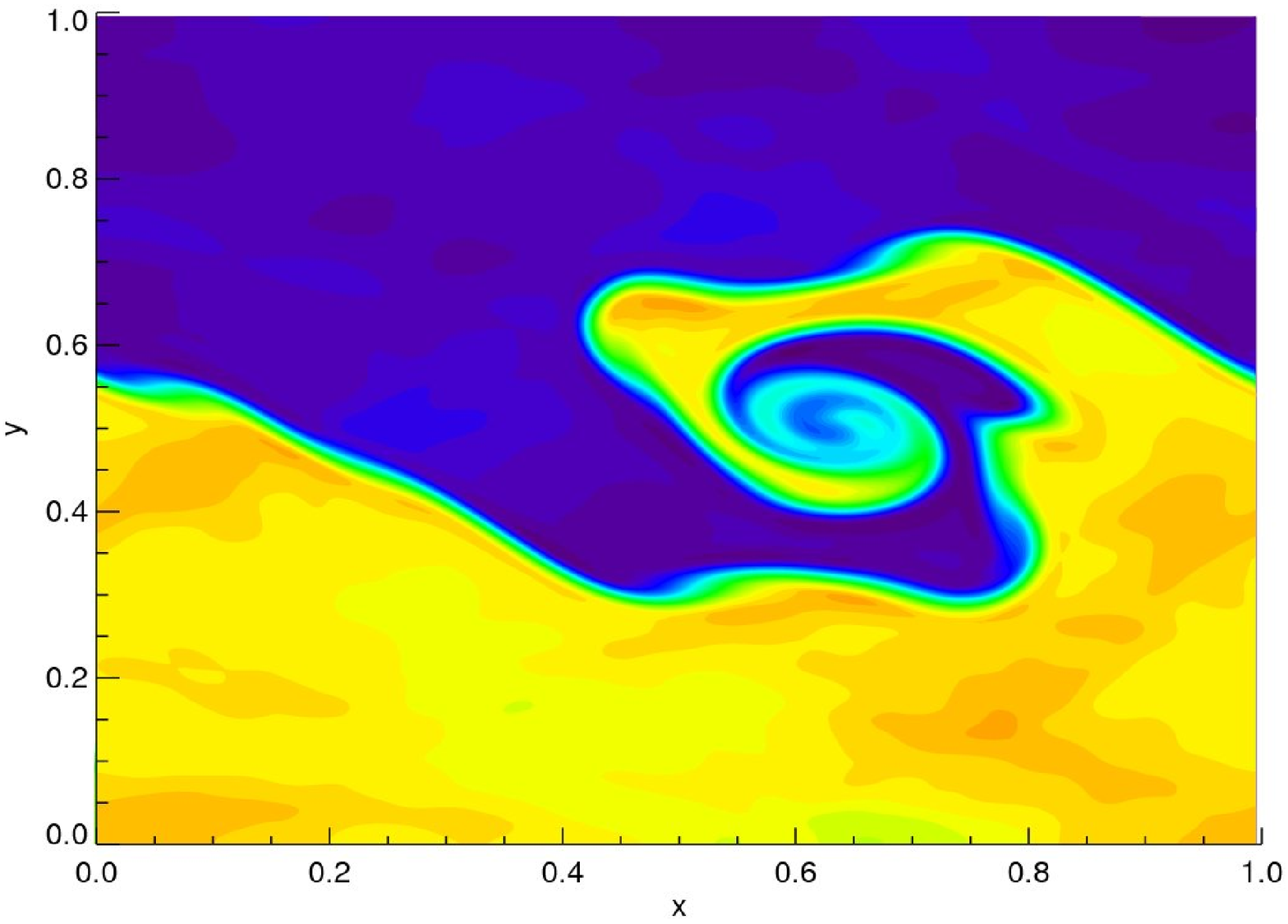}
  \includegraphics{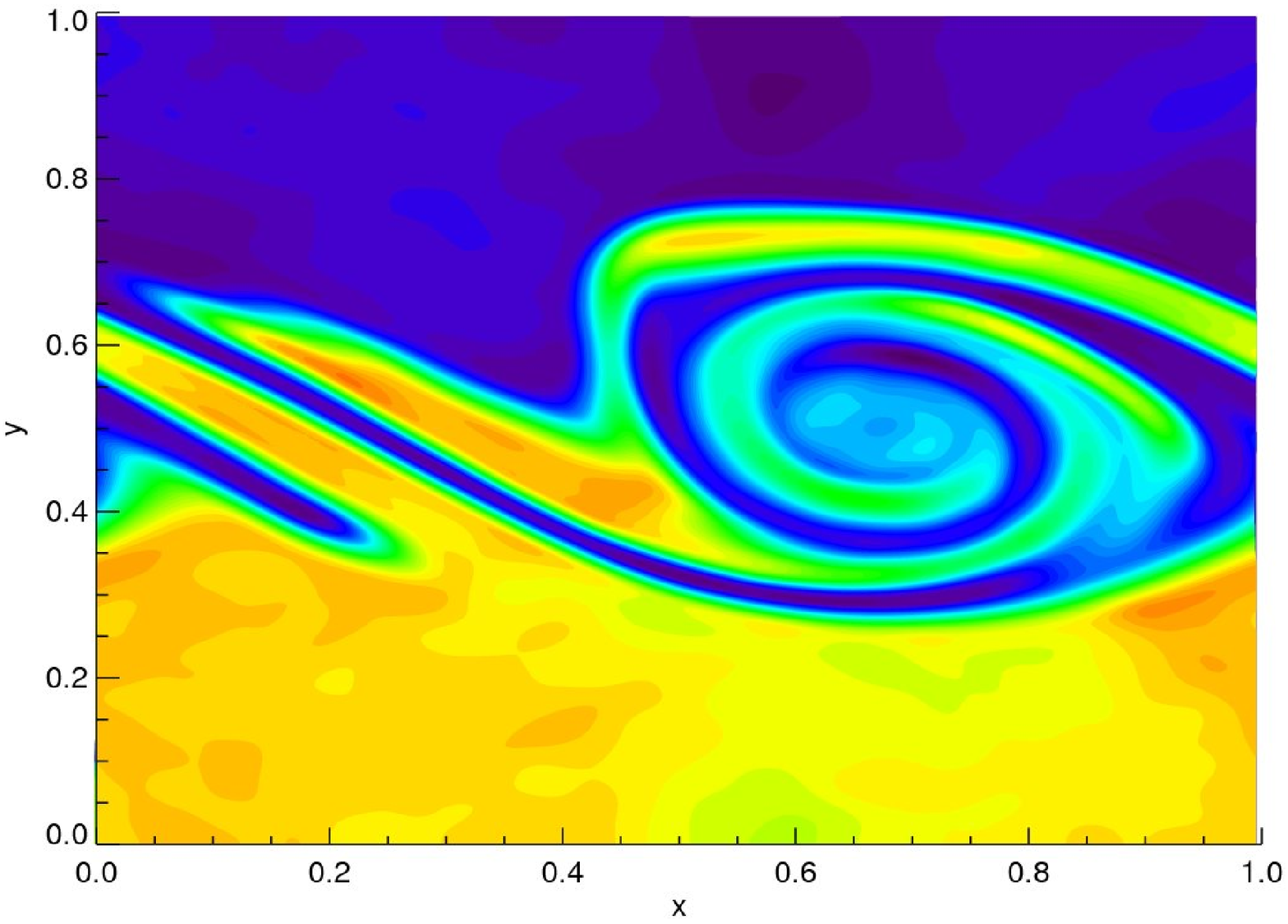}}
\resizebox{0.9\textwidth}{0.215\textwidth}{
  \includegraphics{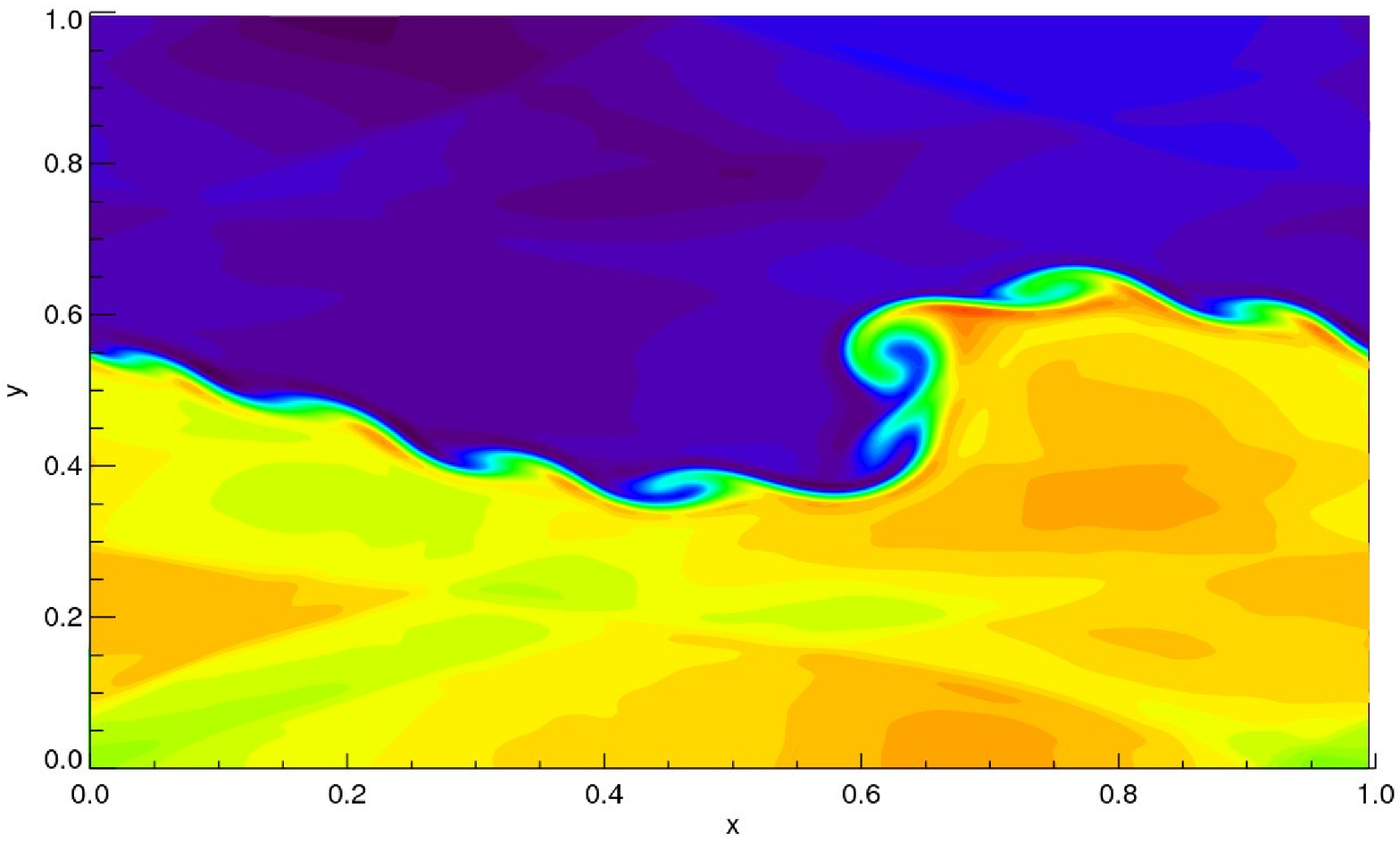}
  \includegraphics{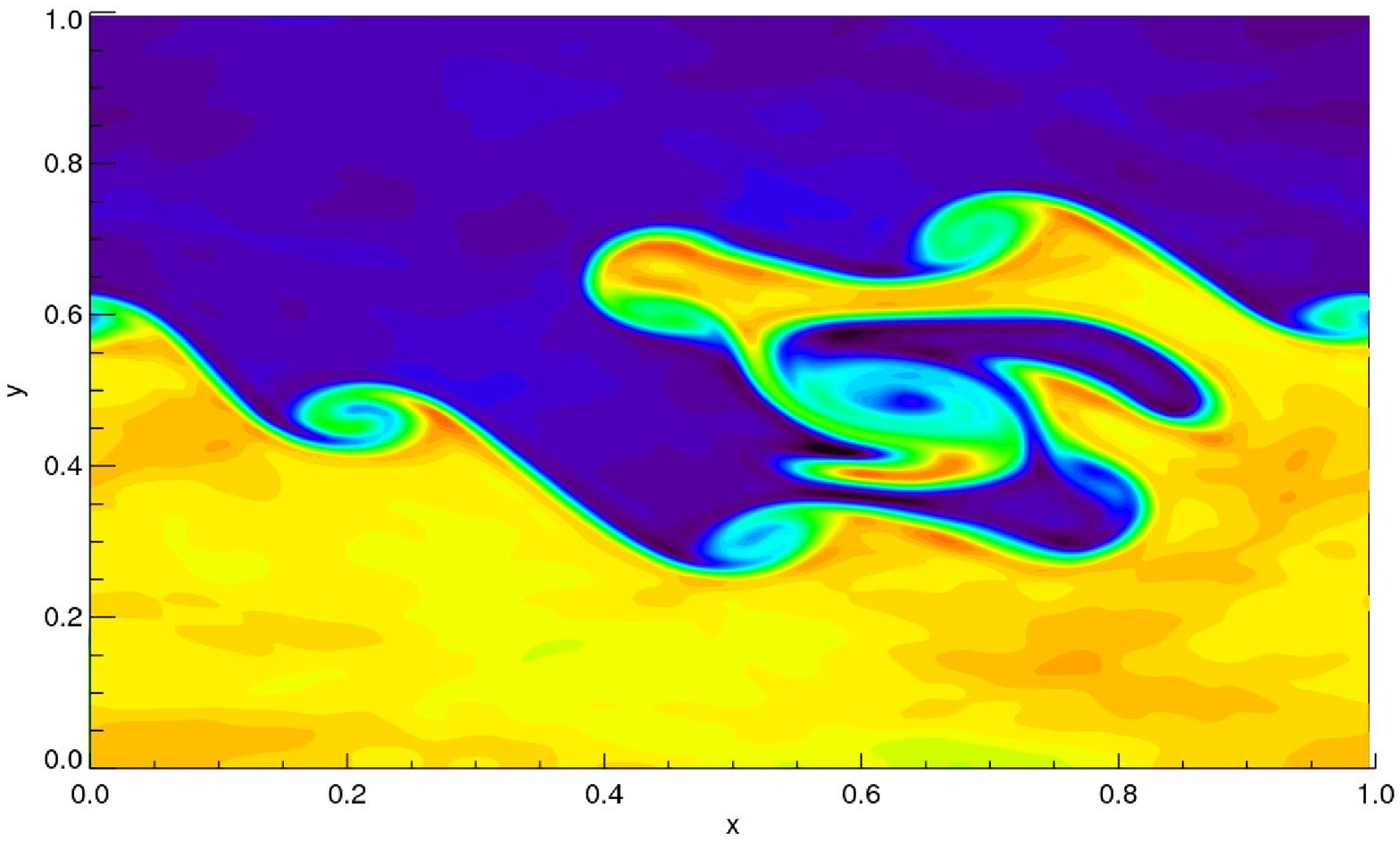}
  \includegraphics{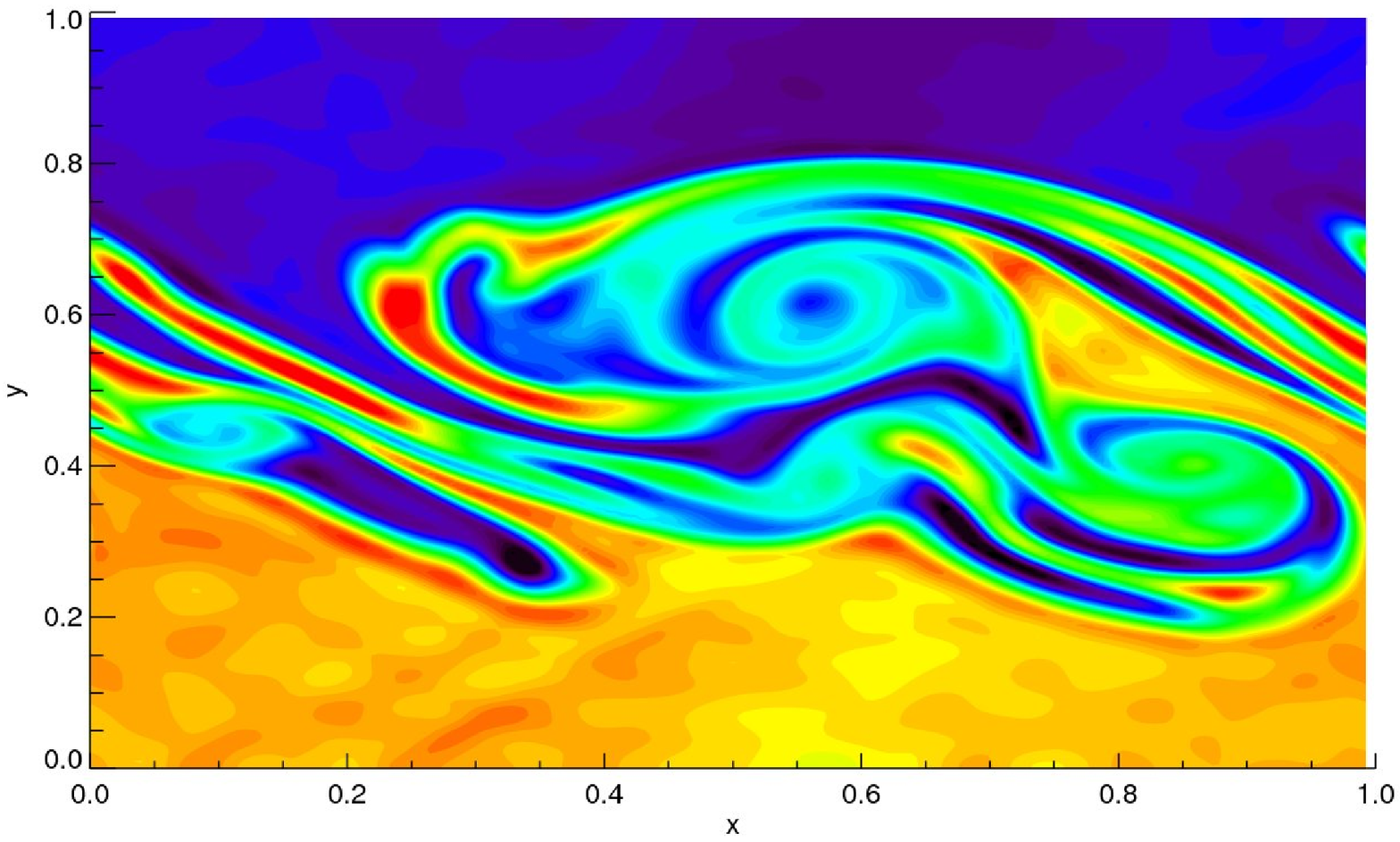}}
\resizebox{0.9\textwidth}{0.215\textwidth}{
  \includegraphics{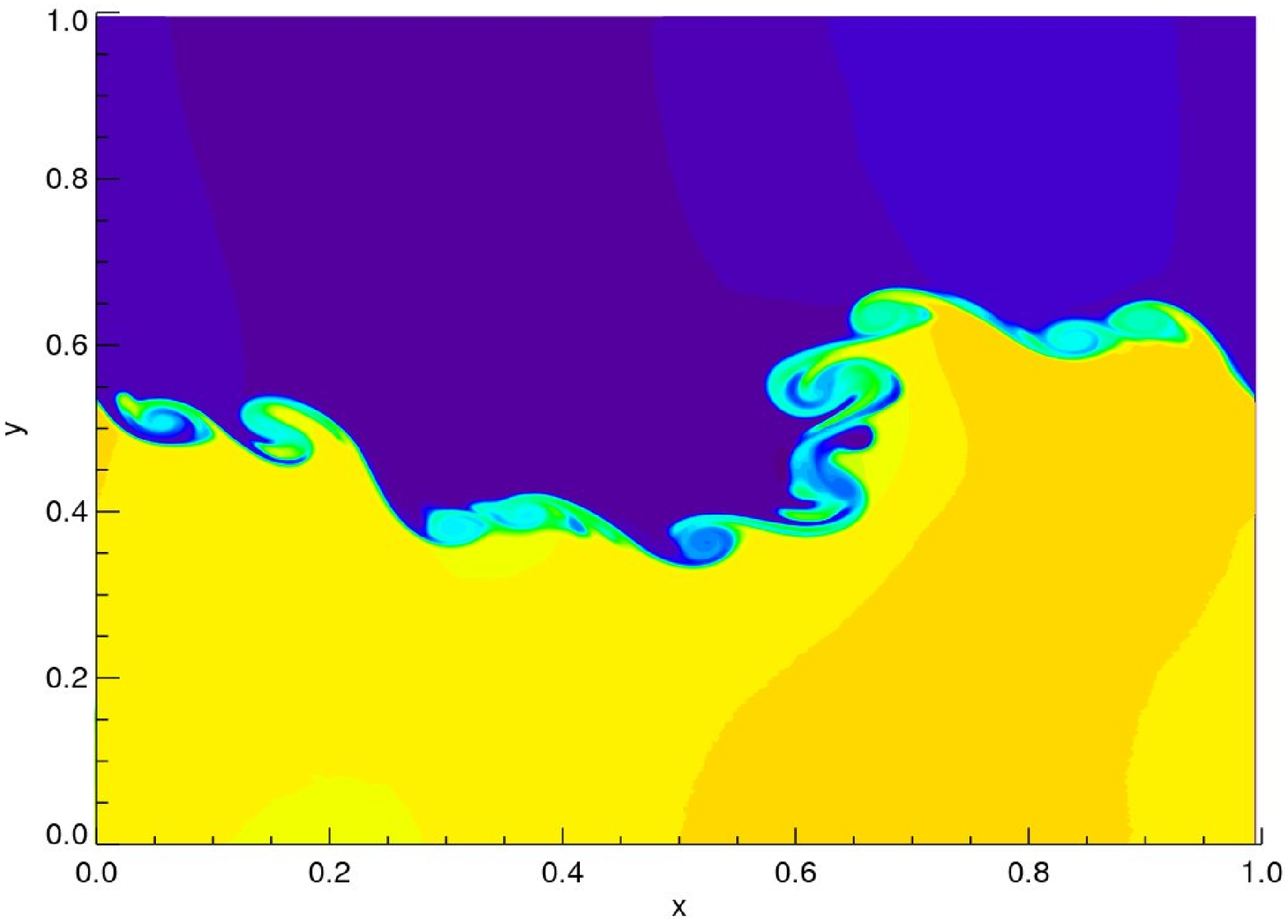}
  \includegraphics{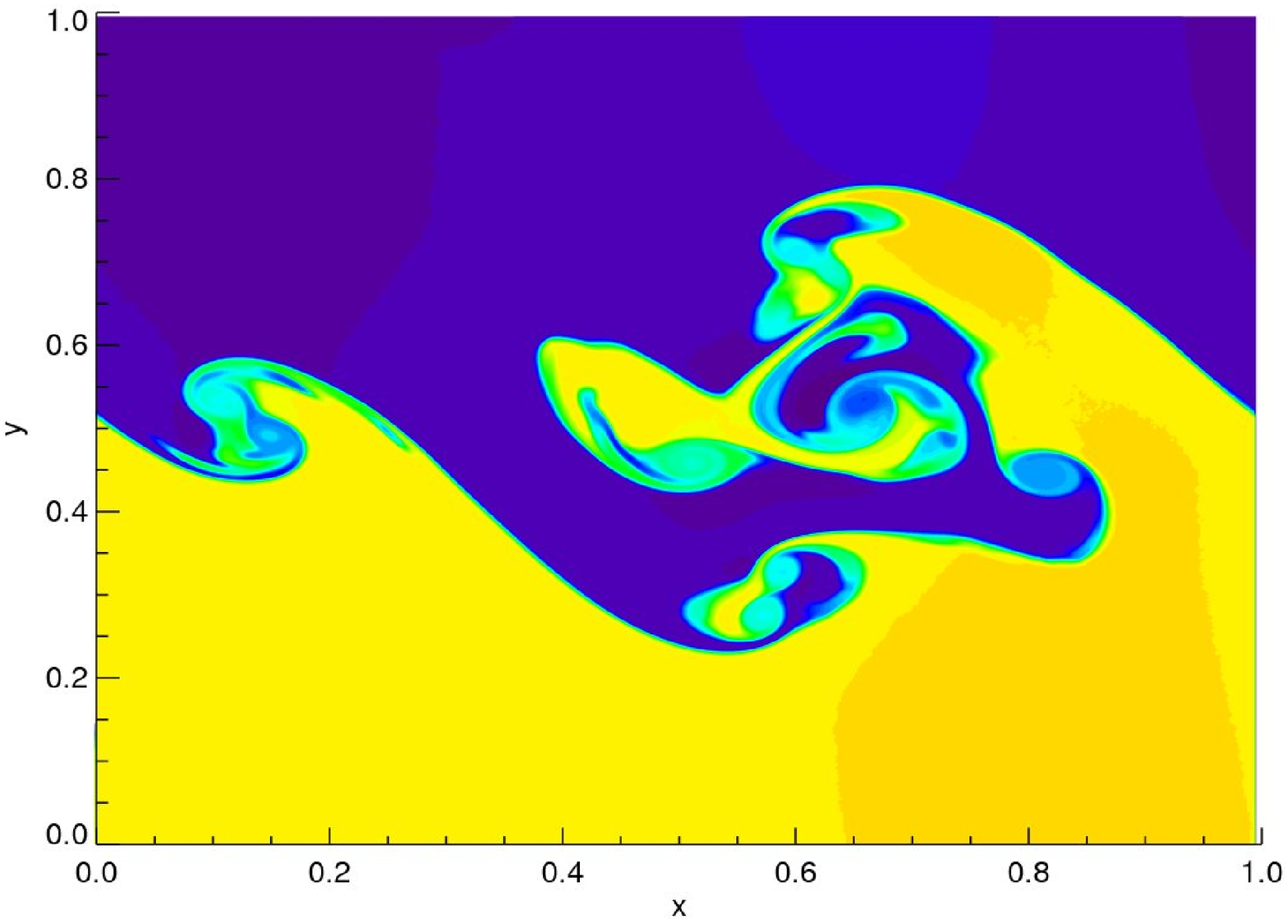}
  \includegraphics{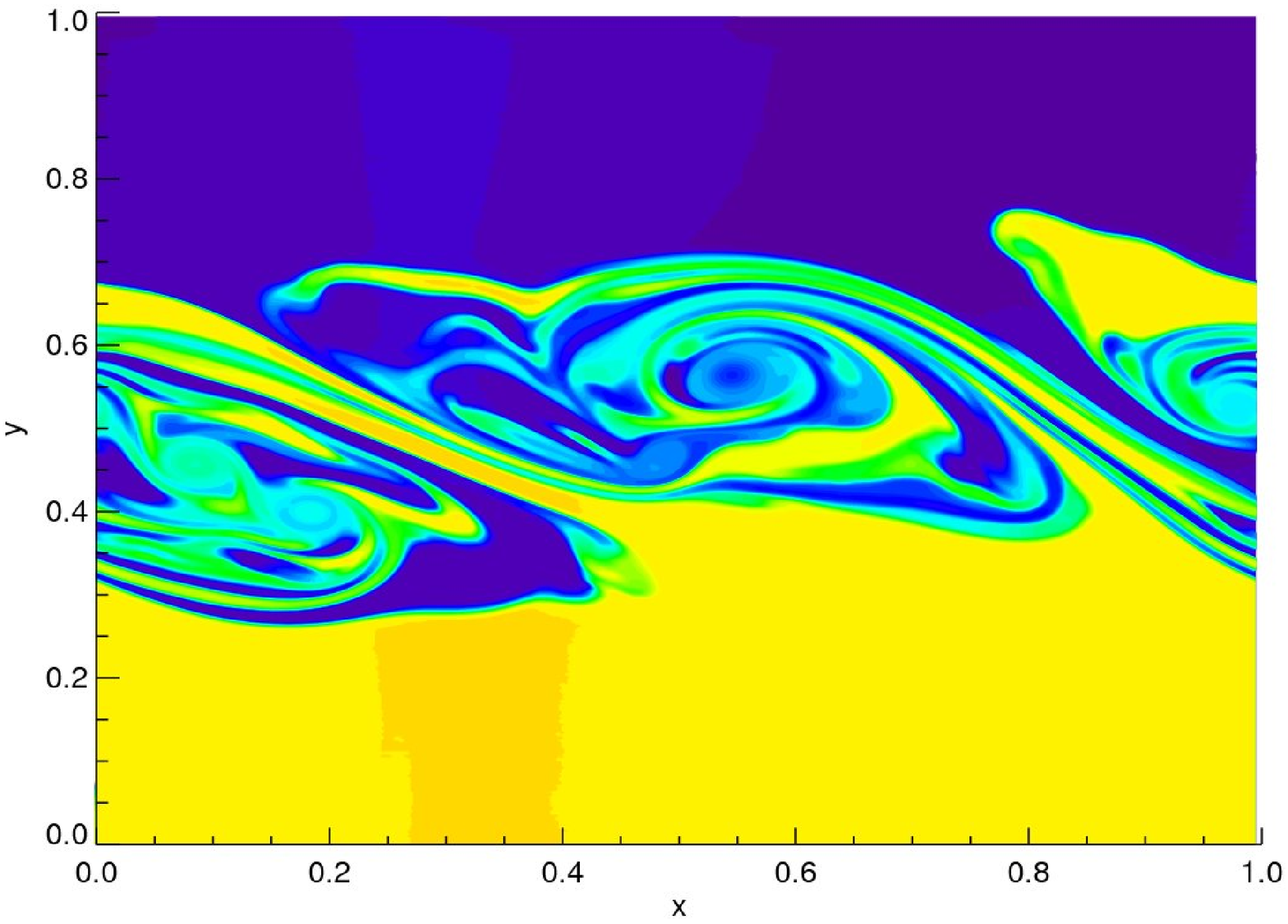}}
\includegraphics{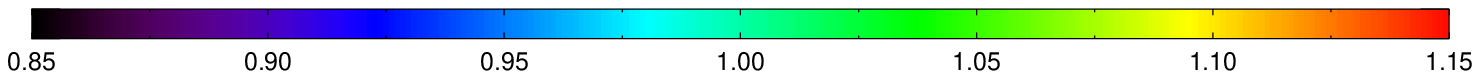}
\end{center}
\caption{\label{fig:khi} Density contours for a KH instability taken
  at $t=0.3$, $t=0.6$, and $t=1.5$.  The first four rows show results from
  RAPID using, in sequence, the Minmod, Van Leer, MC, and mixed limiters.
  The last row shows results from the PPM code.}
\end{figure*}

Here we compare results from the different RAPID limiters and the PPM
code on simulations of the two-dimensional Kelvin Helmholtz (KH)
instability.  While this instability should develop in any two fluids
with a strong enough velocity shear compared to the stratification, it
is difficult to capture accurately in numerical simulations.  Codes 
like the PPM code can overproduce the instability's small-scale
turbulent structure for a given resolution \citep{dwarkadas04}, while a
code with too much diffusion will un\-der-pro\-duce such structure, compared to experiments. SPH codes perform particularly poorly on such
a test when there is a density jump across the shearing region.  We
find that our method produces a wide range of small-scale structure
depending on the chosen limiter, enabling us to control the amount of
small-scale structure in a simulation by choosing an appropriate
limiter scheme.  

The KH instability is initialized on a Cartesian grid of resolution
$N_x\times N_y=400\times 400$ that is periodic in the x direction and
has reflecting boundaries on the top and bottom.  The fluid in the top
half of the box is set moving to the left at about one thirteenth of
the sound speed and the fluid on the bottom is set moving to the right
with the same speed.  The densities of the two regions of fluid are
set to values of $0.9$ and $1.1$ on the top and bottom, respectively,
in order to help visualize the instability.  The interface between the
two fluids is initialized to a sine wave in order to excite the
instability.  Figure \ref{fig:khi} shows results from the
RAPID(MM,VL,MC,MB) and PPM codes.  Each rows has three panels showing
density contour plots taken at $t=0.3$, $t=0.6$ and $t=1.5$,
respectively.  Time is measured in units where the undisturbed fluid
propagates completely across the box in unit time. 

\begin{figure*}
\begin{center}
\resizebox{0.7\textwidth}{0.35\textwidth}{
\includegraphics{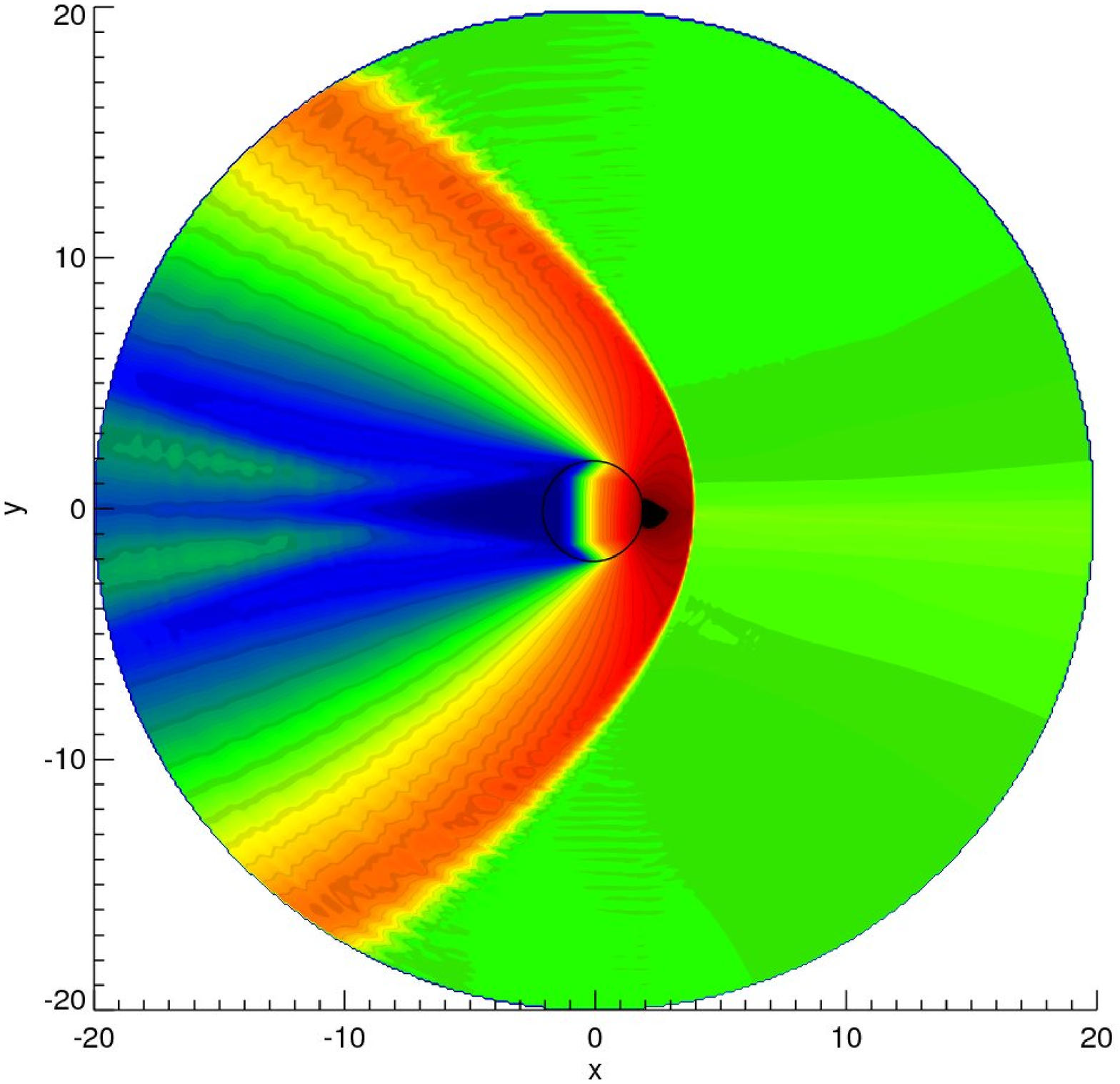}
\includegraphics{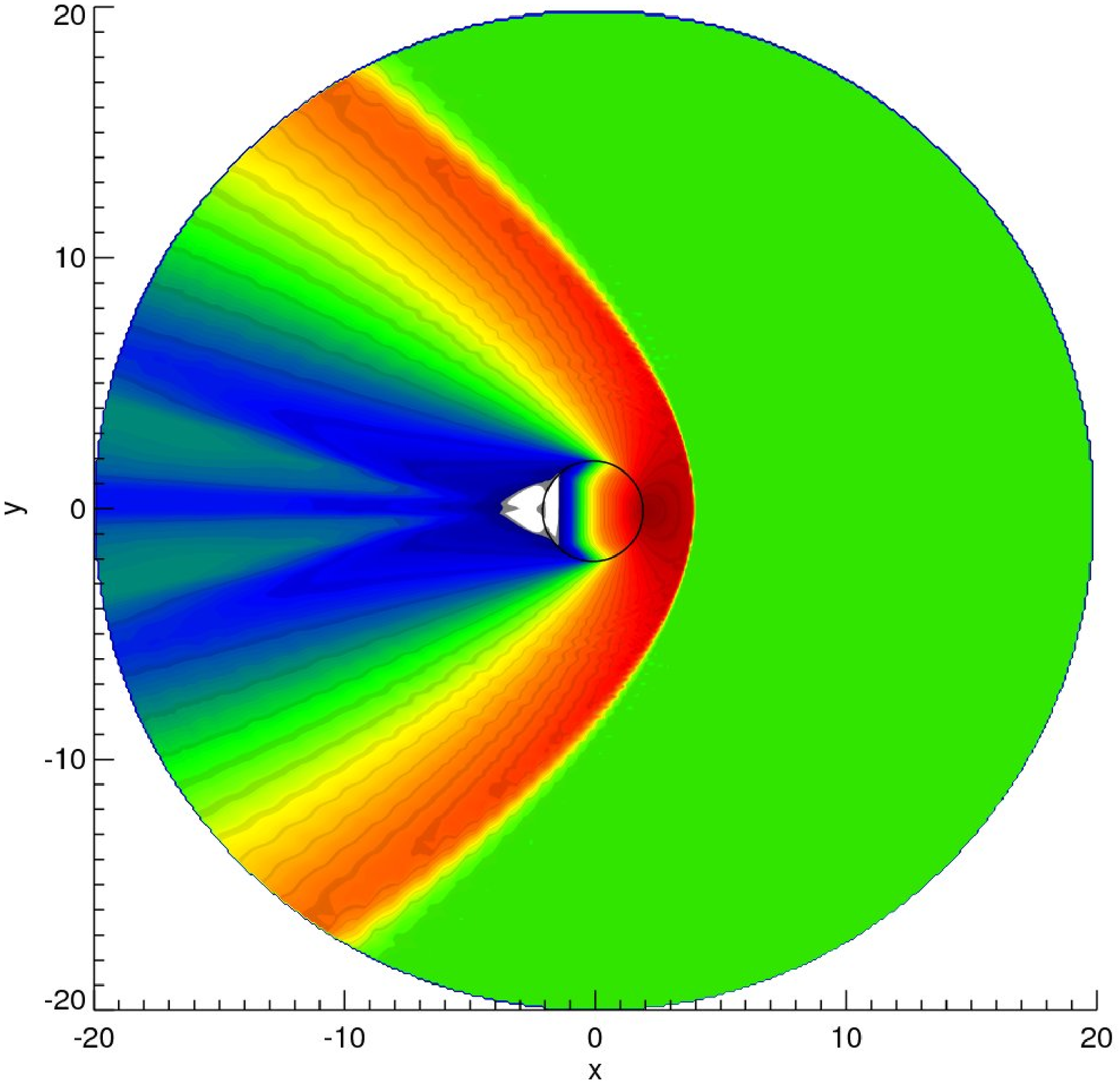}}
\includegraphics[scale=1.00]{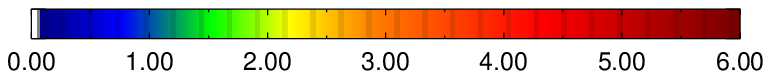}
\end{center}
\caption{\label{fig:cylinder} Density contour plots of the
  steady-state density for the RAPID(VL) code (left) and the PPM code 
  (right).  The black circles represent the boundary of the
  cylinder.} 
\end{figure*}

The Superbee limiter proves too unstable in this test and shortly
after $t=0.3$ the time step becomes unreasonably small.  The other
limiters produce a range of small-scale structure.  The
Minmod limiter produces the least structure, developing only a single
cusp along the interface and only a single, loose cat's eye structure.
In comparison, the structures produced using the Van Leer limiter are
more tightly wound at each time.  The MC limiter begins to show
small-scale KH instabilities forming along the interface at t=0.3.  The
final cat's eye is more tightly wound and shows increased substructure
as well.  The overall trend of increased substructure progresses
through the sequence of limiters until finally the mixed (MB) limiter
produces almost as much substructure as the PPM code.  Indeed at the
times $t=0.3$ and $t=0.6$, the results look quite similar. At time
$t=1.5$, both codes show very complicated interfaces with much mixing
between the two fluid layers.  As in the oblique-shock test, the PPM
code shows evidence of flattening or clipping of the density profile:
the TVD runs exhibit larger variations in the density, not only near
the interface, but also in the bulk regions of the fluids.  By
contrast, the density of the PPM code is very uniform in the bulk
section of each fluid region. 

The KH instability test readily demonstrates the differences between
the different limiters and codes. Given its infinitely sharp interface,
the KH problem is formally ill-posed, and infinitely small
disturbances grow infinitely fast.  Because there exists no
small-scale cutoff for the dynamics, the numerics themselves dictate
the evolution on the smallest scales.  We emphasize that the above
simulations are strictly two-dimensional.  Because the KH instability
can be viewed as the evolution of a single vortex sheet, it behaves
very differently in two dimensions than in three dimensions where
there are extra degrees of freedom along which the vorticity may
evolve.  By simulating a strictly two-dimensional fluid we are not
able to observe its full range of behavior.

\subsection{Supersonic flow around a cylinder} \label{subsec:hydrotests.4}

In order to test the implementation of the Euler equations on a
cylindrical grid, we examine the formation of a bow shock caused by
supersonic flow around a cylinder.  We initialize a cylindrical grid
with a supersonic flow to the left at three times the sound speed. The
grid spans an annular region from $2 \le r \le 20$ and $0 \le q \le
2\pi$ and has a resolution of $N_r \times N_q = 600 \times 150$.  The
initial density and pressure on the grid are uniform: the density is
set equal to a value of five-thirds, and the pressure is set to unity.
The inner boundary of the grid is reflecting, simulating a solid
cylinder around which a bow shock forms.  The outer boundary allows
outflow. Figure \ref{fig:cylinder} shows the results from RAPID(VL)
and PPM simulations.   

The two codes both produce a density peak just in front of the
cylinder with a value close to $6$ (about 6 and 6.4, respectively, for
the PPM and RAPID(VL) codes).  These peak densities are about $90$\% and
$96$\%, respectively, of the maximum-possible post-shock density
$\rho_s = 4\rho_0$ for an adiabatic shock of index $\gamma = 5/3$.
Again RAPID displays small-scale oscillations in front of the shock,
which the PPM code appears to have flattened.  Both codes capture the
smaller tail shocks produced at the rear of the cylinder.

\section{The protoplanet problem} \label{sec:planettests}

Planet-disk interactions in protoplanetary systems are one of the
primary scenarios RAPID is designed to study.  Here we provide a
suite of simulations detailing its performance on a standard
setup for such scenarios.  We compare results for different choices of
solution vector, levels of added viscosity, numerical resolution and
choice of limiters.  The purpose of such a comparison is to be able to
better determine which details observed within a run are physically
realistic and consistent between runs and which are likely due to
numerical artifacts.  

\subsection{Protoplanet problem setup} \label{subsec:planetsetup}

The details of the setup are those used by
\citet{valborro06} and represent what is now a standard problem in
accretion disk theory.  A polar grid is setup with initial conditions
(described below) using a given mesh resolution $N_r\times N_\q$.
The azimuthal range is always taken to be $[-\pi,\pi]$, and unless
otherwise indicated, the radial range is $[0.4a,2.5a]$, where $a$ is
the mean orbital radius of the protoplanet.  A central star is modeled
by calculating its gravitational potential at a grid position
$\br=(r,\q)$ in terms of its mass $M_*$ and location $\br_*$ as
$\phi_* = -GM_*/|\br-\br_*|$.  A protoplanet is represented by the
softened potential 
\begin{equation}
  \phi_p = \frac{-G m_p}{\sqrt{|\br-\br_p|^2 +
      \epsilon^2}}. \label{eq:planetpot}
\end{equation}
The softening length is defined to be $\epsilon=0.6H(a)$, where $H(a)$
is the undisturbed disk's scale height at radius $a$.  A
``Jupiter-mass'' planet is defined to have a mass ratio of 
$\mu = m_p/(m_p+M_*) = 10^{-3}$. 

We assume a locally isothermal equation of state in the disk, set by
the ideal gas law $p=\rho c_s^2$, where the sound speed $c_s^2$ is
set to be a fixed fraction of the Keplerian speed.  This statement is
equivalent to assuming a thin disk, with a scale height given by $H/r
=c_s/v_{_K}$. We adopt the standard value of $H/r = 0.05$ for the disk
thickness. Because the pressure is prescribed in terms of the sound
speed and density distribution, there is no need to solve for the
energy equation. 

Simulations with a single planet are calculated in the frame
corotating with the planet, with the origin at the center of 
mass of the planet and central star.  This choice of origin means that
the star orbits a distance $\mu a$ from the origin, the planet at a
distance $(1-\mu)a$. Additional simulations have been performed in
the corotating frame with the origin held fixed on the star, as well as
in the inertial frame with the origin fixed on the star, or at the
planet and star's center of mass.  All these simulations yield similar
results. 

\begin{figure*}[t]
\begin{center}
\resizebox{0.9\textwidth}{0.21\textwidth}{
  \includegraphics{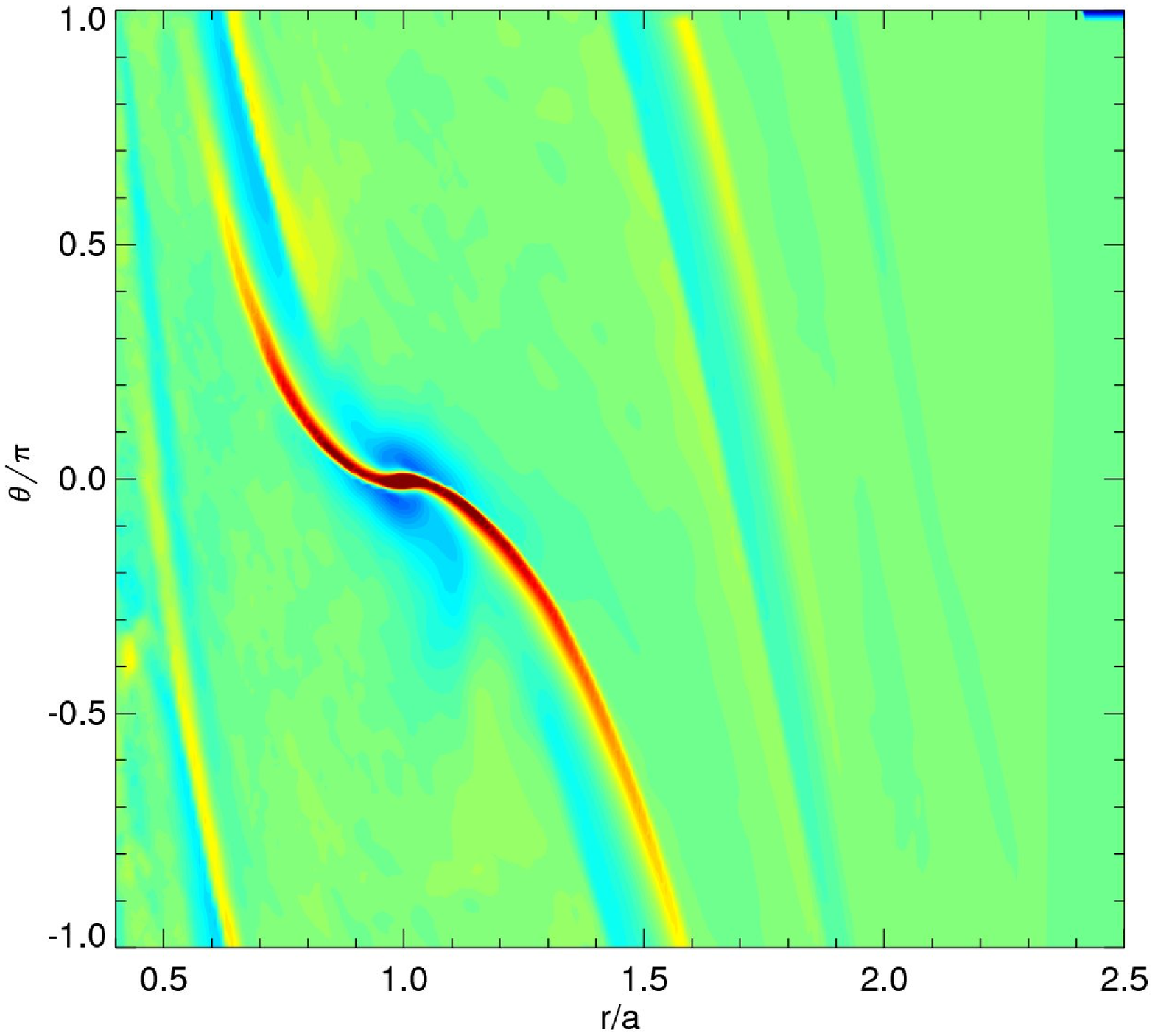}
  \includegraphics{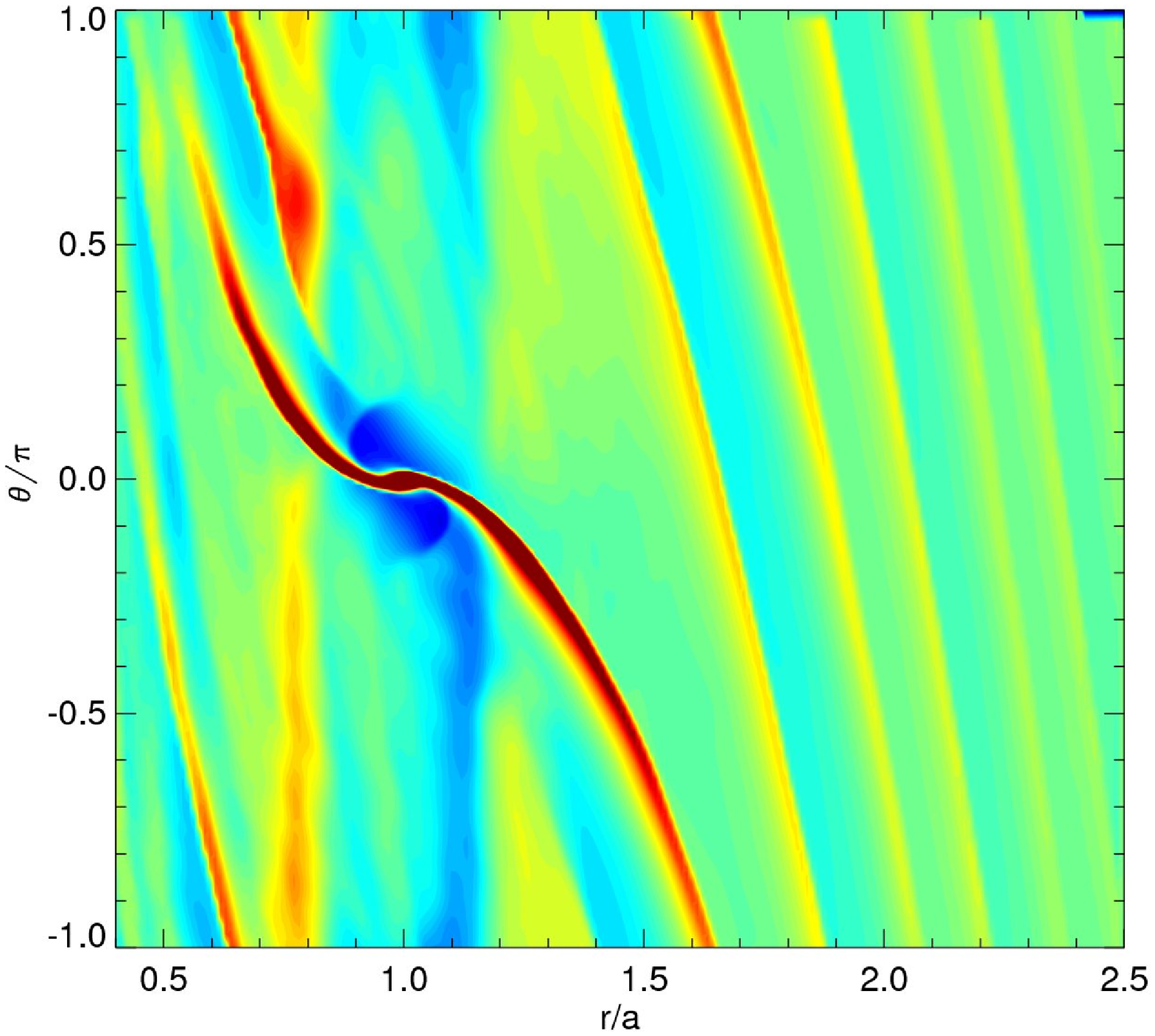}
  \includegraphics{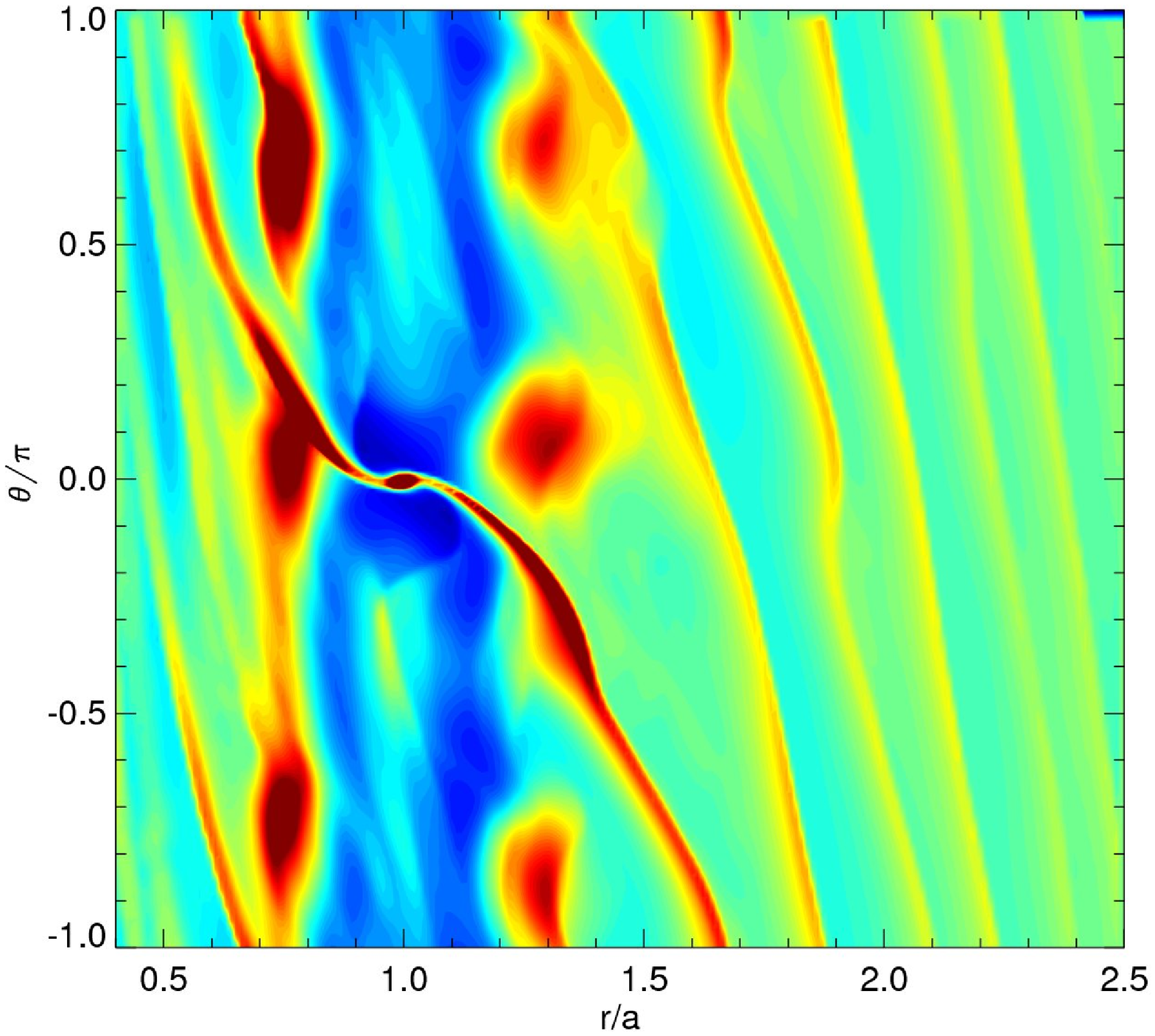}}
\resizebox{0.9\textwidth}{0.21\textwidth}{
  \includegraphics{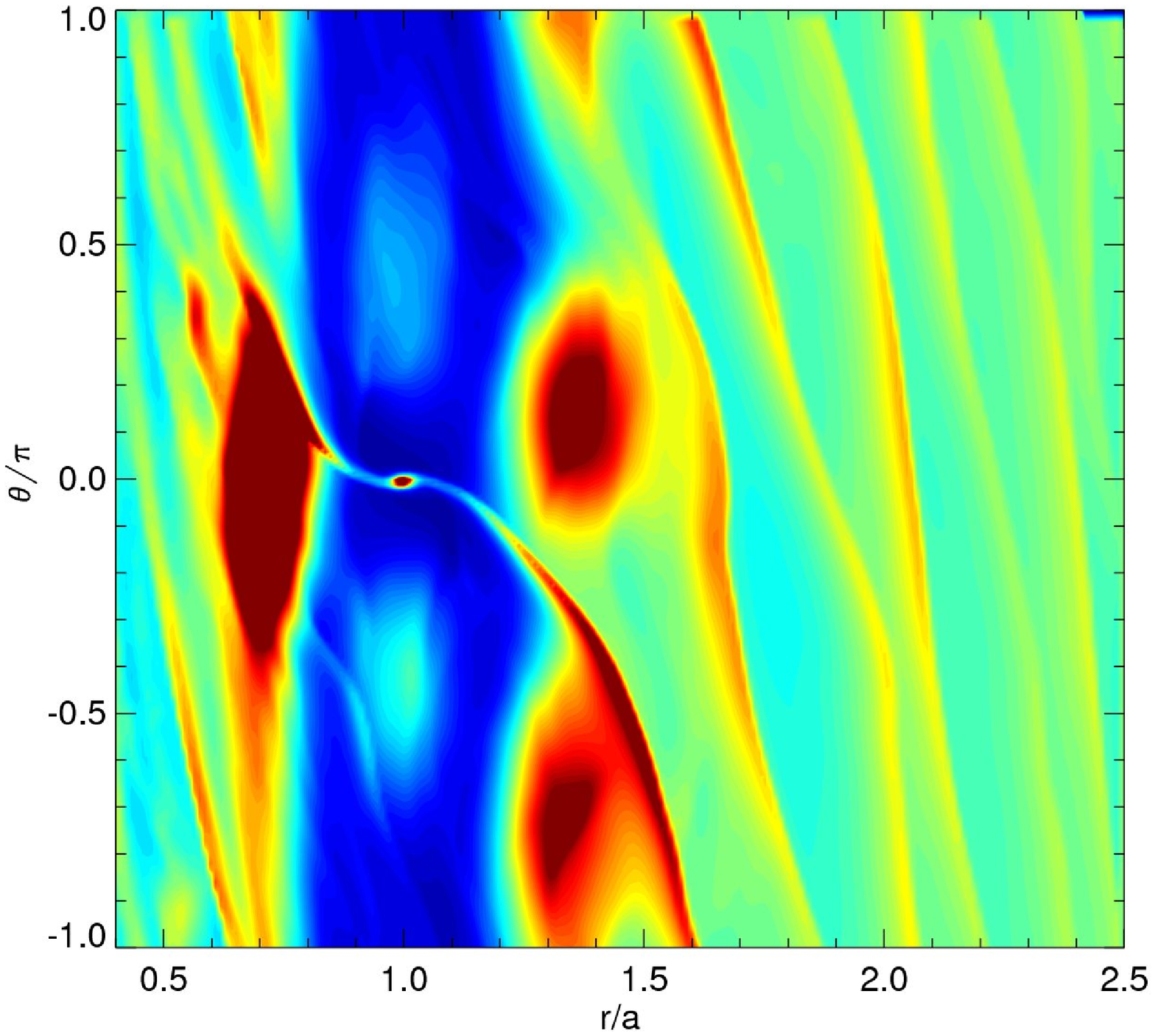}
  \includegraphics{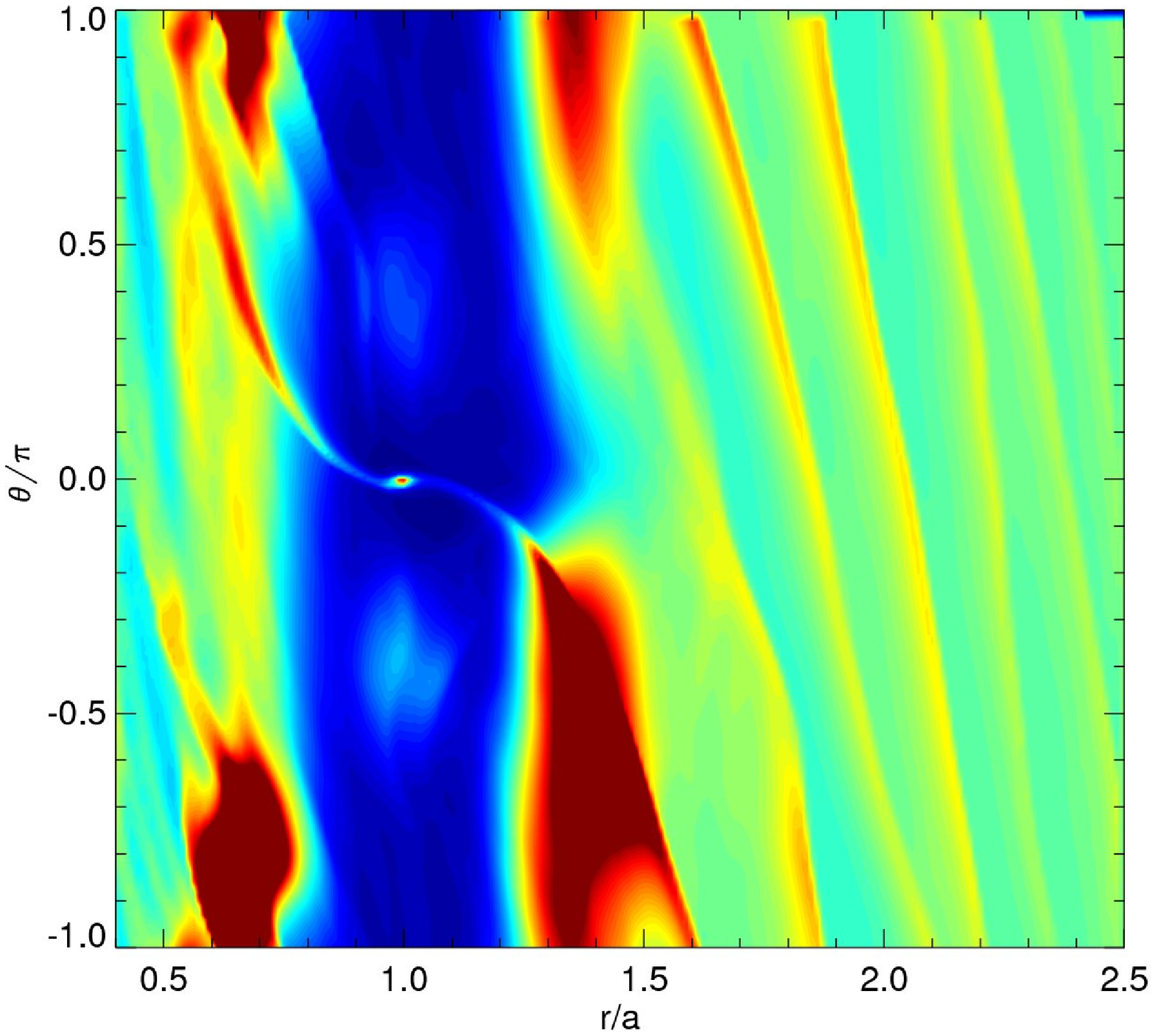}
  \includegraphics{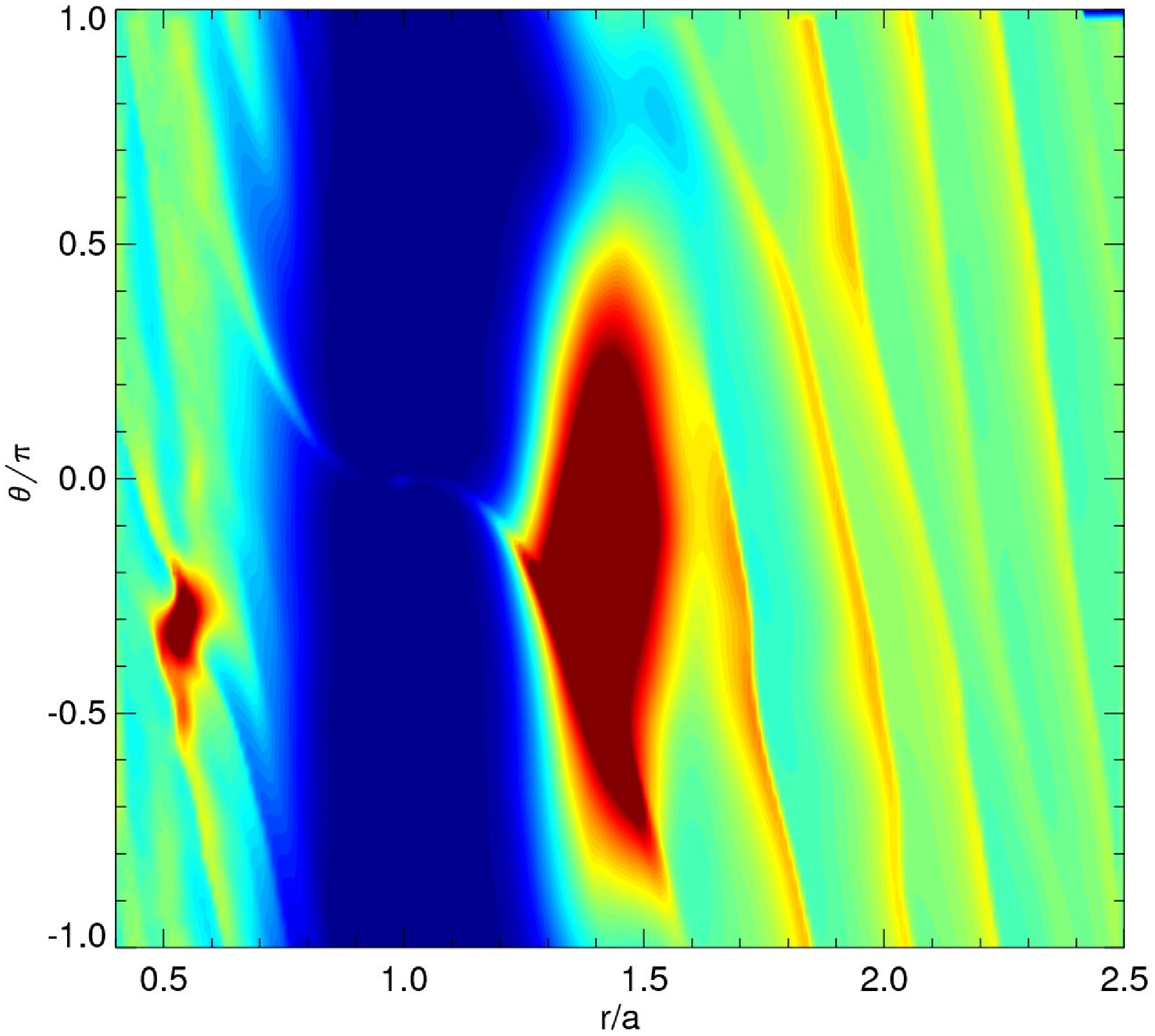}}
\includegraphics{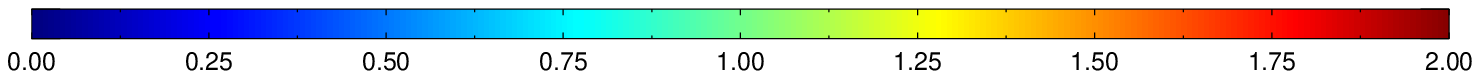}
\end{center}
\caption{\label{fig:SJ.den} Density contours for a standard Jupiter
          mass run using the VL limiter.  Plots in sequence are 
          taken after $5$, $10$, $20$, $50$, $100$, and $300$ orbits.}
\end{figure*}

We use dimensionless units where the unit of mass is taken to be $M_*
+ m_p$.  Length is measured in units of the planet's initial radial
separation from the star $a$, and we set the gravitational constant
$G$ to unity.  Time is measured in units of
\begin{equation}
  \tau = \sqrt{\frac{a^3}{G(M_*+m_p)}}.
\end{equation}
With this definition, one orbital period takes $2\pi$ units of
simulation time.

Fiducial initial conditions are those of a uniform-density Keplerian
disk, which is the equilibrium solution for a single central potential
of mass $M_*$.  We transform the azimuthal velocity into the
corotating frame and correct for pressure support as $\uq =
\sqrt{(GM_*)/r}[(1-(H/r)^2)^{1/2} - (r/a)^{3/2}]$, where $m_H=H/r$ is
the disk thickness.  In order to allow the Keplerian disk to gradually
adjust to the additional potential of a planet, the potential of the
planet is slowly ``turned-on'' according to the prescription
\begin{equation}
  \phi_p(\br,t) = \sin^2 \left [\frac{t}{4\mathcal{N}\tau} \right
  ]\phi_p(\br),
\end{equation}
where $t$ is the simulation time and $\mathcal{N}=10$ is the number of
orbits over which the potential is turned on.  After the prescribed
number of orbits the full value of the potential $\phi_p(\br)$ is
held constant.

\subsection{Standard run} \label{subsec:standardrun}

We present a standard comparison run of the code for a Ju\-pi\-ter-mass
planet, run for 300 orbits using the VL limiter scheme and with
resolution $N_r \times N_\q = 384 \times 384$.  The calculation is perform\-ed
in the corotating frame, advecting the solution set 
$(\rho,\rho u_r, H)$, where ${\cal H}=\rho r(u_\q+r\Omega)$ is the inertial
angular momentum (combined gas and frame momentum). 

In Figure \ref{fig:SJ.den} we show density contours after $5$, $10$,
$20$, $50$, $100$ and $300$ orbits.  Appearance of the spiral arms
occurs very quickly (within a few dynamical times) with two trailing
arms outside the planet's orbital radius and three arms inside the
orbital radius.  They are close to steady-state in the sense that they
occur at fixed locations within the disk when time-averaged over a few
orbits although they exhibit small spatial and temporal oscillations
in simulations with low-viscosity.

\begin{figure*}
\begin{center}
\resizebox{0.9\textwidth}{0.38\textwidth}{
  \includegraphics{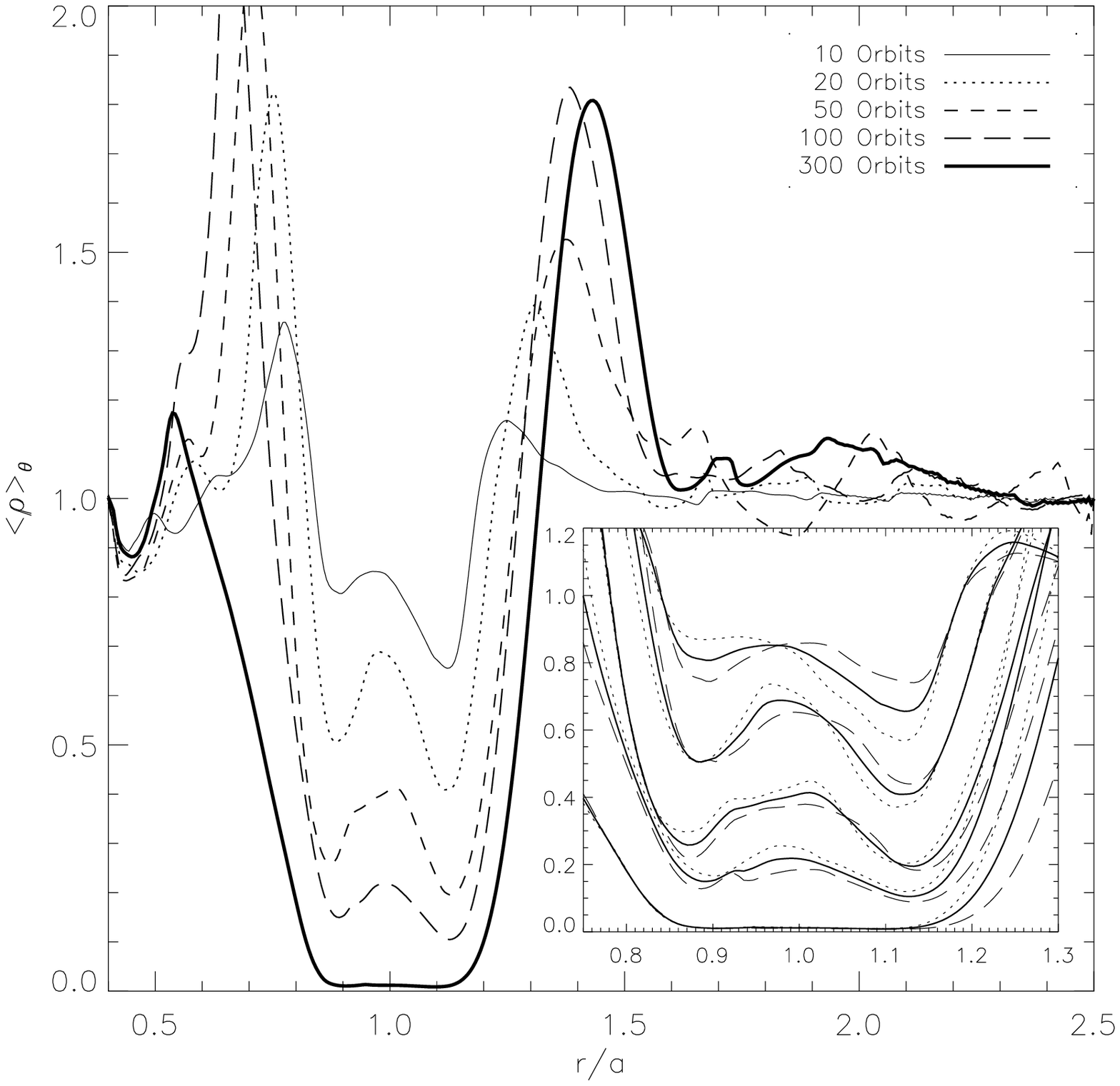}
  \includegraphics{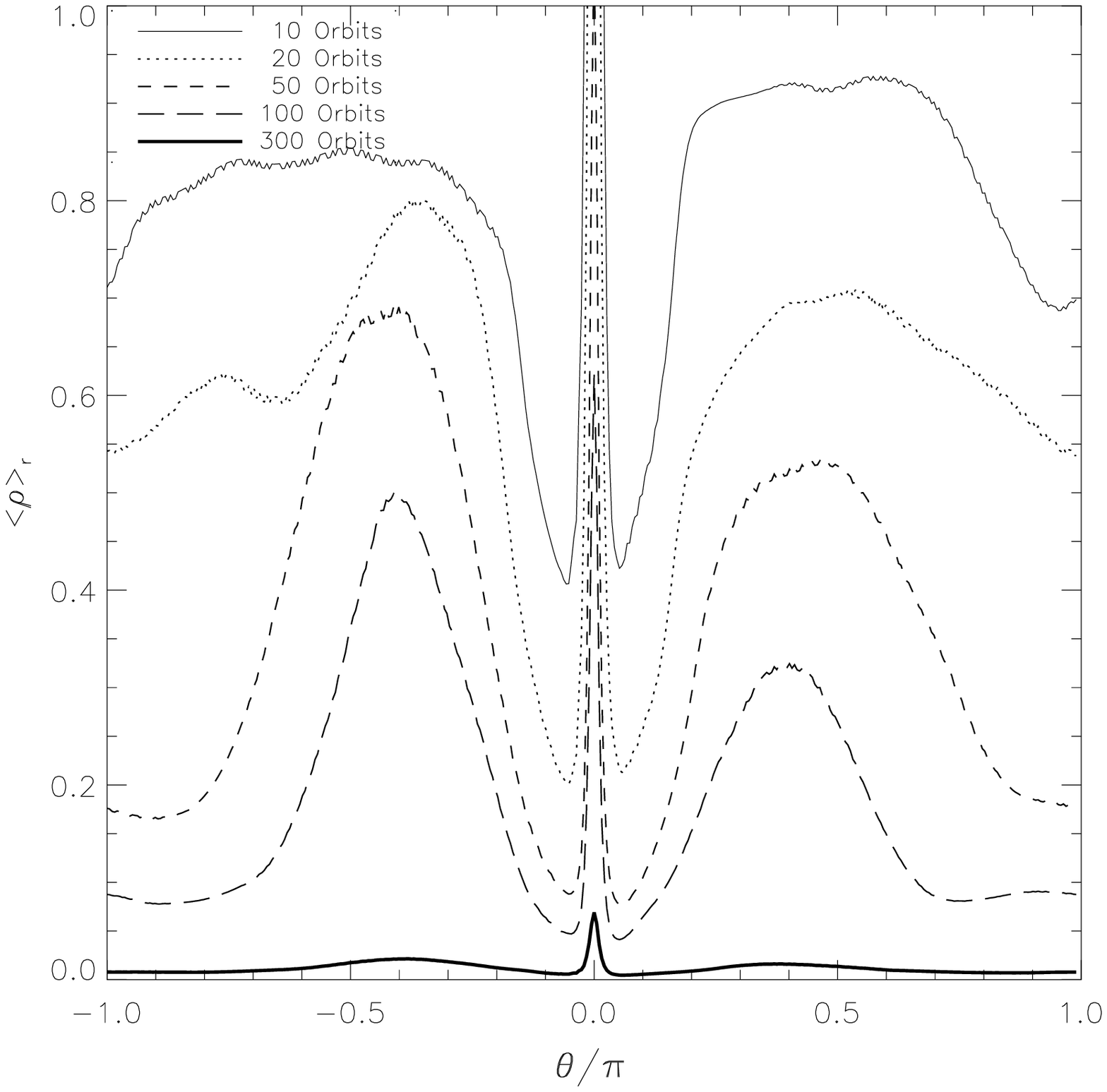}}
\end{center}
\caption{\label{fig:SJ.avg} Left: azimuthally averaged density 
  for the standard Jupiter mass run using the VL limiter.  The inset
  displays the same averaged densities for the 5 different times in
  the solid curves along with averages for $\q>0$ in the dotted curves
  and $\q<0$ in the dashed curves.  Right: radially average density
  within the gap region ($|r-a| \le 0.1a$).  Note the asymmetry
  between the $L_4$ and $L_5$ equilibrium points.}
\end{figure*}

As the simulation progresses, the planet begins to clear gas from its
orbit but not uniformly.  Gas is more readily cleared in two orbital
tracks, $1-2$ Hill radii ($R_H \equiv (\mu/3)^{1/3}$) to the inside
and the outside of the planet's orbit.  These locations are the 
approximate distance at which the averaged torque density due to
neighbouring resonances peaks (around resonance order, $m=10$; see
\citet{ward96} for details).  There is a further asymmetry in the
efficiency of the gas clearing for locations trailing and leading the
planet as demonstrated in Figure \ref{fig:SJ.avg}. It shows
the density, averaged along the full azimuth direction, at five
different times during the simulation.  Especially during the initial
formation of the gap, the total density in the trough outside the planet's
orbit is much lower relative to the density in the trough inside its
orbit.  In the inset, solid lines show the averaged density as before,
while the broken lines show the averages separated into halves for
$\q>0$ (dashed) and $\q < 0$ (dotted).  While the gap 
region leading the planet seems to clear approximately equally inside
and outside the planet's orbit, the region trailing the planet to the
outside clears more quickly than elsewhere, and the region trailing the
planet to the inside clears more slowly.

Note that there are regions of fluid within the gap which persist over
time.  These regions surround the $L_4$ and $L_5$ Lagrangian
equilibrium points located at $\q = \pm \pi/3$.  In Figure
\ref{fig:SJ.den} these are the circularly shaped overdensities which
remain within the gap.  We illustrate the evolution of these regions
in Figure \ref{fig:SJ.avg}.  The density plotted has been radially
averaged at each azimuth from $r=0.9a$ to $r=1.1a$.  Again there is a
leading-trailing asymmetry.  This asymmetry has been observed in most
planet codes to varying degrees \citep[see][]{valborro06}.

In addition to the above libration islands at the $L_4$ and $L_5$
points, there are large over-densities which begin to develop due to
the generation of vortices to either side of the gap region.
Initially several small vortices develop at roughly the same radii
but spaced in azimuth.  These vortices appear as roughly concentric
overdensities in the density contour plots (in Fig.\ \ref{fig:SJ.den}
at 20 orbits there are two such regions at both $r/a=0.75$ and
$r/a=1.3$).  As the simulation progresses the vortices grow and begin
to merge, depending on their radial location within the disk.  In
disks with large viscosity ($\nu\gtrsim 10^{-5}$), the vortices do
not form.  Similar structures have been observed in other codes at low
viscosity.  The vortices are possibly the result of Rossby-wave
instabilities \citep[][and references
  therein]{li00,li01,lovelace99,papaloizou89}. 

\begin{figure}[!b]
\begin{center}
\includegraphics[scale=0.44]{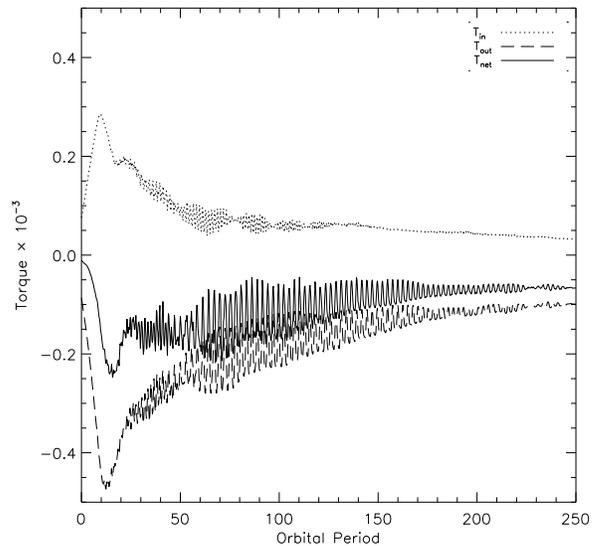}
\end{center}
\caption{\label{fig:SJ.torque}  Time evolution of the torque on the
  planet's orbit due to the disk.  The torque is broken up into
  components from the disk material inside (dotted) and outside
  (dashed) the planet's orbit.  Also plotted is the total net torque
  (solid).  Torque shown excludes material within one Hill radius,
  $R_H= (\mu /3)^{1/3}$, of the planet.} 
\end{figure}

In Figure \ref{fig:SJ.torque} we present the total torque summed over
various regions of the disk, showing its evolution over the course of
the simulation.  A running average over a period of $10$ orbits has
been performed to smooth out some of the oscillations.  As per the
treatment in \citet{valborro06}, material within the Hill sphere is
excluded, mimicking the effect of the torque-cutoff.  The
gas within this region feels the softened gravitational potential of
the planet, rather than the long-range singular potential.  As
theoretically predicted \citep{goldreich80,ward86}, the torque inside
the planet's orbit is positive (transferring angular momentum to the
planet), while that outside the planet's orbit is negative (angular
momentum is transferred from the planet to the exterior disk).  Also
as predicted, there is an asymmetry in the magnitudes of these torques
\citep{ward96}.  The net torque on the planet is negative and would
cause it to migrate inwards, were its orbit not held fixed.

The initially smooth and more broad-scale oscillations of the total torque
(on a time scale of $\sim 20$ orbits) are consistent with Phase I
evolution, as described by \citet{koller03}.  The frequency of the
subsequent rapid variations that develop matches the inverse period
of the large vortex outside the planet's orbit as measured in the
frame of the planet.

\subsection{Effects of viscosity} \label{subsec:viscosity}

\begin{figure*}[t]
\begin{center}
\resizebox{0.9\textwidth}{0.22\textwidth}{
  \includegraphics{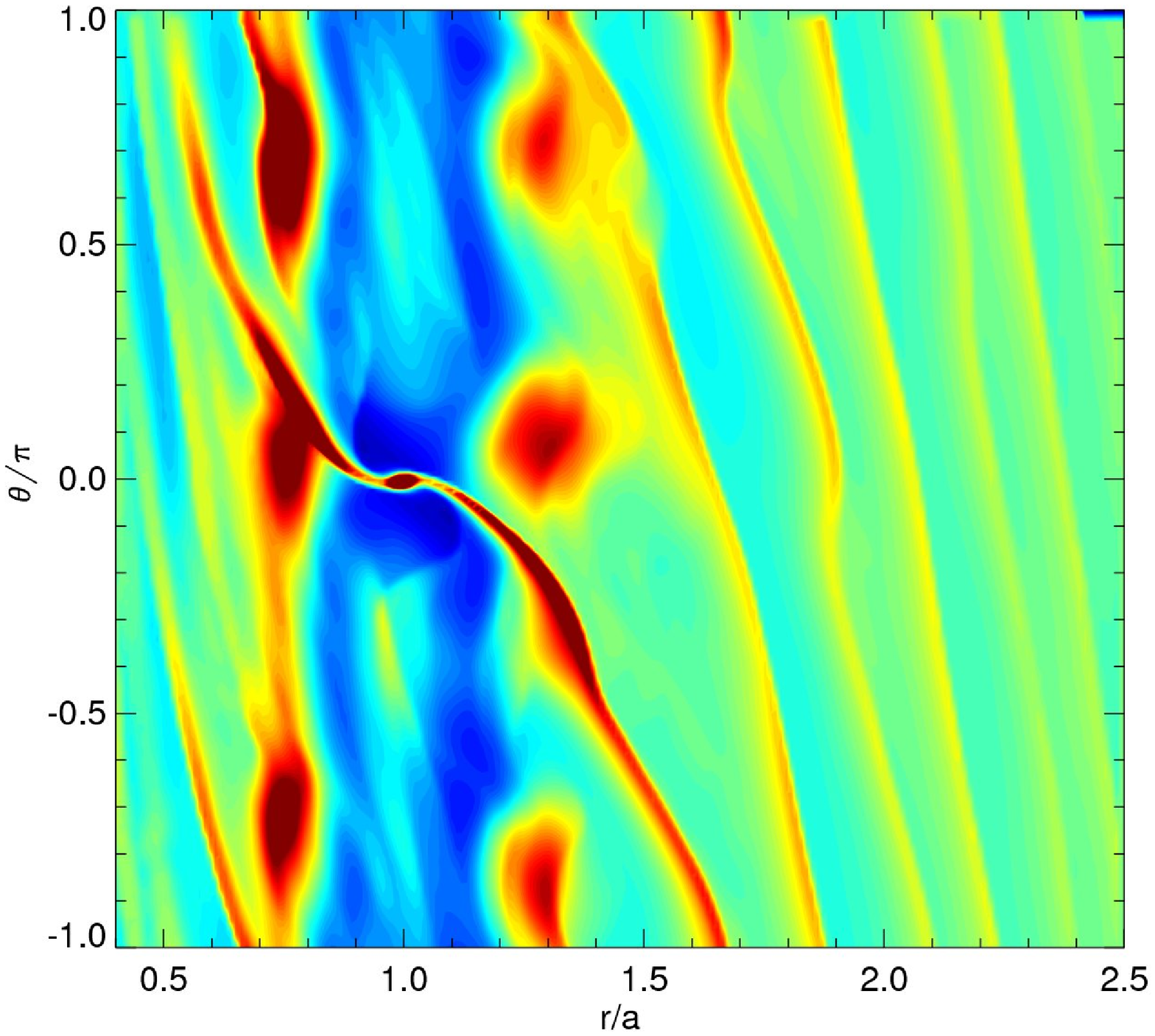}
  \includegraphics{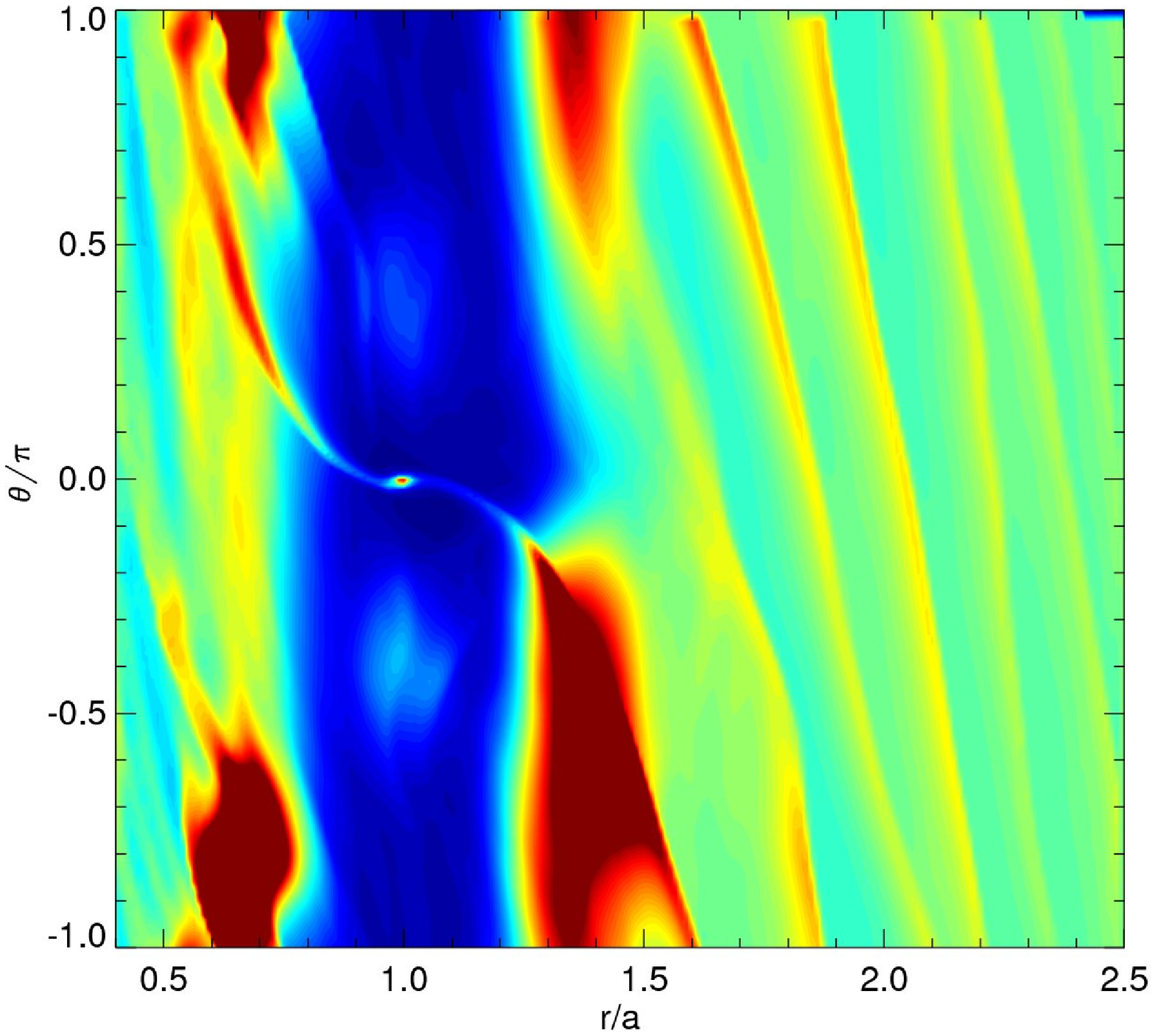}
  \includegraphics{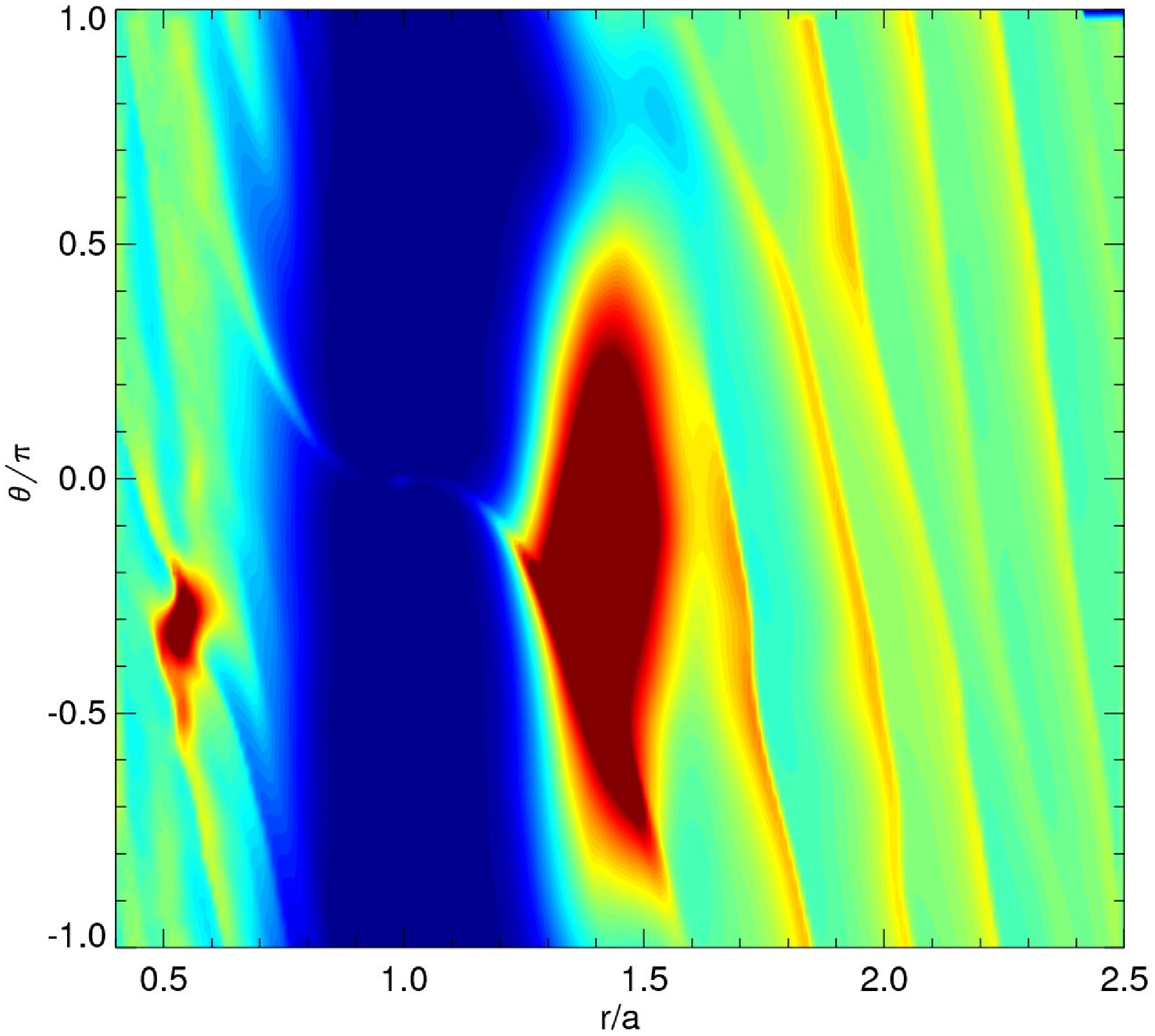}}
\resizebox{0.9\textwidth}{0.22\textwidth}{
  \includegraphics{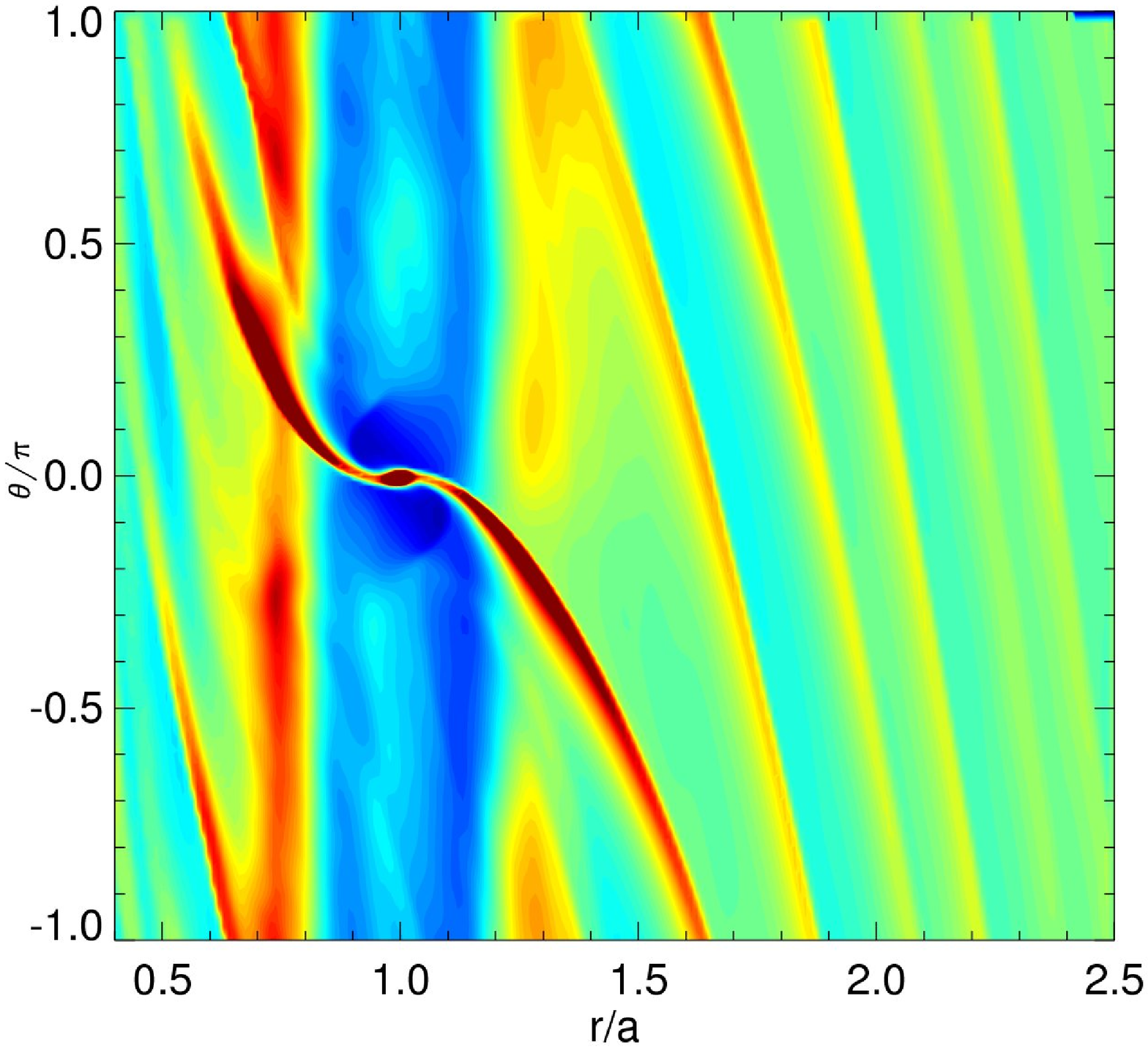}
  \includegraphics{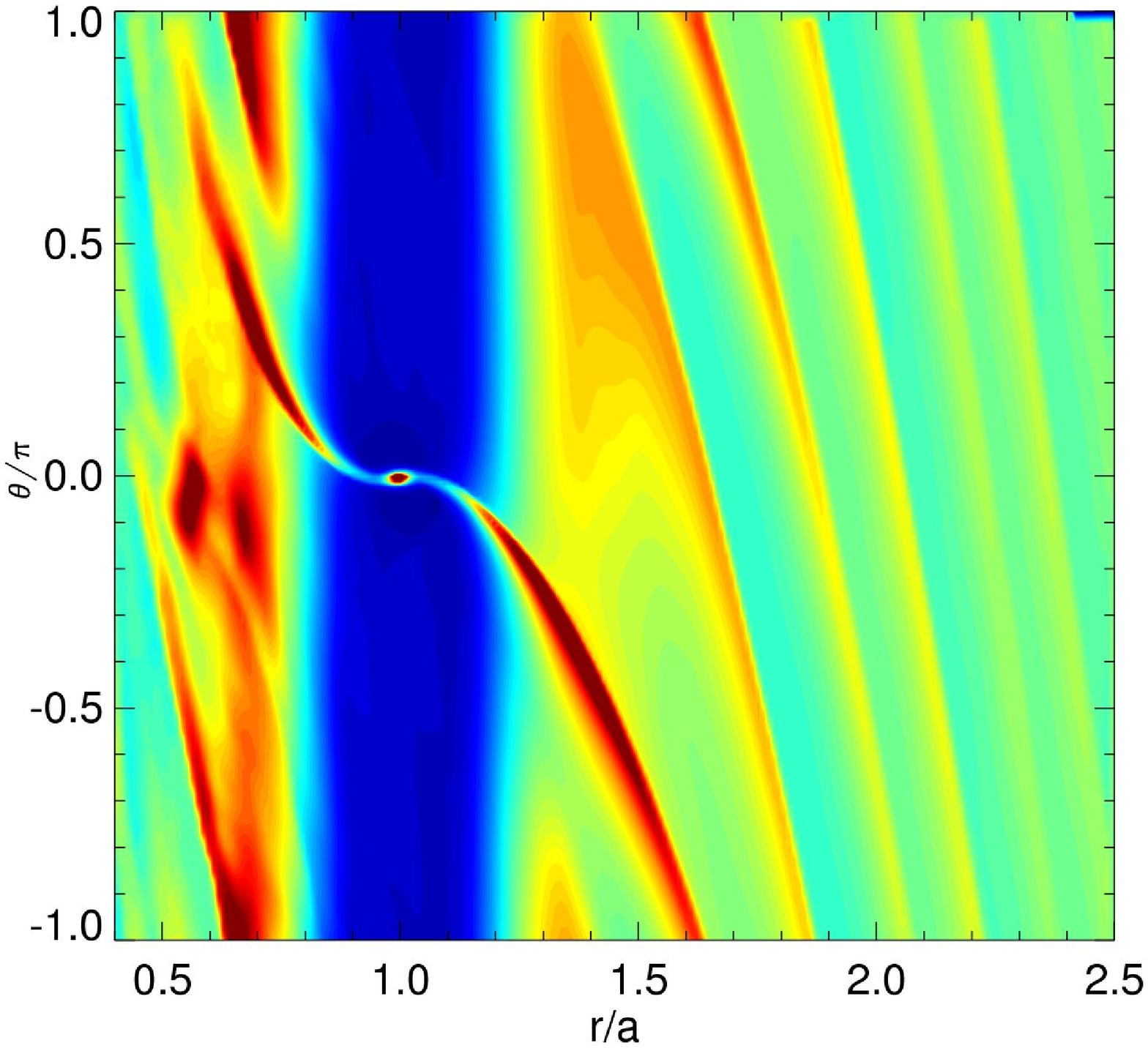}
  \includegraphics{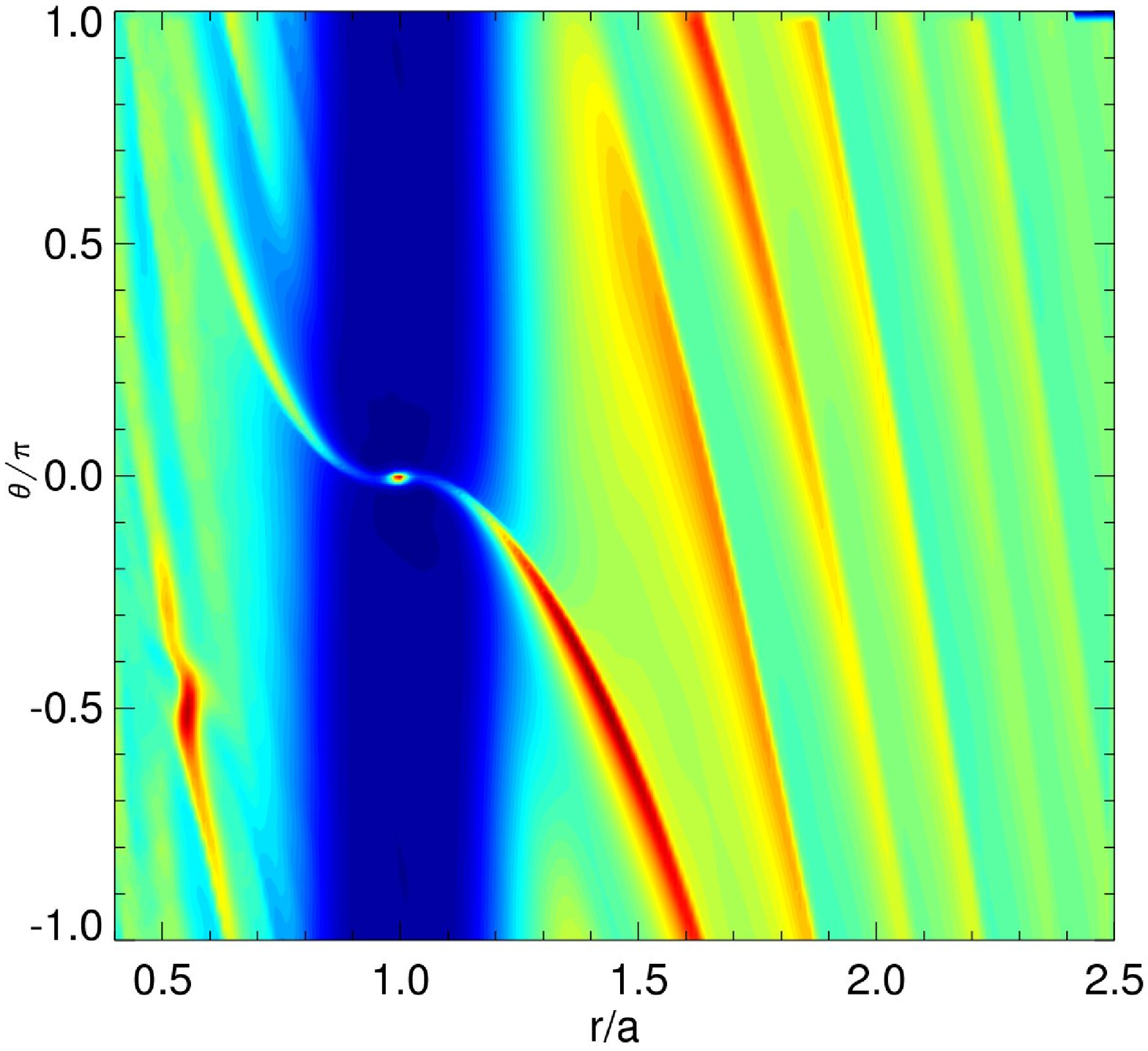}}
\includegraphics{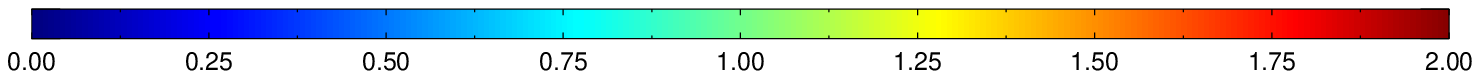}
\end{center}
\caption {\label{fig:SJv50.den} Snapshots of the density after $20$,
  $100$ and $300$ orbits, respectively from left to right. The top row
  shows results from the standard (inviscid) run.  The second row shows
  those from the viscous run ($\nu = 10^{-5}$,$\alpha \approx 0.004$).}  
\end{figure*}

\begin{figure*}[!h]
\begin{center}
\resizebox{0.9\textwidth}{0.4\textwidth}{
\includegraphics{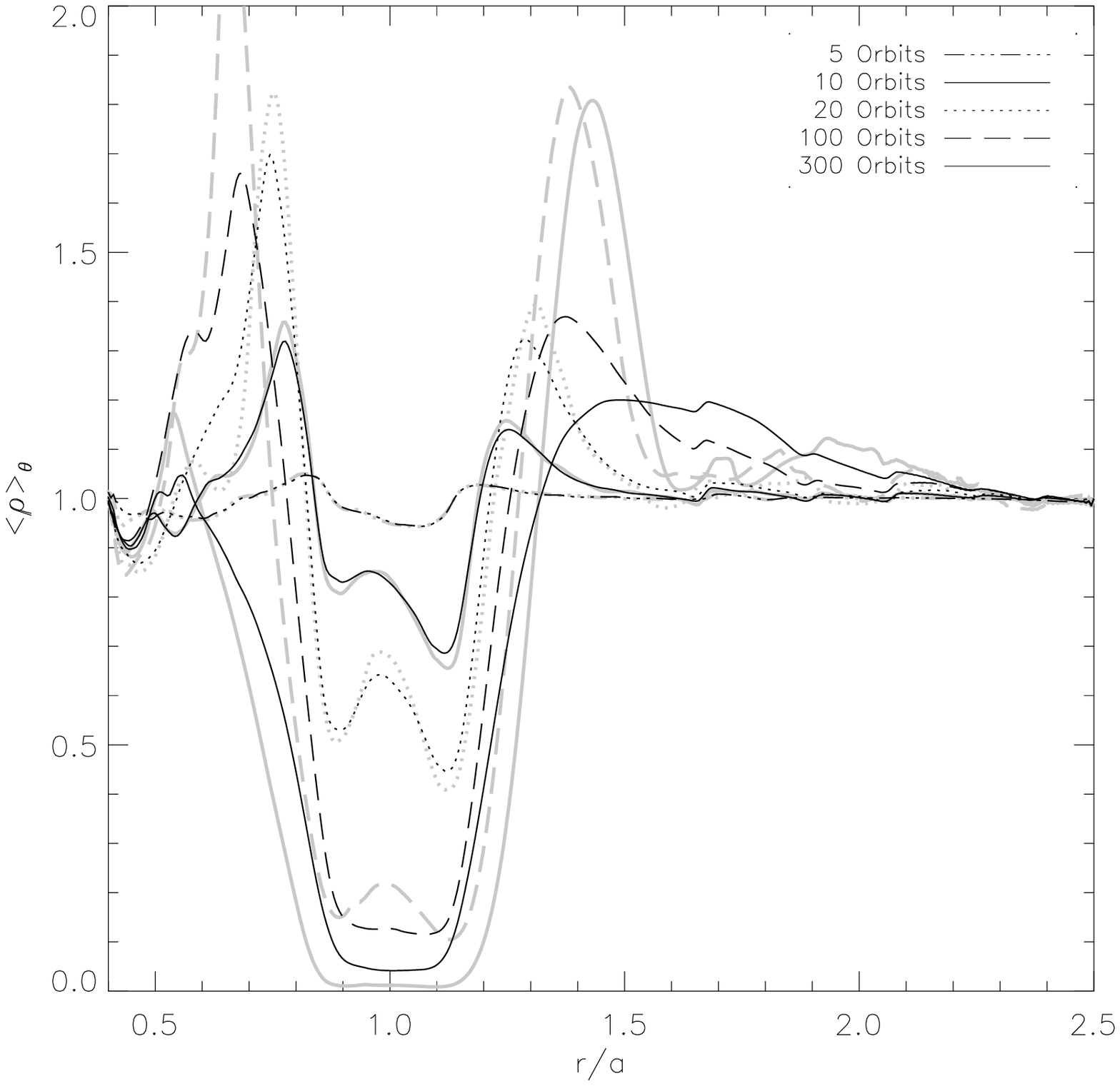}
\includegraphics{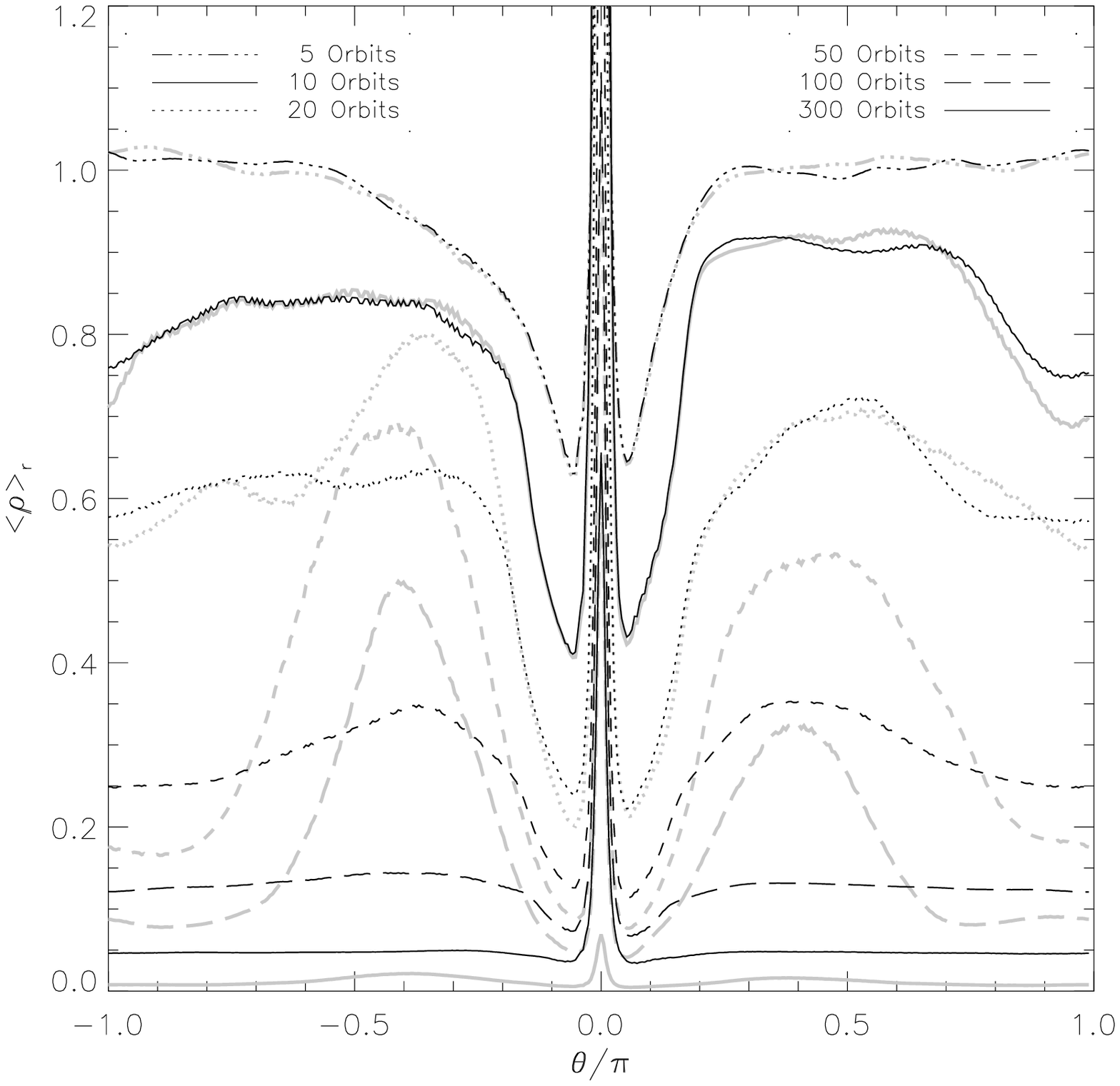}}
\caption{\label{fig:SJv50.avg} Comparisons of the azimuthally averaged
  density and the radially averaged density in the gap region.  The
  dark lines have $\nu=1\times 10^{-5}$; the pale lines have no added
  viscosity.  The viscosity makes the islands of fluid surrounding the
  $L_4$ and $L_5$ points unmaintainable.}
\end{center}
\end{figure*} 

The presence of physical viscosity tends to smooth perturbations of
physically conserved quantities.  We parametrize the viscosity as a
uniform alpha-disk model, whereby the viscosity coefficient $\nu$
and turbulent efficiency $\alpha$ are related to one another by $\nu = \alpha
(H/r)^2 \sqrt{GM_*r}$.  We substitute this relationship for the
parameter $\nu$ in our implementation of viscosity (see
eq. [\ref{eq:viscosity}]).  For a range of $\alpha$-values taken
to be $\alpha = 10^{-2}-10^{-3}$ \citep{hartmann98} and $H/r=0.05$,
one finds $\nu \approx 10^{-4.5}-10^{-5.5}$. Figure
\ref{fig:SJv50.den} shows the evolution of the density for a
simulation with $\nu=1\times 10^{-5}$ (at $r=1$;  all subsequent
values of $\nu$ are quoted for $r=1$).  While the evolution of the
spiral arms occurs on the same timescale as in the inviscid run, many of
the structures in the simulation are no longer present when viscosity
is added: the $L_4$ and $L_5$ libration islands are less marked and
the vortex lines seen previously are absent. Figure
\ref{fig:SJv50.avg} compares the radially and azimuthally averaged
densities at several orbital times for runs with and without added
physical viscosity.  While it appears that there are libration islands and
vortices in the viscous run which begin to develop, their radial and
azimuthal structure is smoothed out by the viscosity. Note that if the
added physical viscosity is reduced in  order by another
half-magnitude, the libration islands and vortices are once again
present (see below). 

\subsubsection{Calibration of viscosity} \label{subsubsec:viscositycalbration}

\begin{figure*}
\begin{center}
\resizebox{0.9\textwidth}{0.35\textwidth}{
\includegraphics{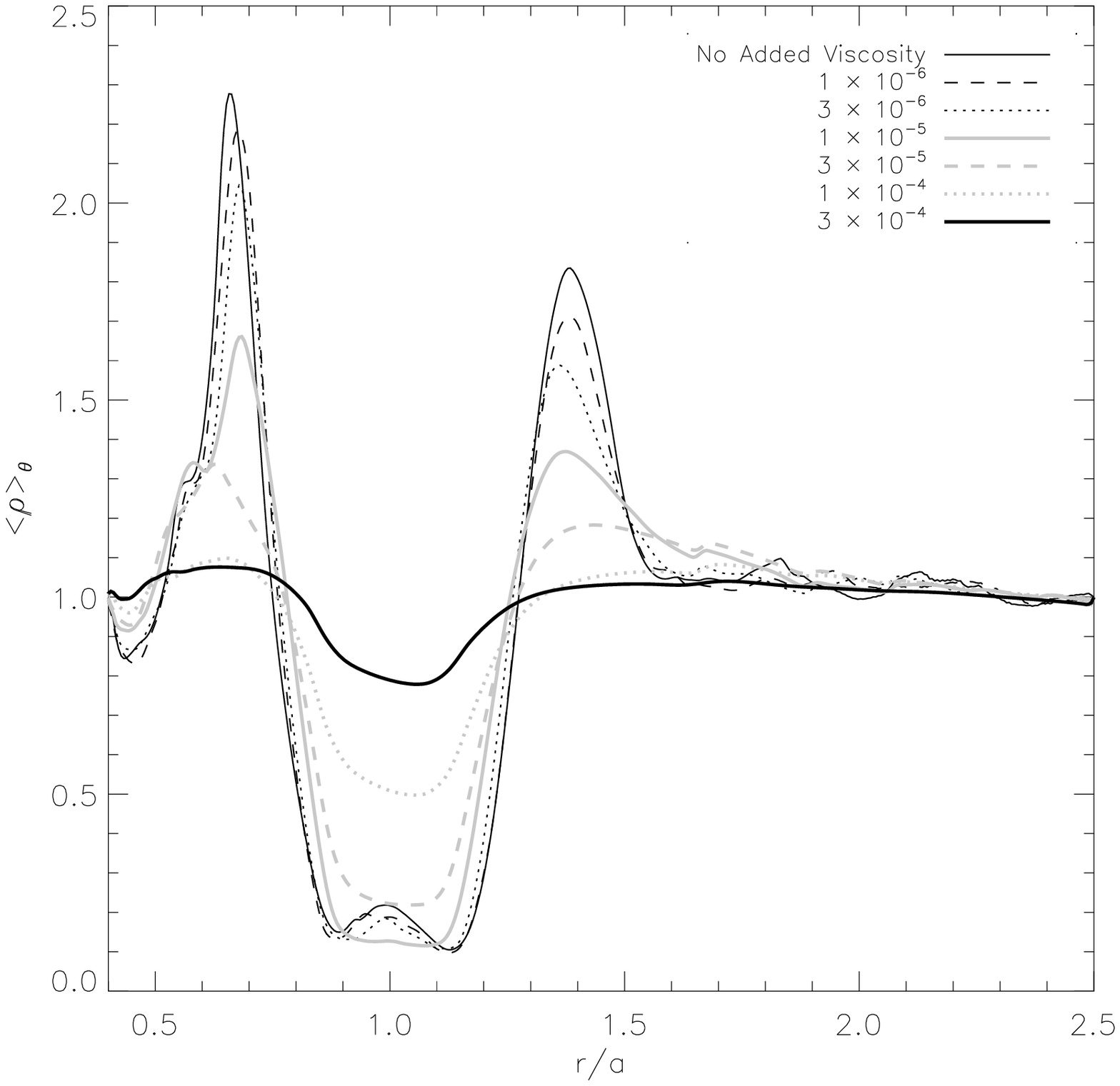}
\includegraphics{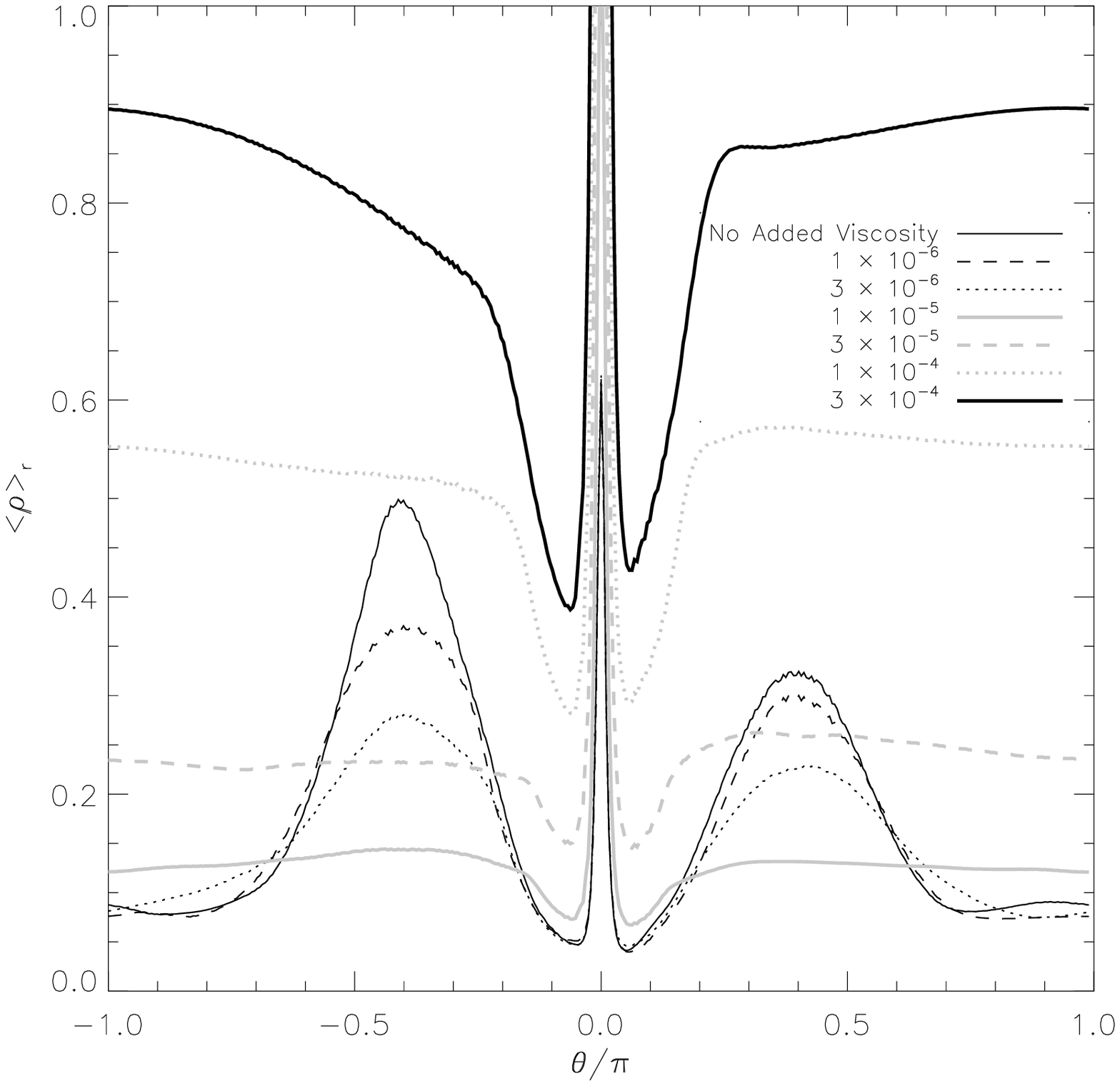}}
\end{center}
\caption {\label{fig:SJvall.avg} Comparisons of the azimuthally
  averaged density and the radially averaged density in the gap region
  at different values of added physical viscosity.  All results use
  the VL limiter and are taken at $100$ orbits.  The addition
  of physical viscosity, even at the level of $\sim 10^{-6}$, alters
  the results of the simulation, especially within the gap region,
  suggesting the numerical viscosity of the simulation is at the same
  level or lower. }
\end{figure*} 

Any numerical algorithm exhibits numerical viscosity due to the
combined results of diffusive and dispersive errors (discussed in \S
\ref{subsec:wavesplitsolution}).  While physical viscosity is
characterized by the form of the stress-strain tensor (in a Newtonian
fluid there is a presumed linear relationship between stress and
strain), an algorithm's numerical viscosity will differ from the
physical viscosity, not only in the amplitude or spatial dependence of
the viscosity coefficient, but also in the relationship between the
stress and strain. Except for the most diffusive schemes, the
numerical viscosity of an algorithm usually displays a nonlinear
dependence on the velocity gradient. These higher order terms tend to
introduce dispersion. 

Despite this potential incongruity between physical and numerical
viscosity, it is useful to have an estimate for the value of the
viscosity coefficient at which the two viscosities may be considered
approximately equal in their effects.  In order to determine this
value, we compare the results of several simulations with various
levels of added physical viscosity (implemented as described in \S
\ref{subsec:viscosityimplementation}).  By reducing the value of the
viscosity coefficient $\nu$ to a point where the results of the
simulations are approximately the same irrespective of its addition,
we obtain an estimate for the numerical viscosity present in the
simulation.  We note that the measure of viscosity obtained in this
manner depends on the particulars of the setup.  In higher resolution
runs, for example, we would expect the actual numerical viscosity
exhibited to be smaller than the value we obtain from lower resolution
runs.  Likewise, we note it would likely be different in three
dimensional simulations where there exist extra degrees of freedom.

Figure \ref{fig:SJvall.avg} shows the results of decreasing the value
of the physical viscosity coefficient from $\nu=3\times 10^{-4}$ to
$\nu=1\times 10^{-6}$ in roughly half-magnitude increments.  These
simulations are performed using the VL limiter scheme run for $100$
orbits.  Increasing the level of viscosity present narrows the gap
width, decreases its depth, and softens the density gradient at its
edges.  Even when introduced at a level of $~ 10^{-6}$ there is a
difference in the averaged density profile, especially in the gap
region.  This suggests the numerical viscosity of the code is of
approximately the same magnitude or less.  Results using other
limiter schemes are analogous and suggest a similar level of diffusion
with more or less dispersion.  They are discussed further in Section
\ref{subsec:limiters}.  

\begin{figure}[b]
\begin{center}
\includegraphics[scale=0.44]{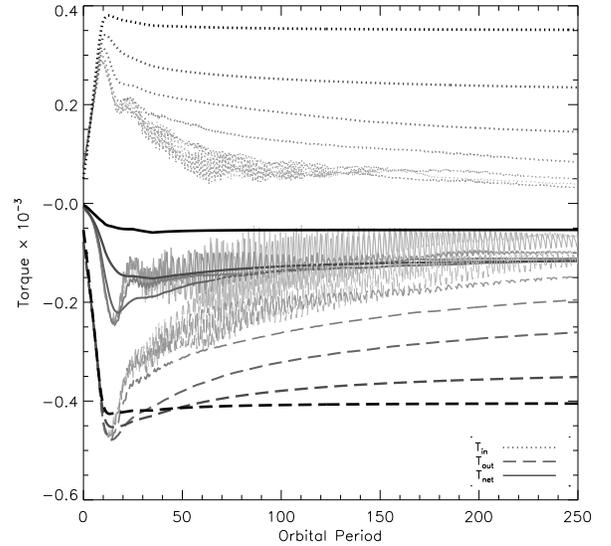}
\end{center}
\caption {\label{fig:SJvall.torque} Time evolution of the torque on
  the planet from the disk.  The torque is broken up into
  components from the disk material inside (dotted) and outside
  (dashed) the planet's orbit. Also plotted is the total net torque (solid).
  The torque calculated is only that from material further out than
  one Hill radius from the planet.  The palest lines have no added 
  viscosity and from there the viscosity increases in half-magnitude
  increments from $\nu= 1\times 10^{-6}$ to $3\times 10^{-4}$ for the
  heavy black lines.} 
\end{figure} 

\begin{figure*}[!t]
\begin{center}
\resizebox{0.9\textwidth}{0.21\textwidth}{
  \includegraphics{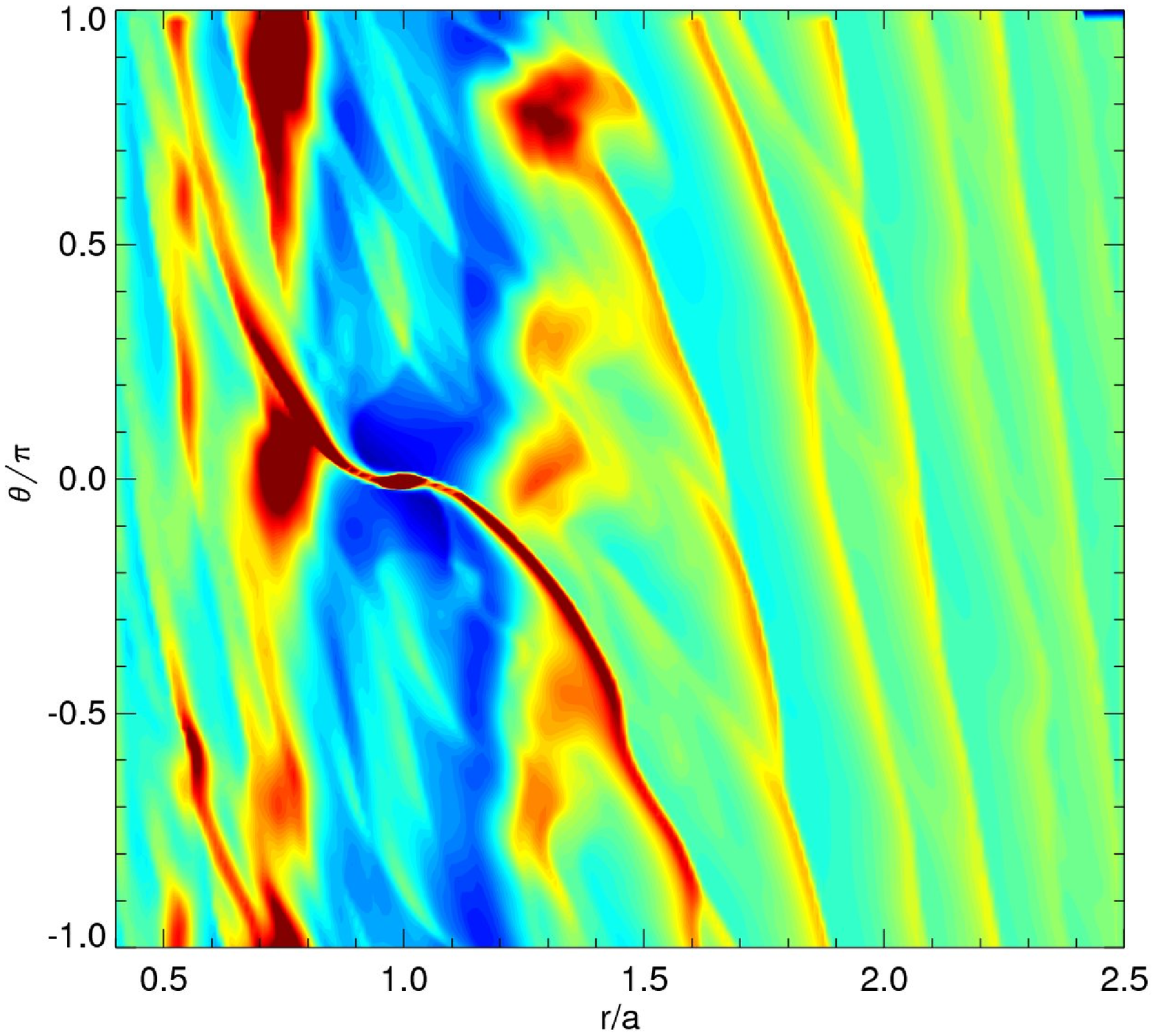}
  \includegraphics{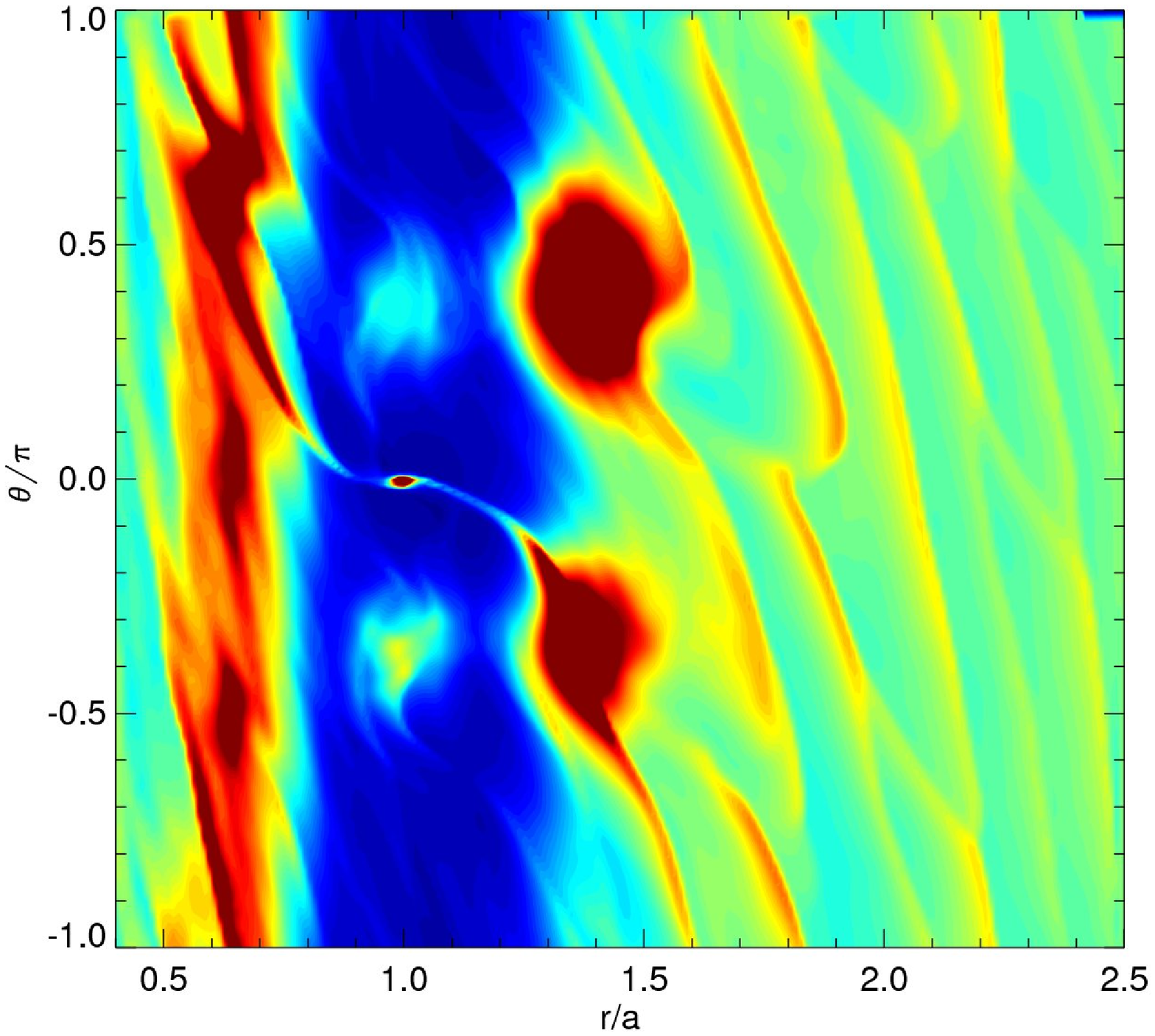}
  \includegraphics{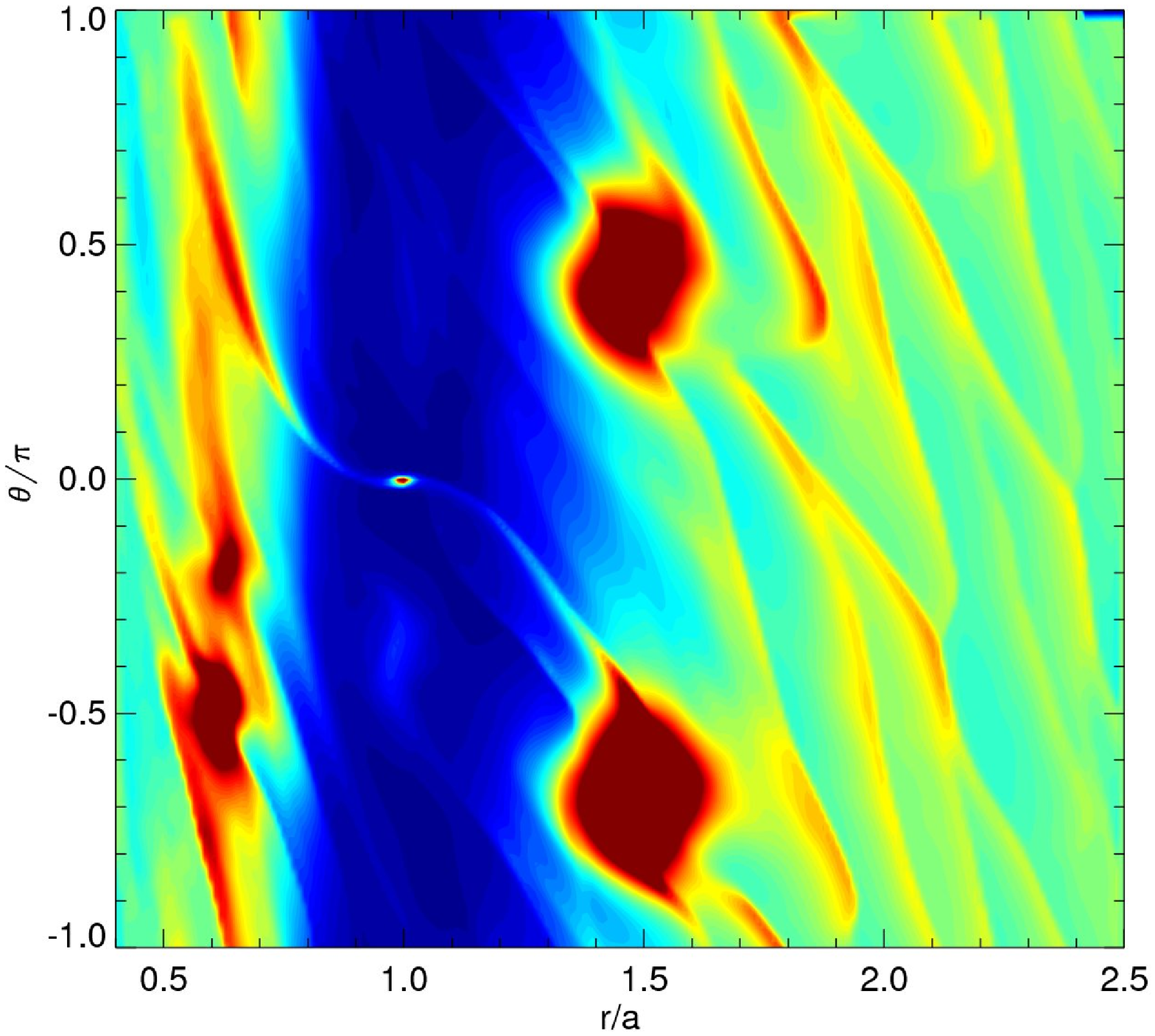}}
\resizebox{0.9\textwidth}{0.21\textwidth}{
  \includegraphics{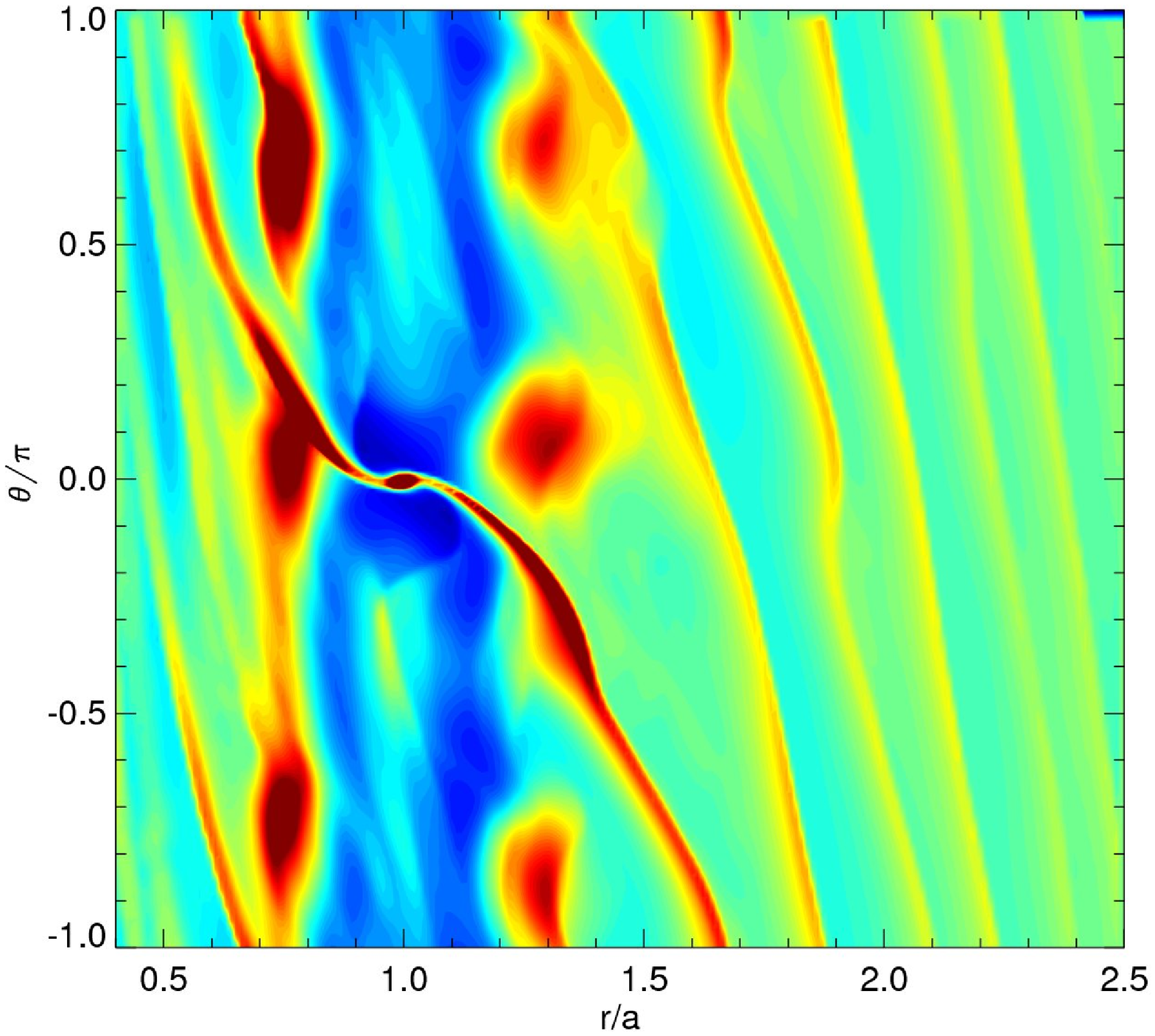}
  \includegraphics{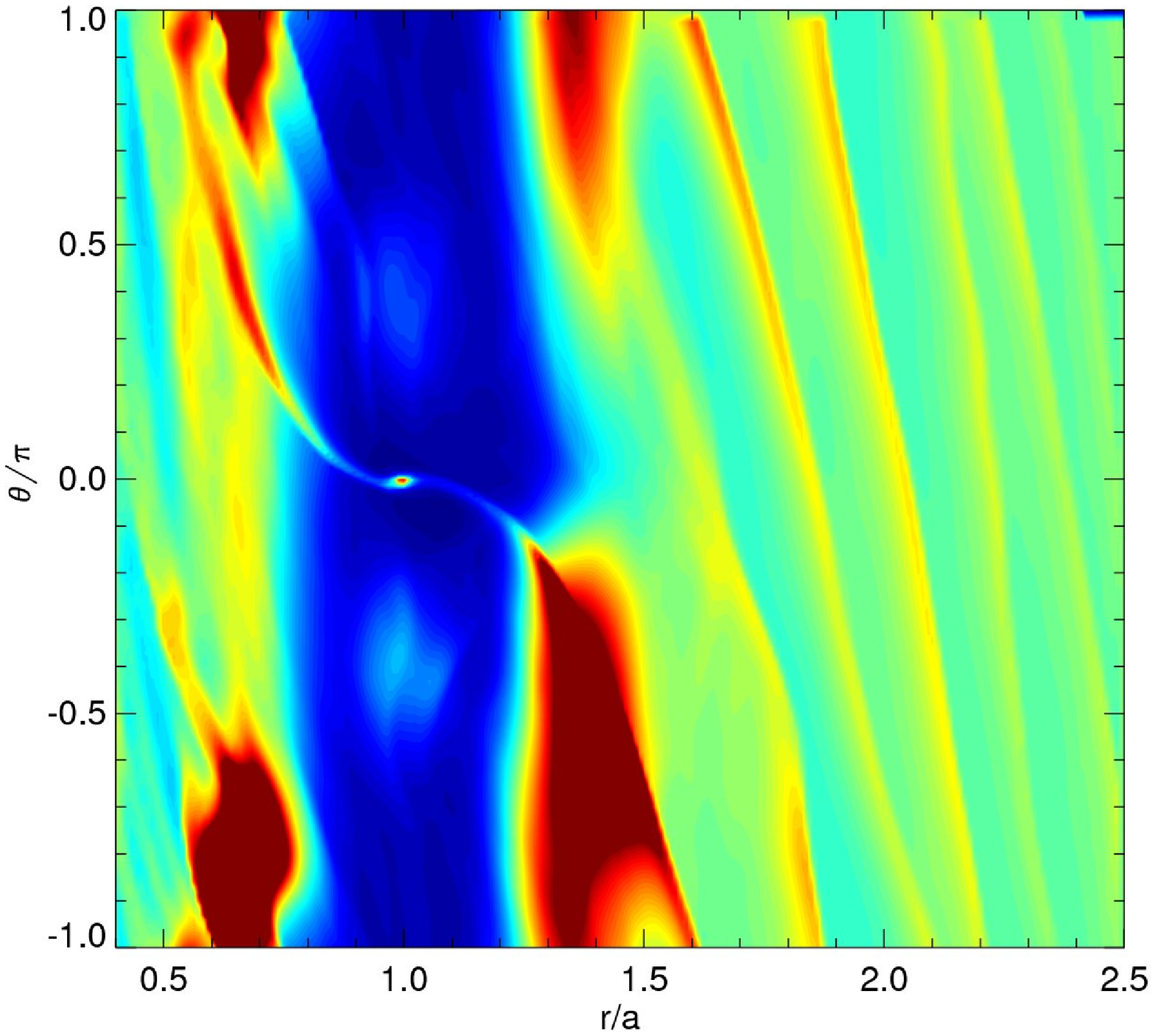}
  \includegraphics{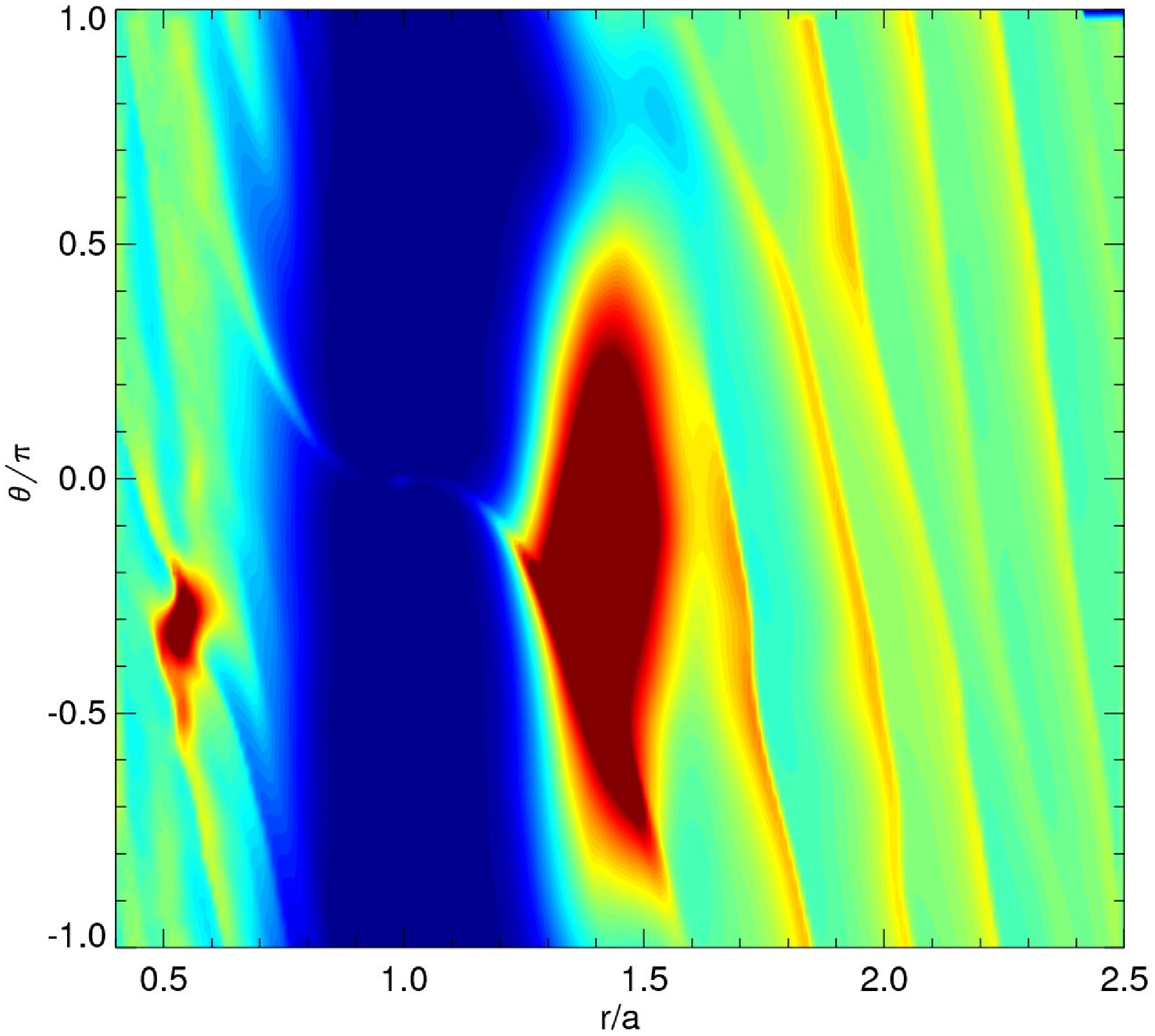}}
\resizebox{0.9\textwidth}{0.21\textwidth}{
  \includegraphics{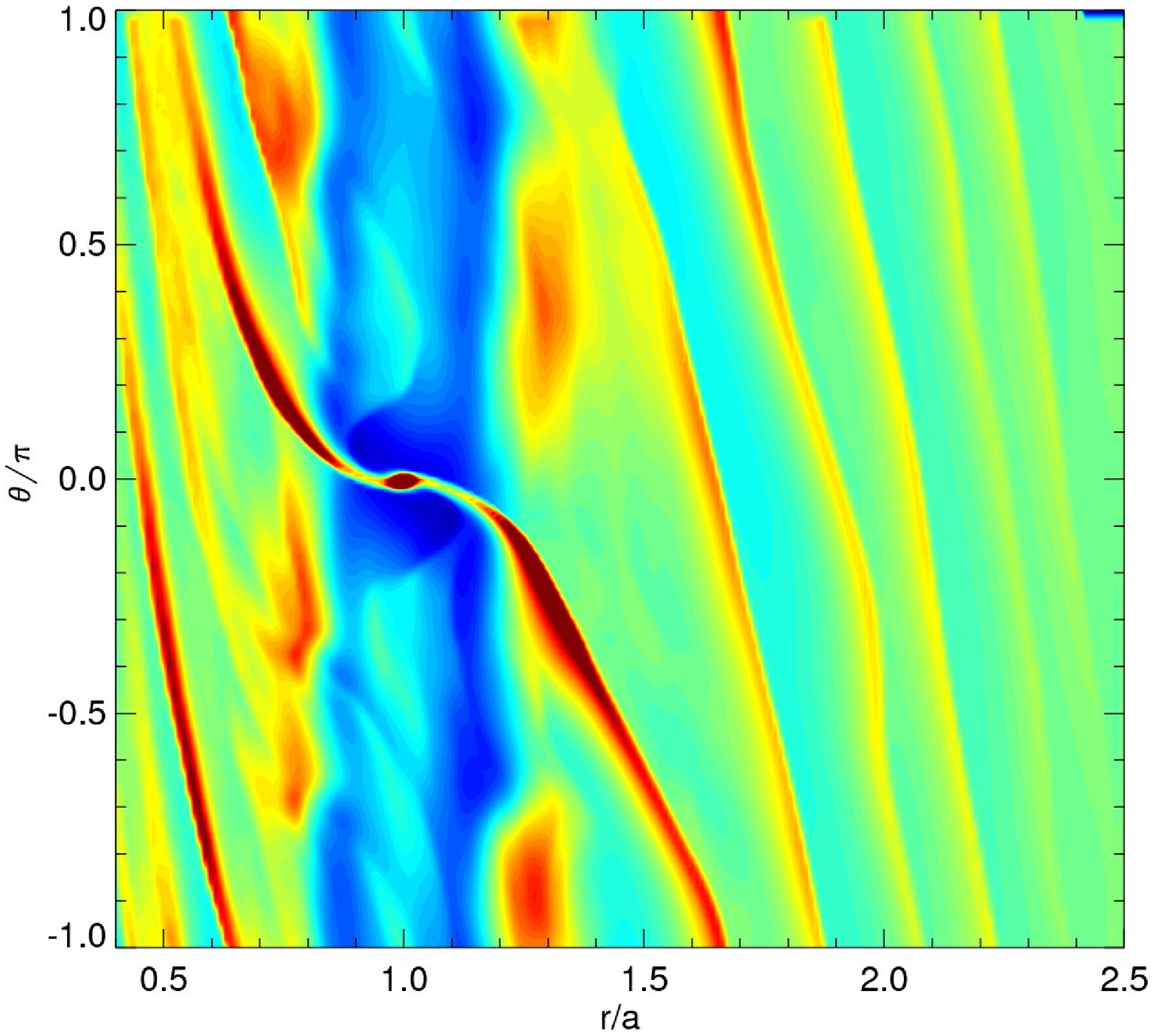}
  \includegraphics{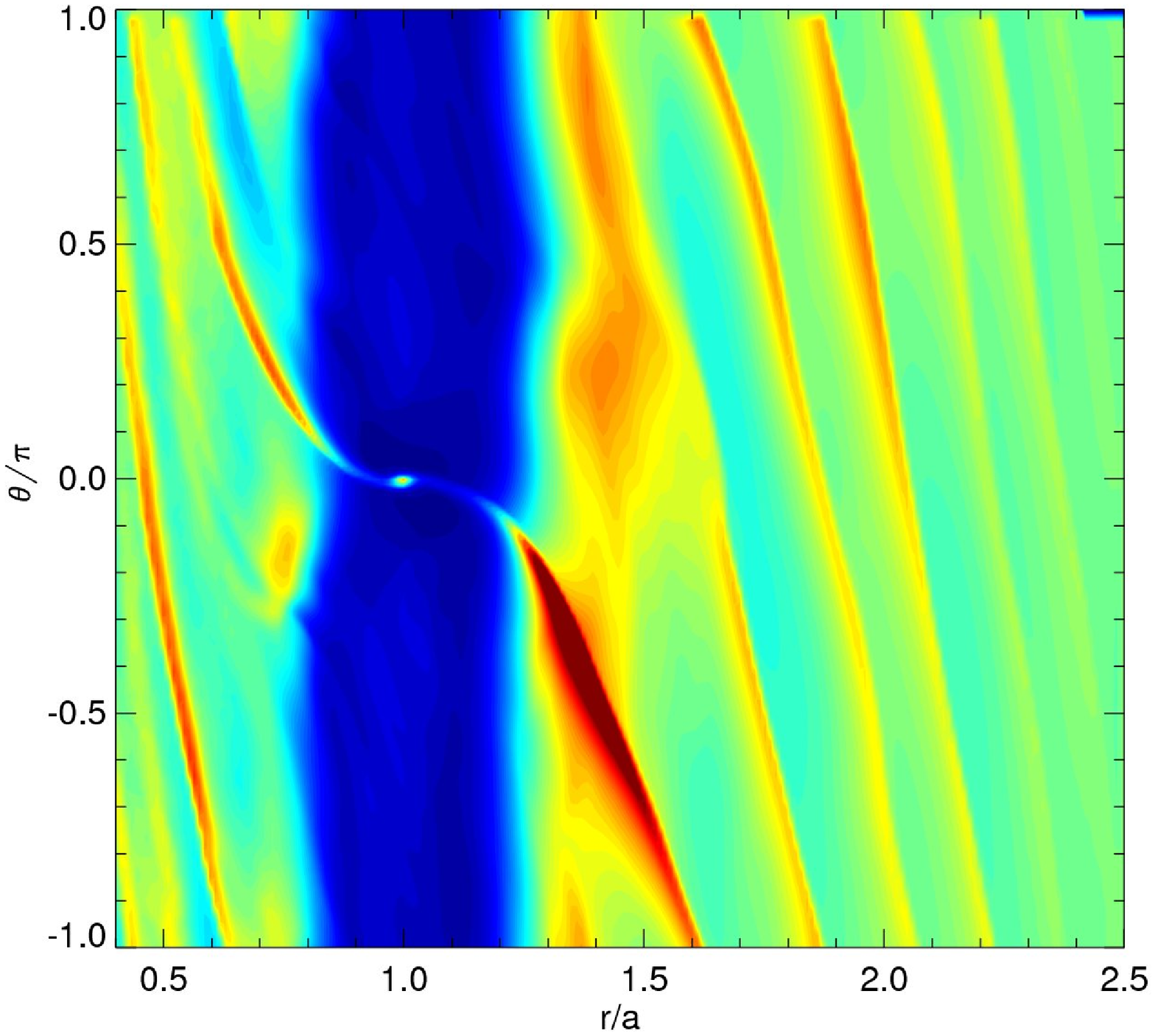}
  \includegraphics{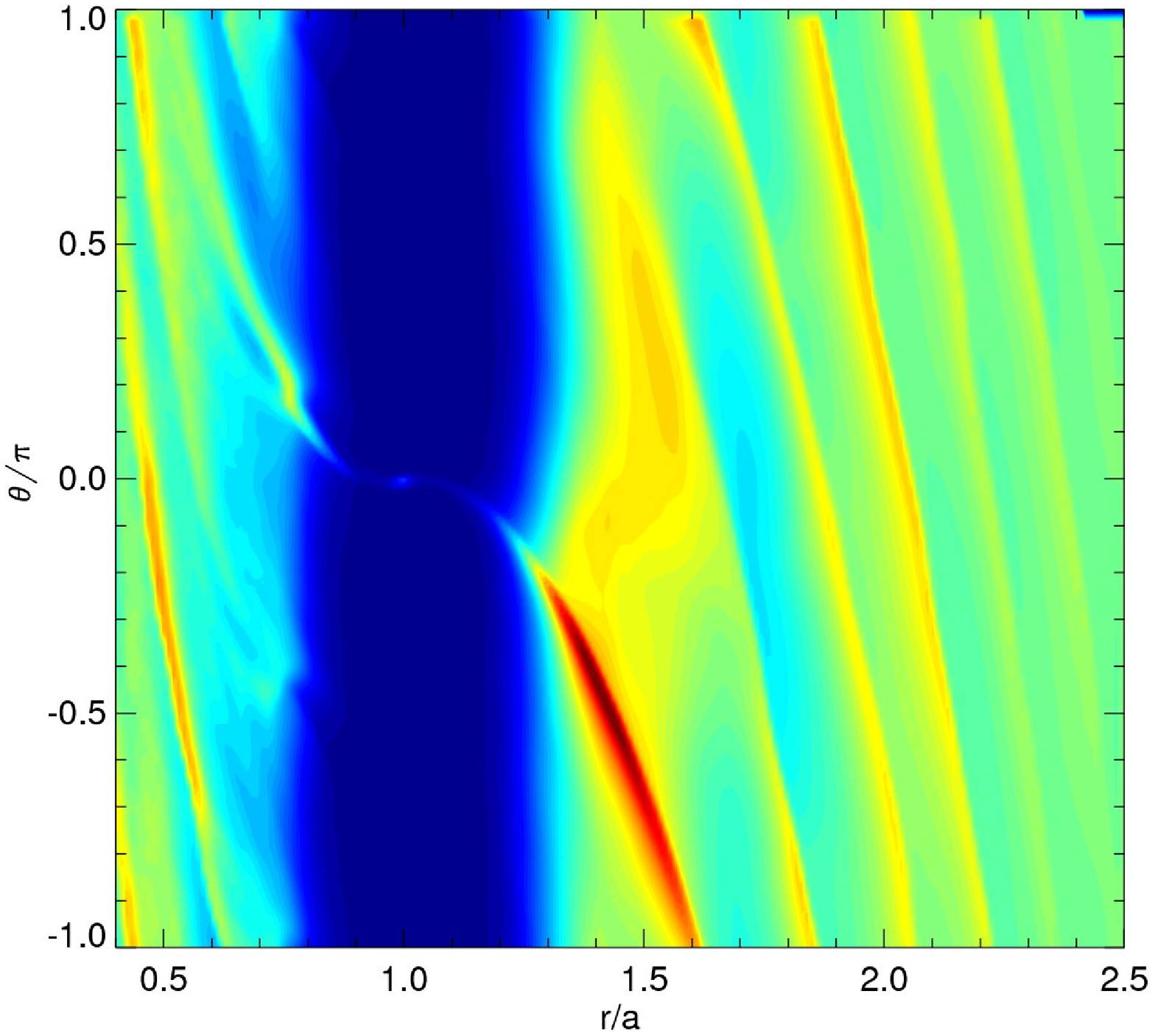}}
\resizebox{0.9\textwidth}{0.21\textwidth}{
  \includegraphics{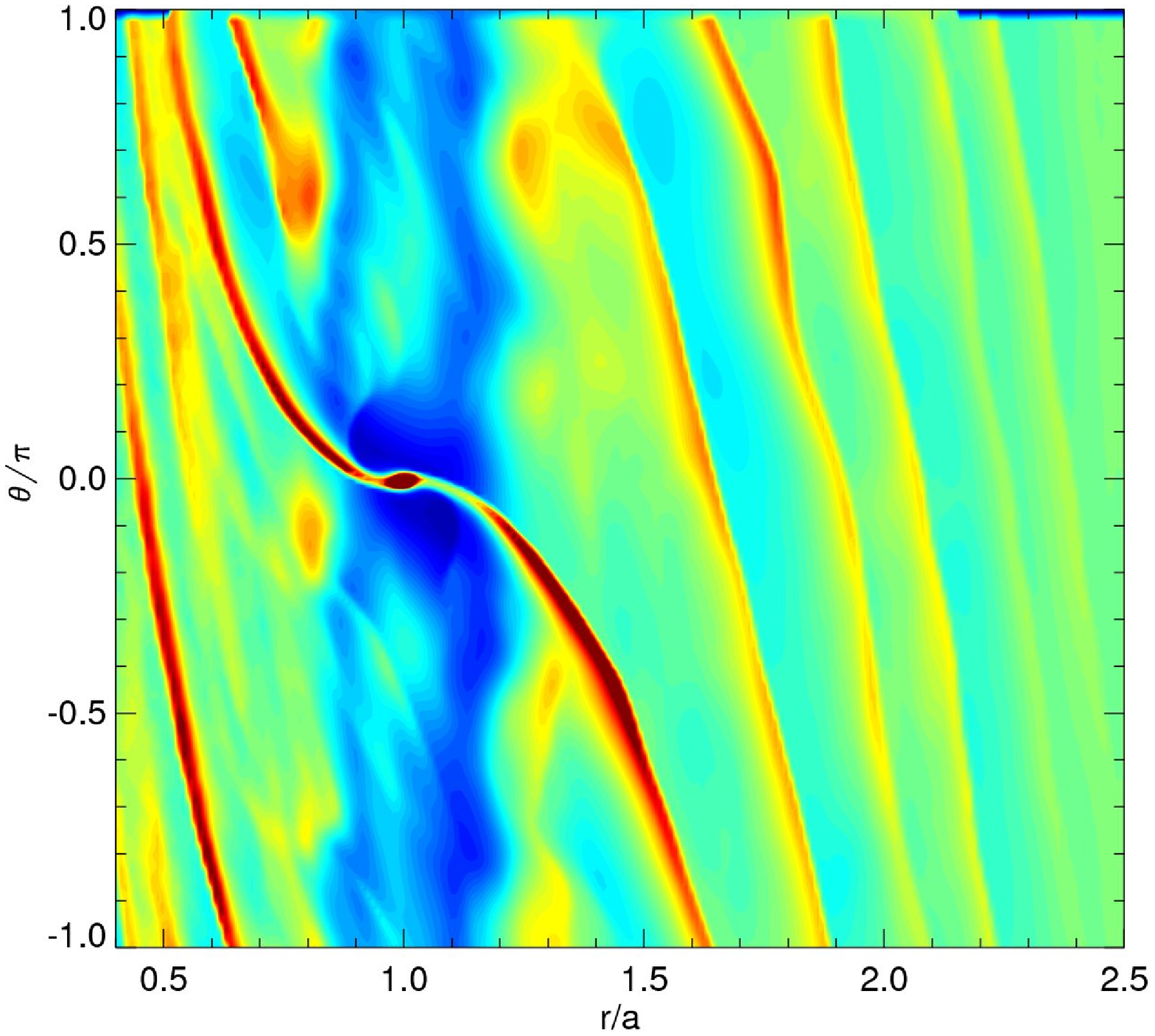}
  \includegraphics{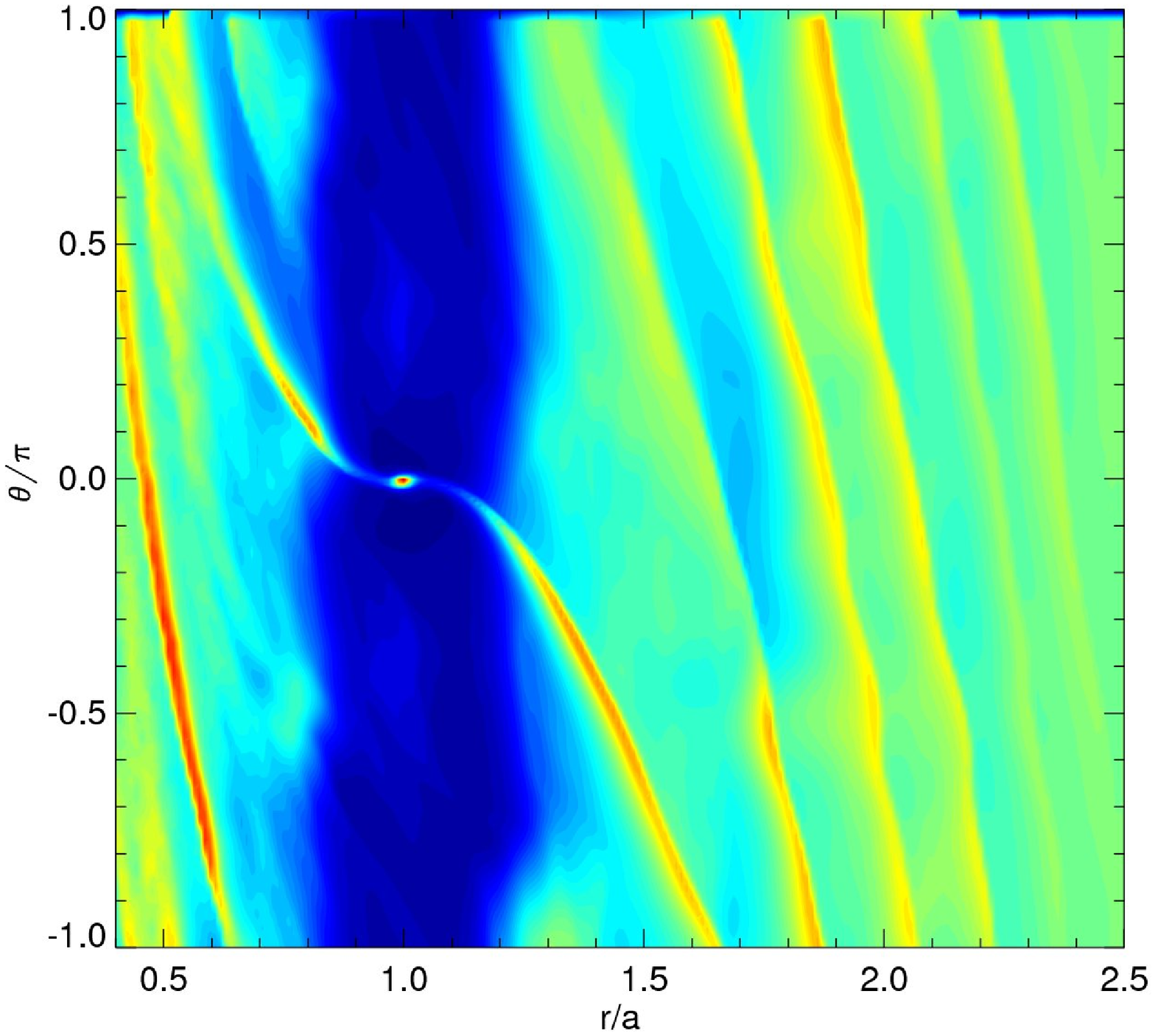}
  \includegraphics{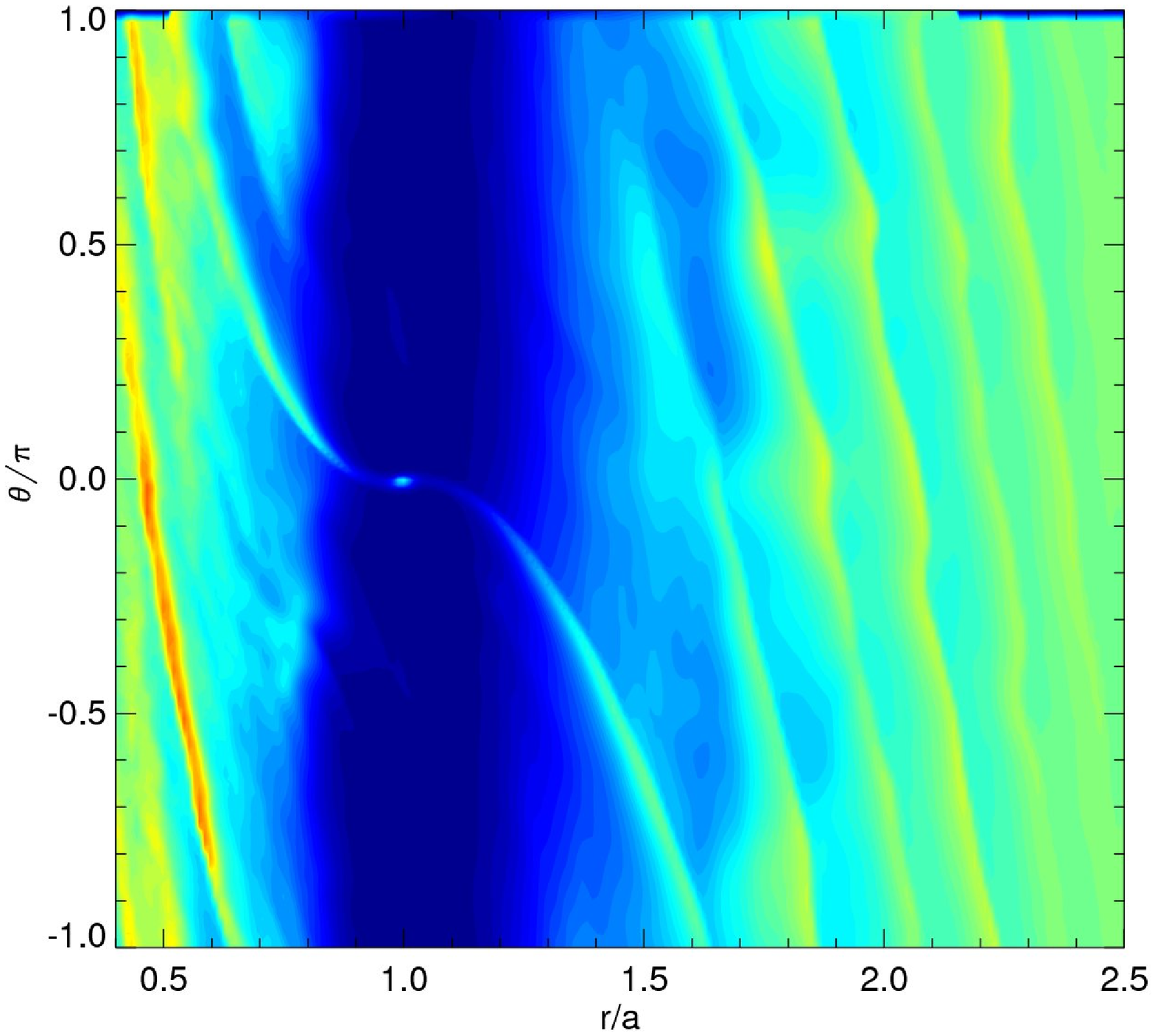}}
\includegraphics{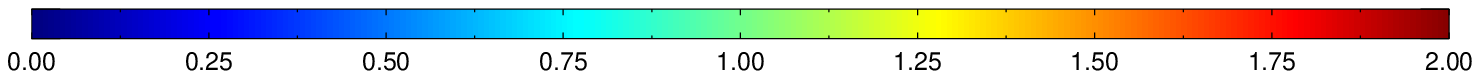}
\end{center}
\caption{\label{fig:SJ0Hhu.den} Density contours after $20$, $100$ and
  $300$ orbits, respectively from left to right. The top row shows
  results from a run which does not use the fast-advection algorithm
  while the second row shows results from the standard run (which
  implements the fast-advection algorithm); both of these two runs use
  the standard solution set $(\rho,\rho u_r, {\cal H})$. The third and 
  fourth rows show results from runs using the fast-advection algorithm
  but which use the solution sets, $(\rho, \rho u_r, {\cal H}/r)$ and
  $(\rho, \rho u_r, \rho u_\q)$, respectively.  The two vortices still
  present in the last plot of the top row have merged into a single
  vortex by 400 orbits as in the standard run.} 
\end{figure*}

In Figure \ref{fig:SJvall.torque} we show the variation of the
torques with viscosity.  All the torques are similar for the three
lowest values of added viscosity.  Only at values of $\nu=1\times
10^{-5}$ or larger are the differences discernible---the torques from
the inner and outer parts of the disk both increase in magnitude, but
the net torque decreases for large enough viscosities. In addition,
the rapid oscillations damp beyond $\nu=10^{-5}$ because the large
outer vortex is no longer able to form.  The increase of material
within the gap region with larger viscosity could explain the increase
in magnitude of the inner and outer torques.  With a large enough
viscosity, the asymmetry of the density profile causing the inner and
outer torques on the planet are smoothed out, and the net torque
decreases.

\subsection{Influence of fast-advection algorithm and choice of solution vector} \label{subsec:influences}

\begin{figure*}
\begin{center}
\resizebox{0.9\textwidth}{0.4\textwidth}{
\includegraphics{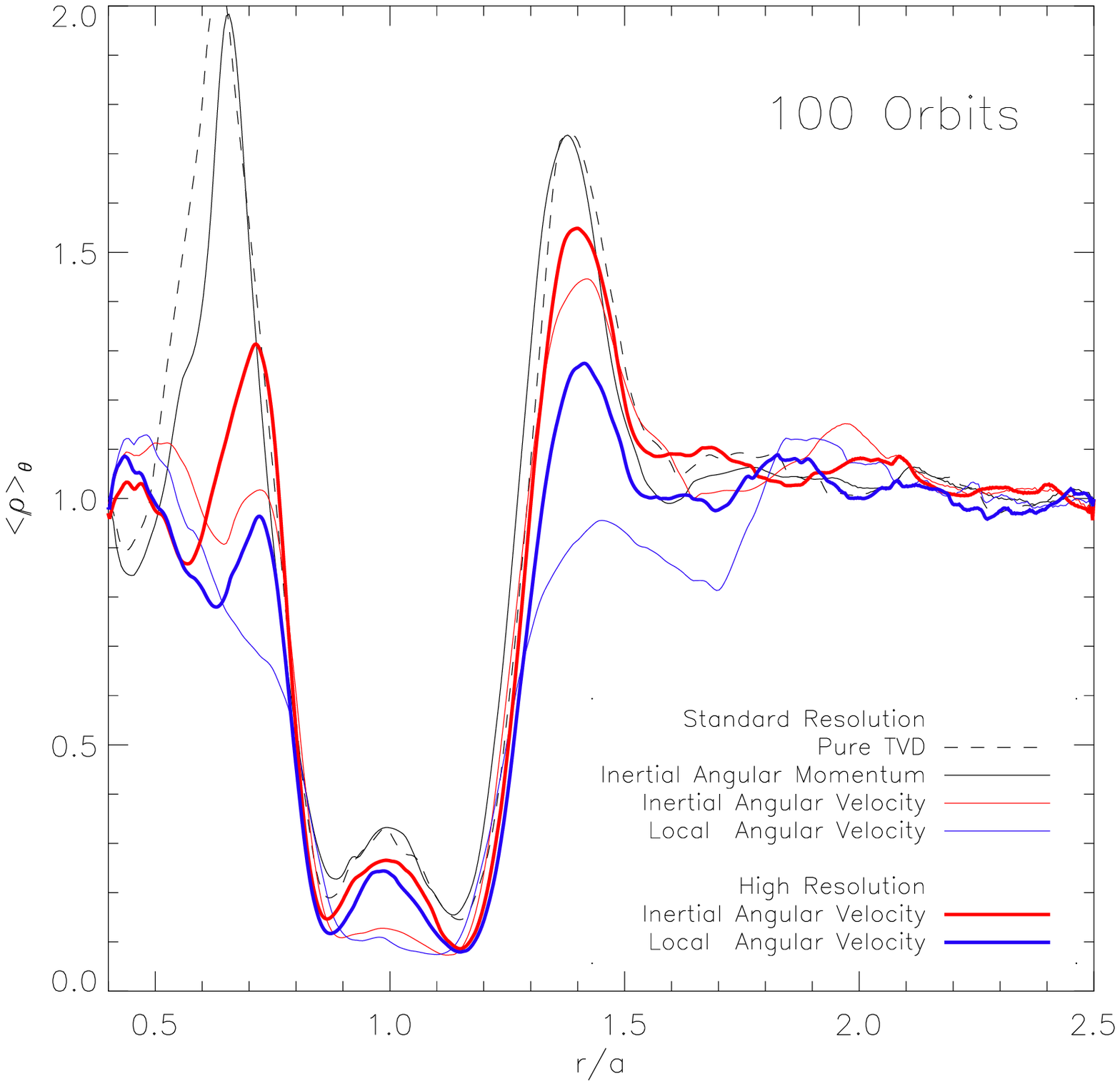}
\includegraphics{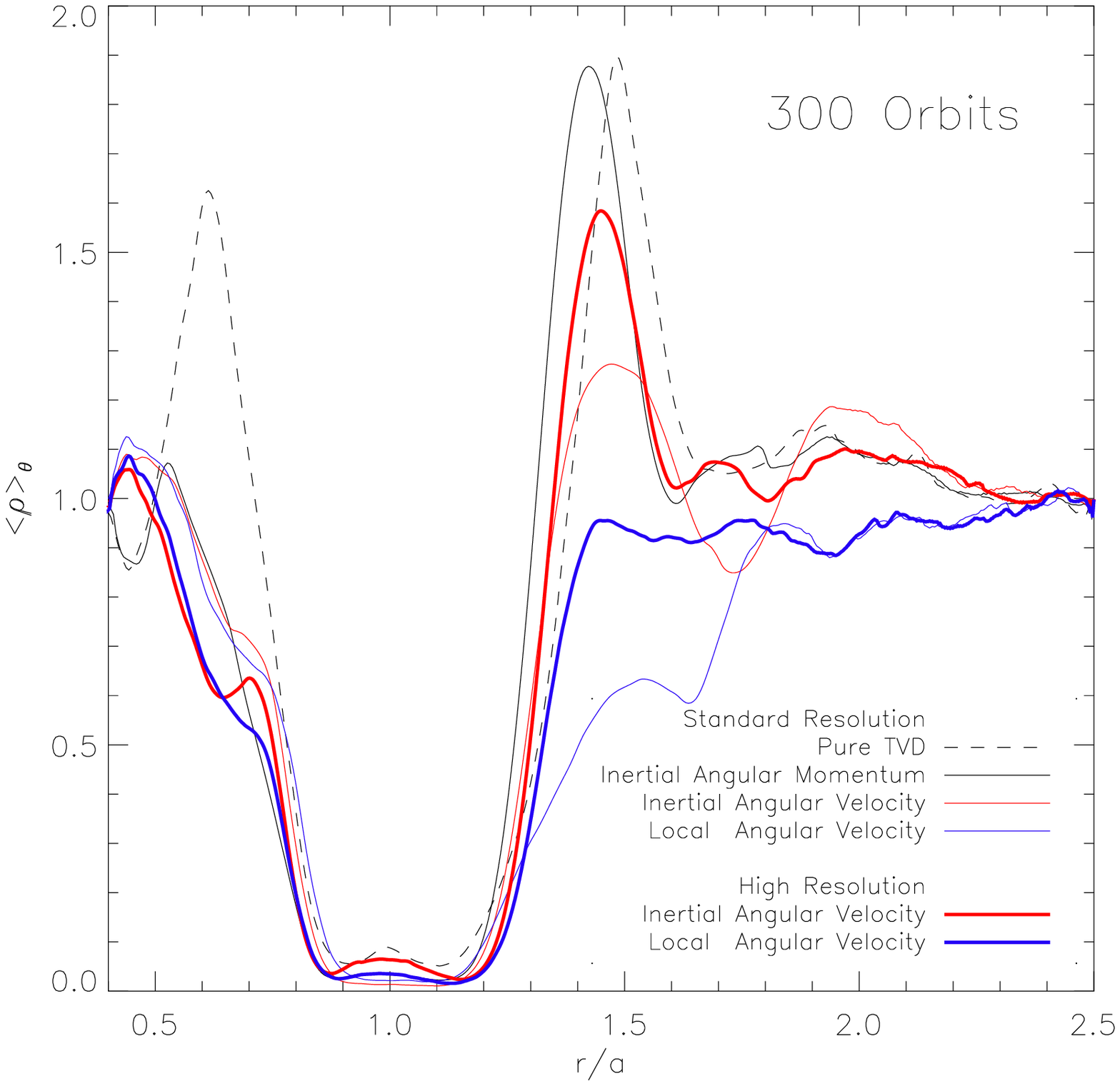}}
\end{center}
\caption{\label{fig:SJ0Hhu.avg} Azimuthally averaged density after 100
  orbits (left) and 300 orbits (right).  Runs with standard resolution
  (fine lines) show progressive numerical diffusion as in Figure
  \ref{fig:SJ0Hhu.den}.  High resolution runs (heavy lines) for the
  two alternate choices of solution variables show markedly less
  diffusion.} 
\end{figure*} 

Using the fast-advection algorithm reduces the required
simulation time by approximately the ratio of the residual azimuthal
velocity to the full azimuthal velocity, but this reduction comes at the
expense of increased diffusion introduced by the transport of
quantities by the background flow.  This increased diffusion
influences the formation of intermittent structures such as the
vortices and libration islands observed in the standard run.

Further differences may also appear as a result of the choice of
solution variables.  In particular, \citet{kley98} showed that
advecting the inertial angular momentum ${\cal H} = \rho r (u_\q + r\Omega)$
(that of both the corotating fluid and the frame), as written in
equation (\ref{eq:cyl.momq}), produces better results than just
advecting the angular velocity in the corotating frame $\rho u_\q$
(unless otherwise marked, $u_\q$ refers to the fluid velocity in the
corotating frame).  In practise this distinction is between accounting
for the Coriolis and centripetal accelerations using the Euler
equation solver or accounting for them as source terms.  We perform a
comparison among three simulations that use the solution sets
$(\rho,\rho u_r, {\cal H})$, $(\rho,\rho u_r, {\cal H}/r)$, and
$(\rho,\rho u_r, \rho u_\q)$.  The first of these sets uses the
inertial angular momentum as the choice of angular variable, the
second uses the inertial angular velocity as the choice of
angular variable and the third uses the corotating frame (local)
angular velocity as the choice of angular variable.  We argue that the
difference in the results caused by the choice of solution variables
does not reflect differences in accuracy, but rather differences in
the amount of numerical viscosity that is present in each of the
simulations.

In Figure \ref{fig:SJ0Hhu.den} we compare the density after $20$,
$100$ and $300$ orbits for the standard run (which makes use of the
fast-ad\-vec\-tion algorithm and which uses the inertial angular momentum
as the angular variable) against one run which does not make use of
the fast-advection algorithm, and against two additional runs which
do but which implement the two alternate choices of angular variables.
The standard run using the fast-advection routine requires $7.8$ times
fewer iterations to reach $100$ orbits and finishes $6.6$ times
faster.  While all simulations properly capture the locations of the
spiral arms, the overall level of detail present and both the
strength and number of vortices present decrease in each successive
row.  This observation suggests that the use of the fast-advection
algorithm introduces extra diffusion into the simulation, and that the
two alternate choices of angular variables also yield more diffusion. 

In Figure \ref{fig:SJ0Hhu.avg} we compare results for the azimuthally
averaged density in the gap region after 100 and 300 orbits.  The
progressive increase in diffusion caused by the FARGO algorithm and
then by the alternate choices of solution vectors is apparent.  Also
shown for the two alternate choices of solution vectors are results
from simulations run at resolution $N_r \times N_\q =768 \times 1252$.
These high resolution results suggest that the shallow gap profile
seen in the runs using the inertial and corotating angular velocities
is due to high levels of diffusion and numerical viscosity in the
simulation.  When run at higher resolution, they show profiles much
closer to that of the standard run.  Comparison with Figure
\ref{fig:SJvall.avg} suggests that the numerical viscosity using the
alternate solution sets $(\rho, \rho u_r, {\cal H}/r)$ or $(\rho,
\rho u_r,\rho u_\q)$ is at least an order of magnitude higher.  This
interpretation of the results, which suggests that the alternate
solution variables are simply more numerically diffusive, differs from
that argued by \citet{kley98} in a similar analysis.

\begin{figure}[b]
\begin{center}
\resizebox{0.45\textwidth}{0.45\textwidth}{\includegraphics{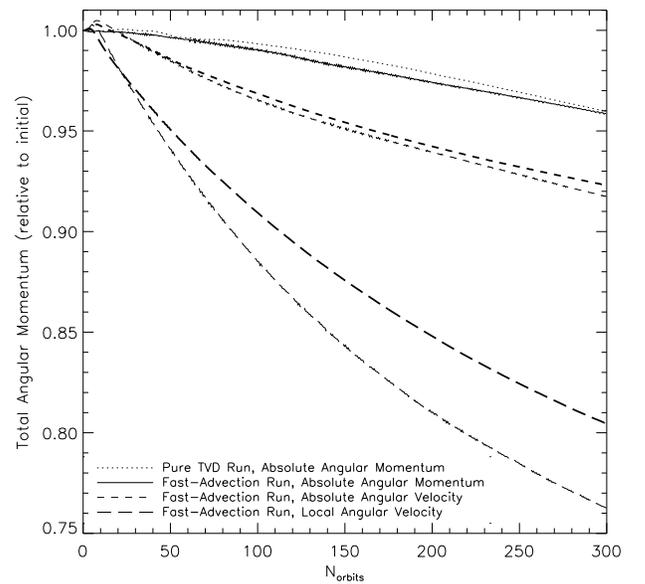}}
\end{center}
\caption{\label{fig:SJ0Hhu.Mcon} Conservation of angular momentum
  measured relative to the initial amount present in the entire
  disk. Heavy lines show results from the high resolution runs with the
  alternate solution variables.}
\end{figure} 

\begin{figure*}[t]
\begin{center}
\resizebox{0.9\textwidth}{0.22\textwidth}{
  \includegraphics{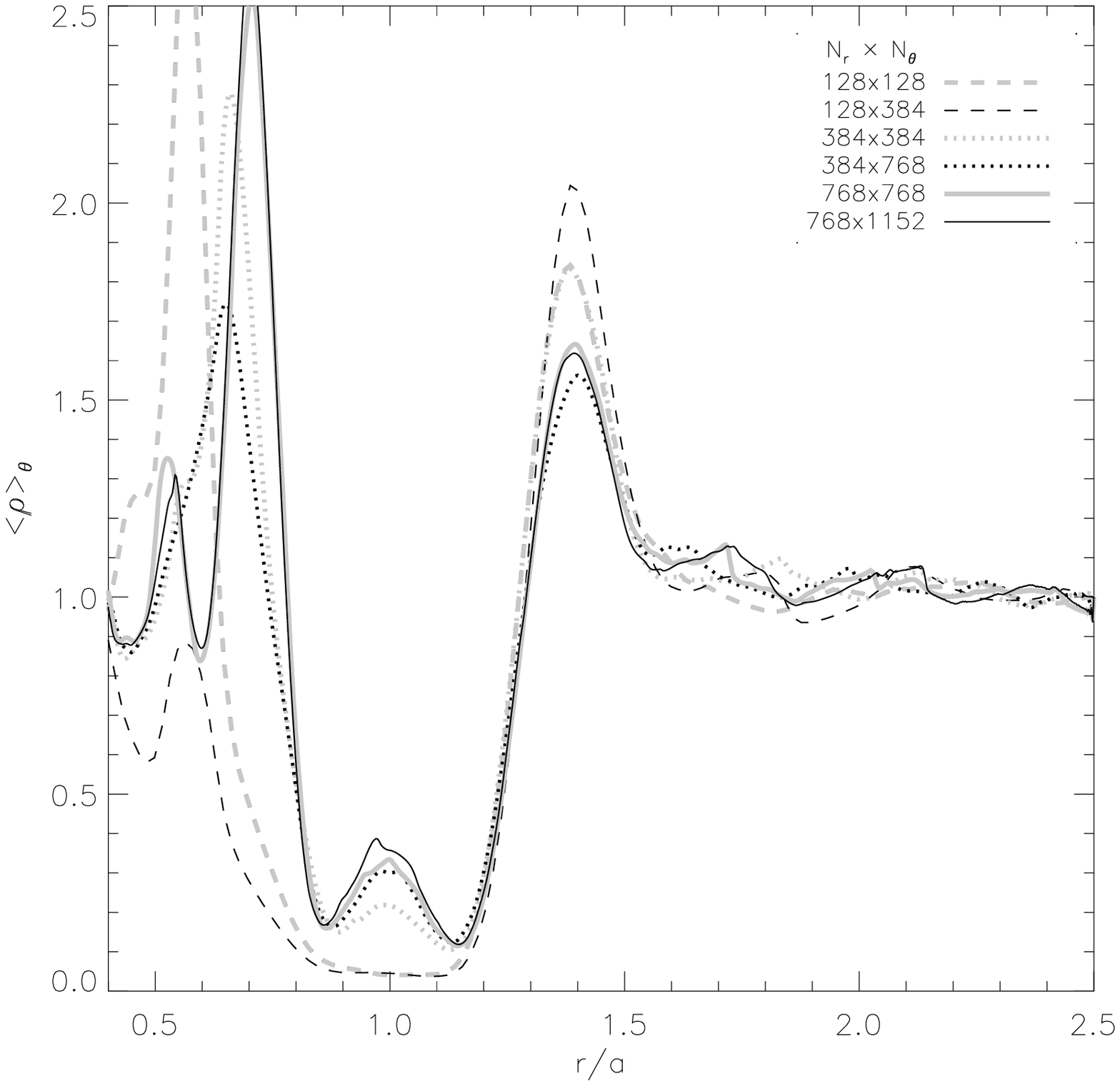}
  \includegraphics{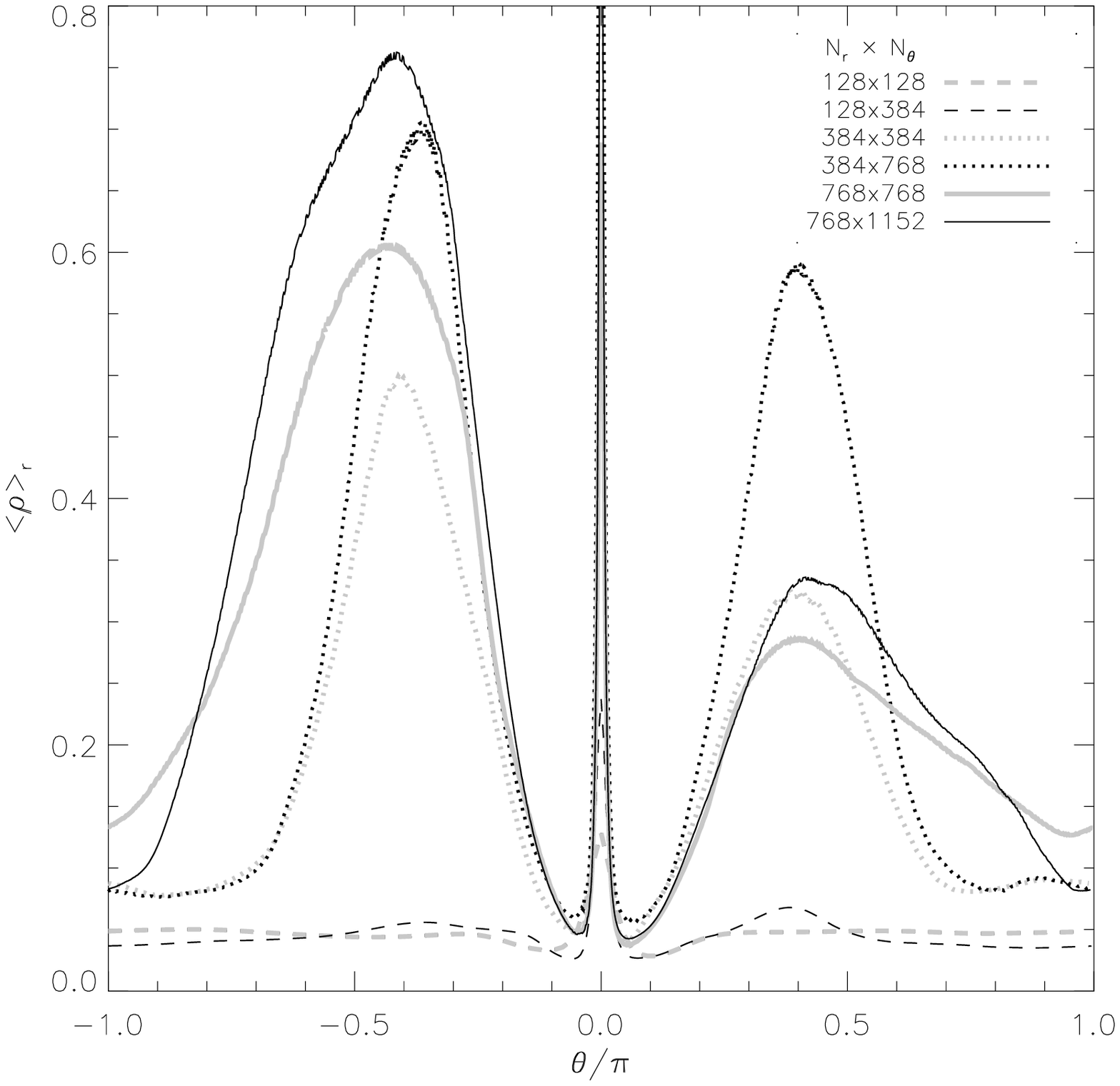}
  \includegraphics{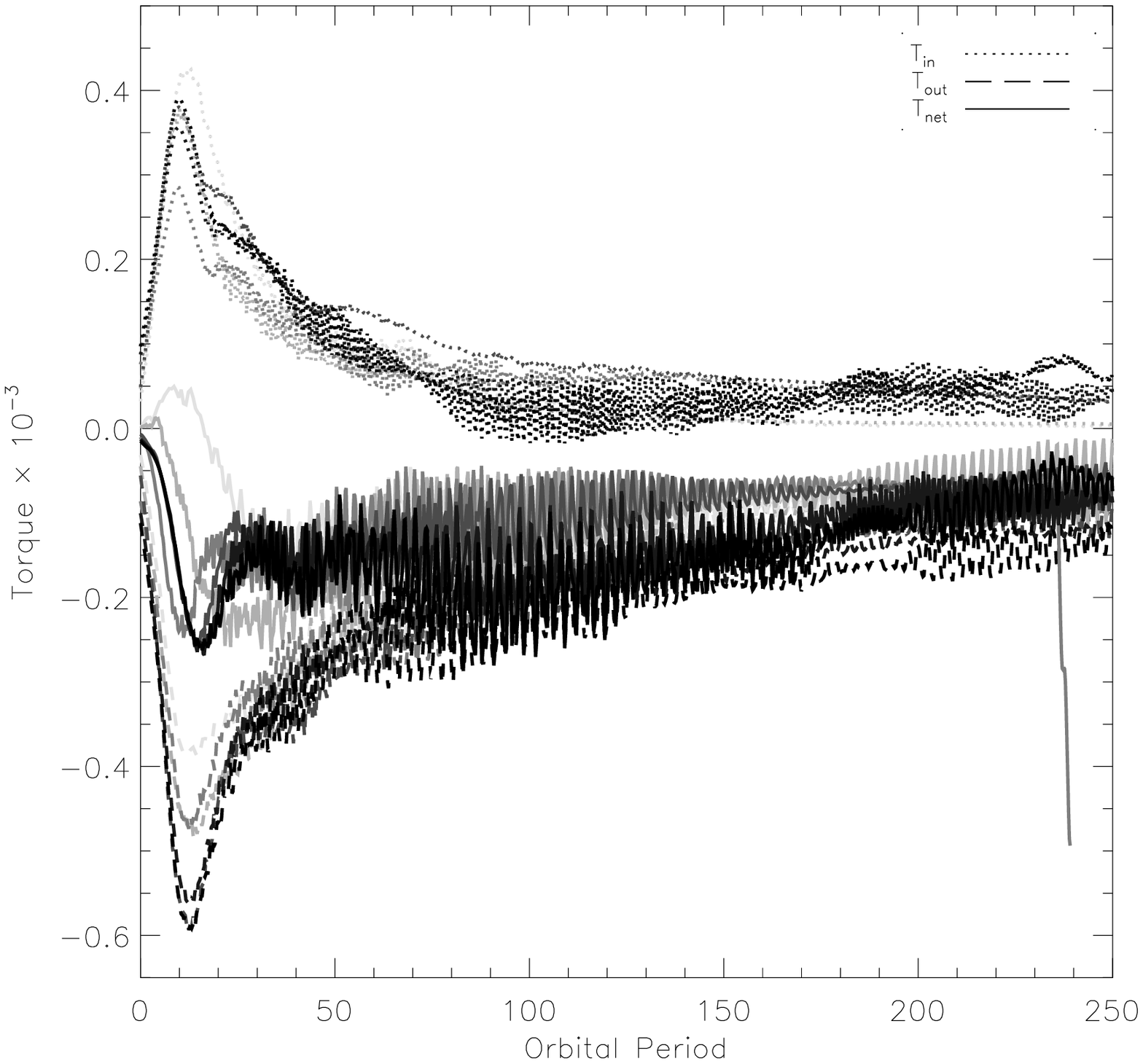}}
\resizebox{0.9\textwidth}{0.22\textwidth}{
  \includegraphics{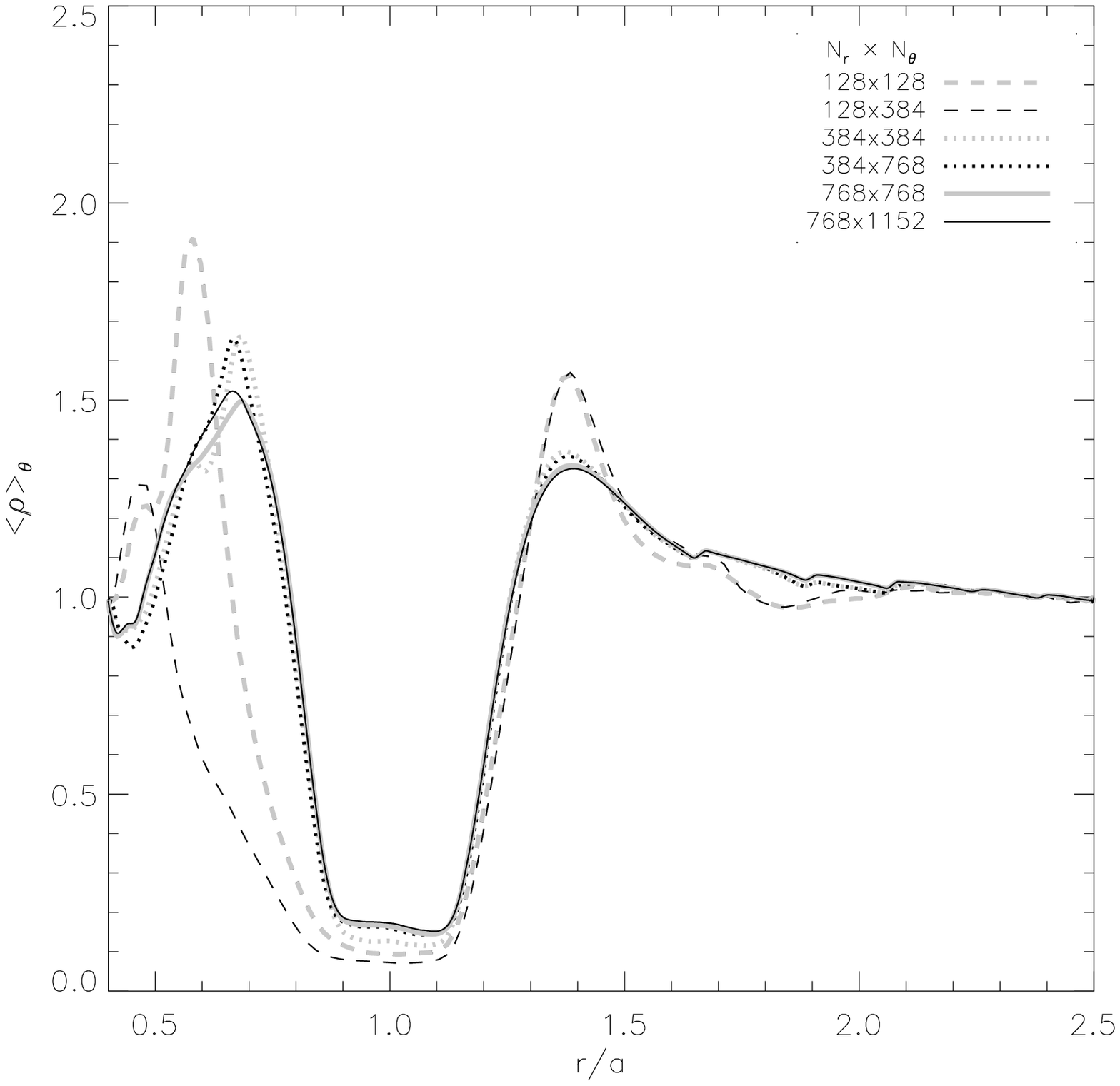}
  \includegraphics{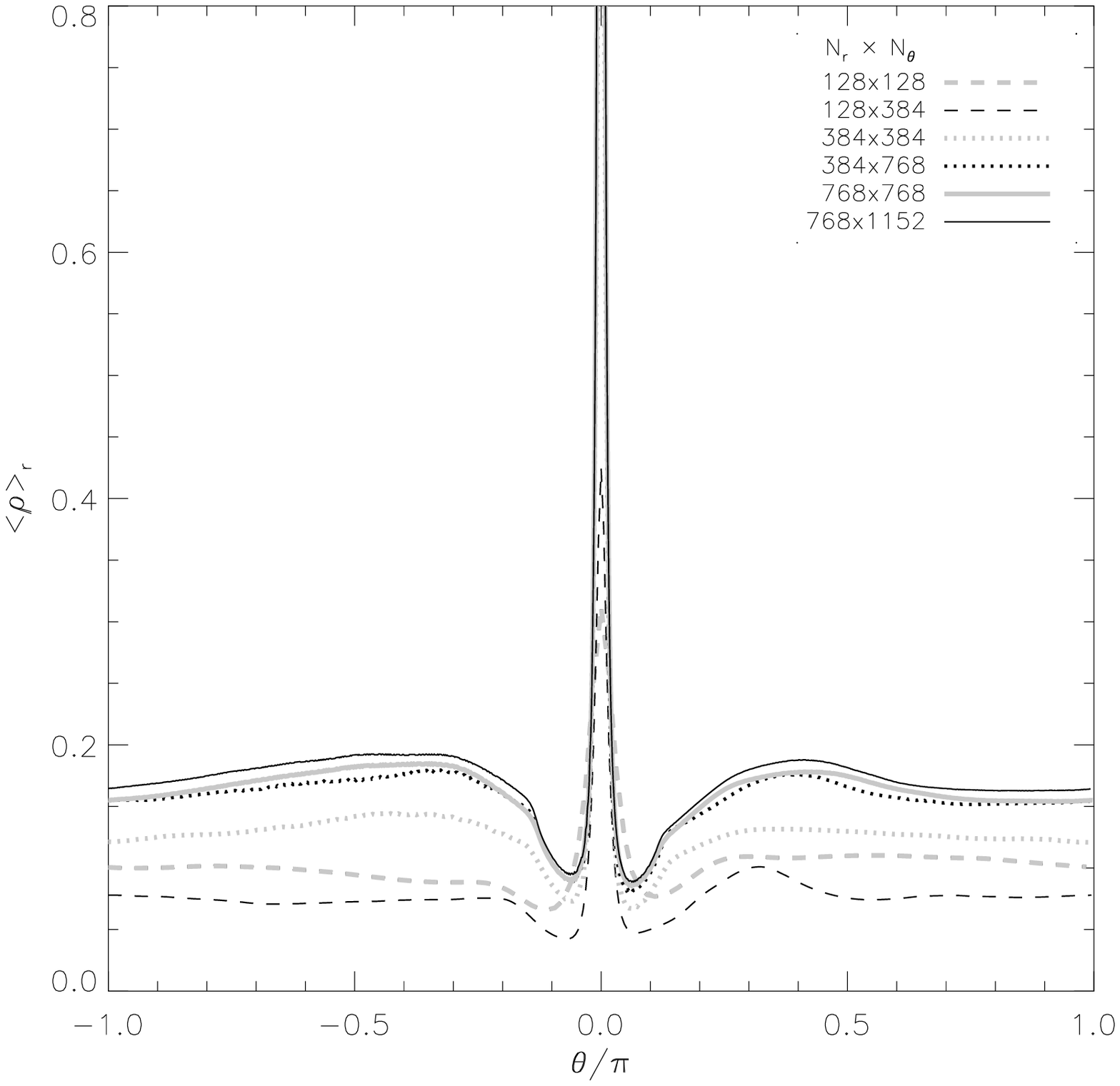}
  \includegraphics{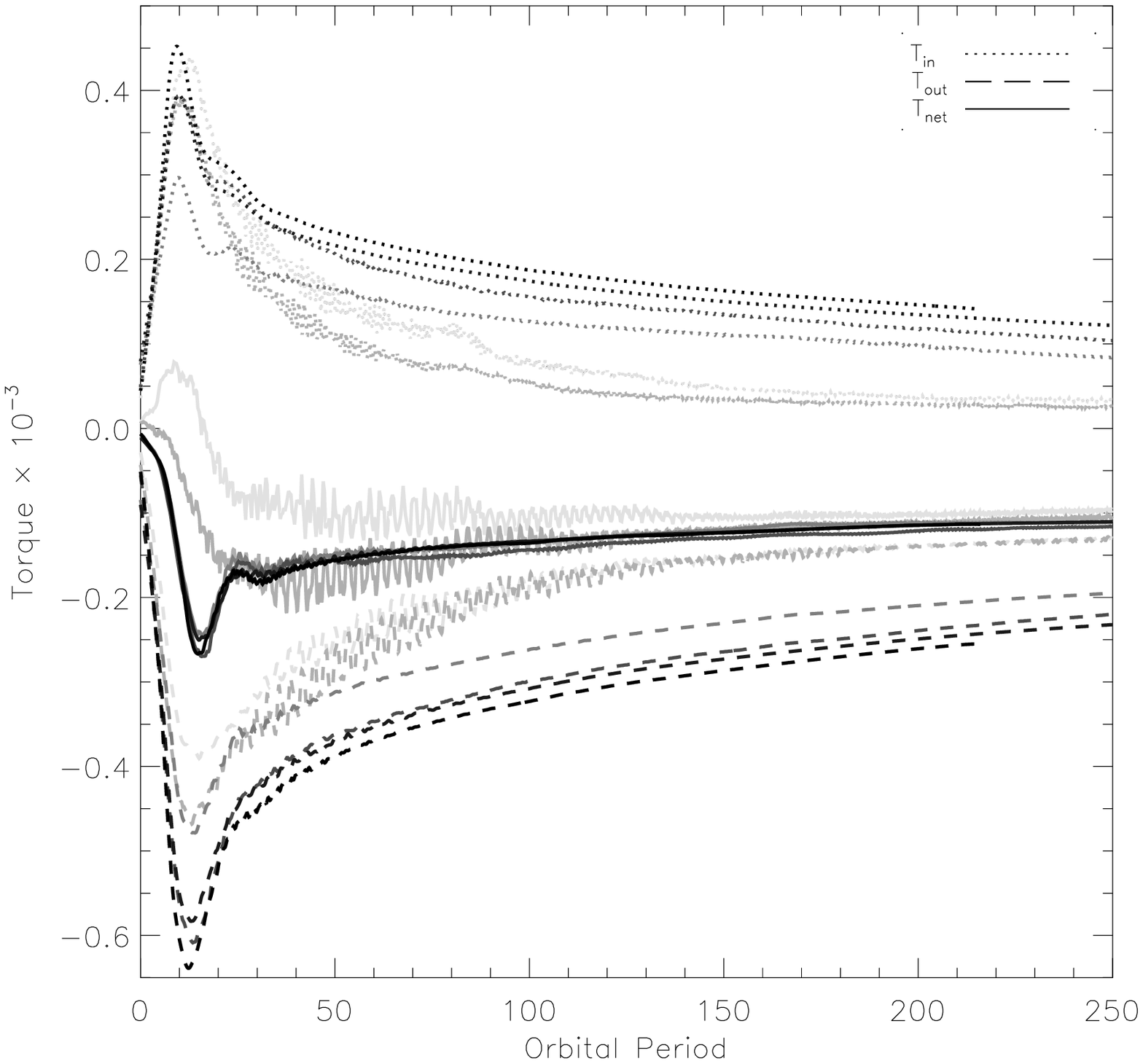}}
\end{center}
\caption {\label{fig:SJres.avg} Left to right: azimuthally averaged
  density, radially averaged density in the gap region, and time
  evolution of the torque for different resolutions.  All
  results use the VL limiter and density averages are taken at $100$
  orbits.  Results in the top row have no added viscosity, those in
  the bottom row have $\nu=10^{-5}$.  The torque is broken up into 
  components from the disk material inside (dotted) and outside
  (dashed) the planet's orbit.  Also plotted is the total net torque
  (solid).  The palest lines show the lowest resolution of the five
  runs.  The darkest lines show the highest resolution.} 
\end{figure*} 

\begin{figure*}[!h]
\begin{center}
\resizebox{0.9\textwidth}{0.22\textwidth}{
  \includegraphics{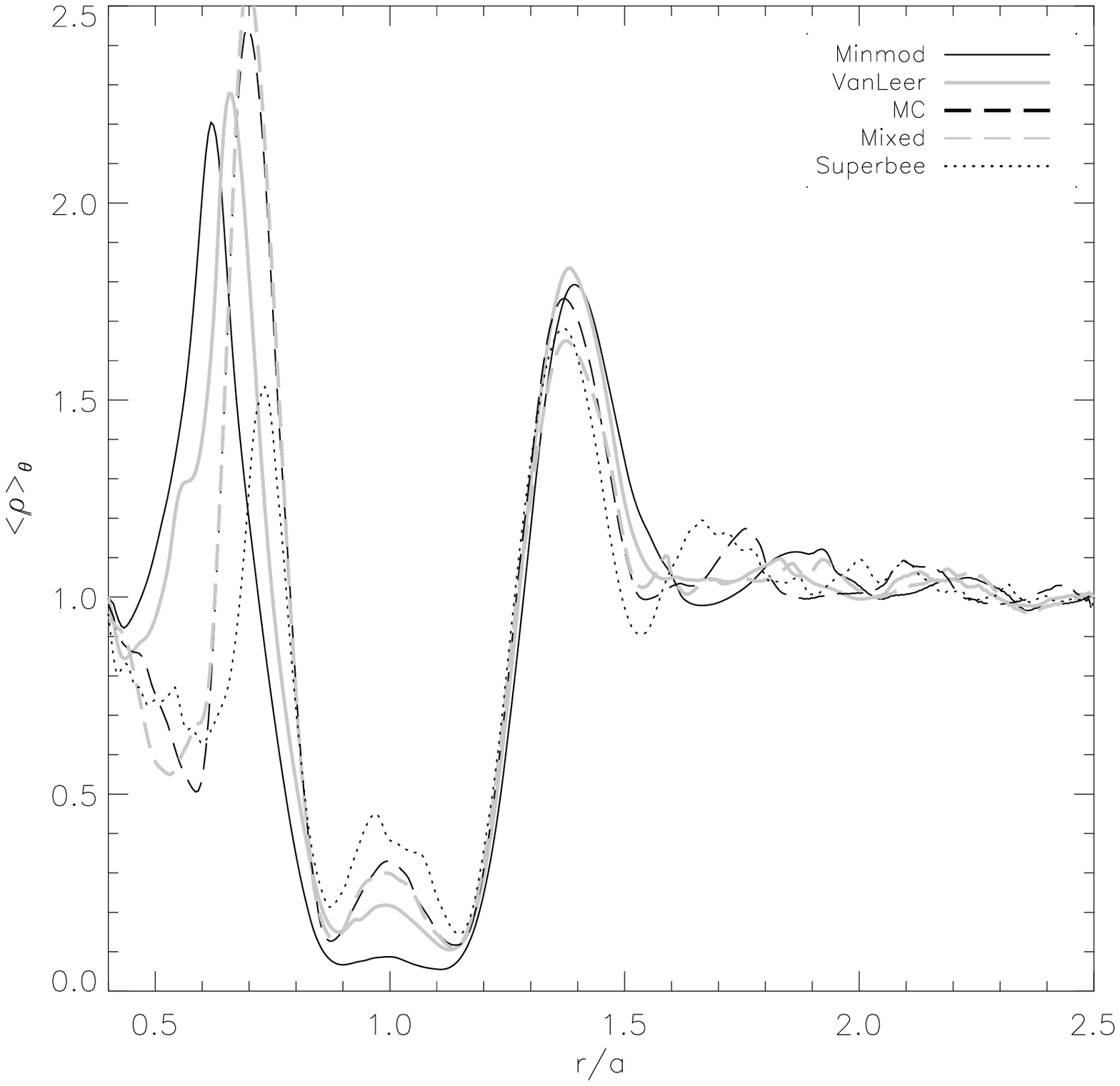}
  \includegraphics{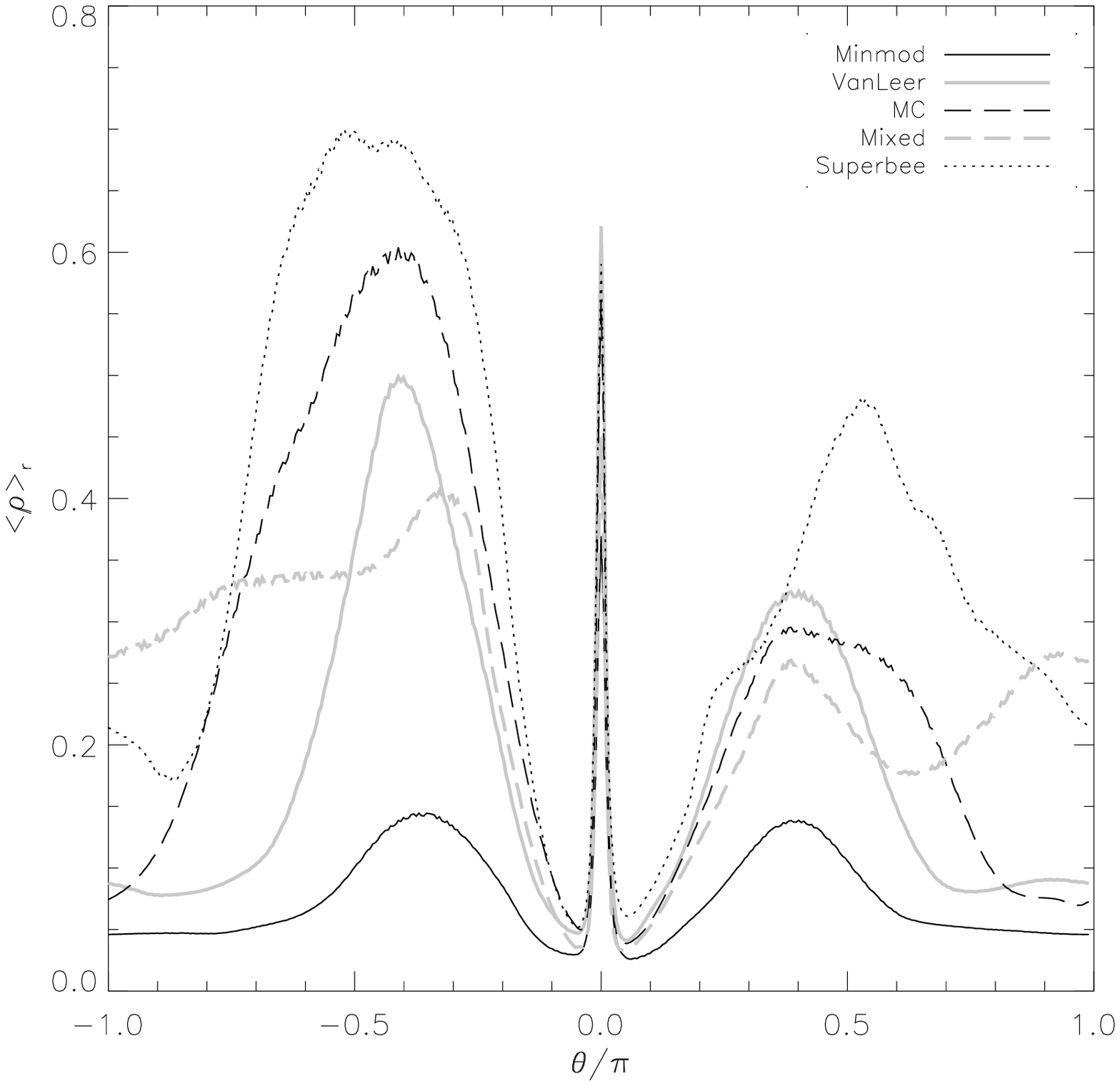}
  \includegraphics{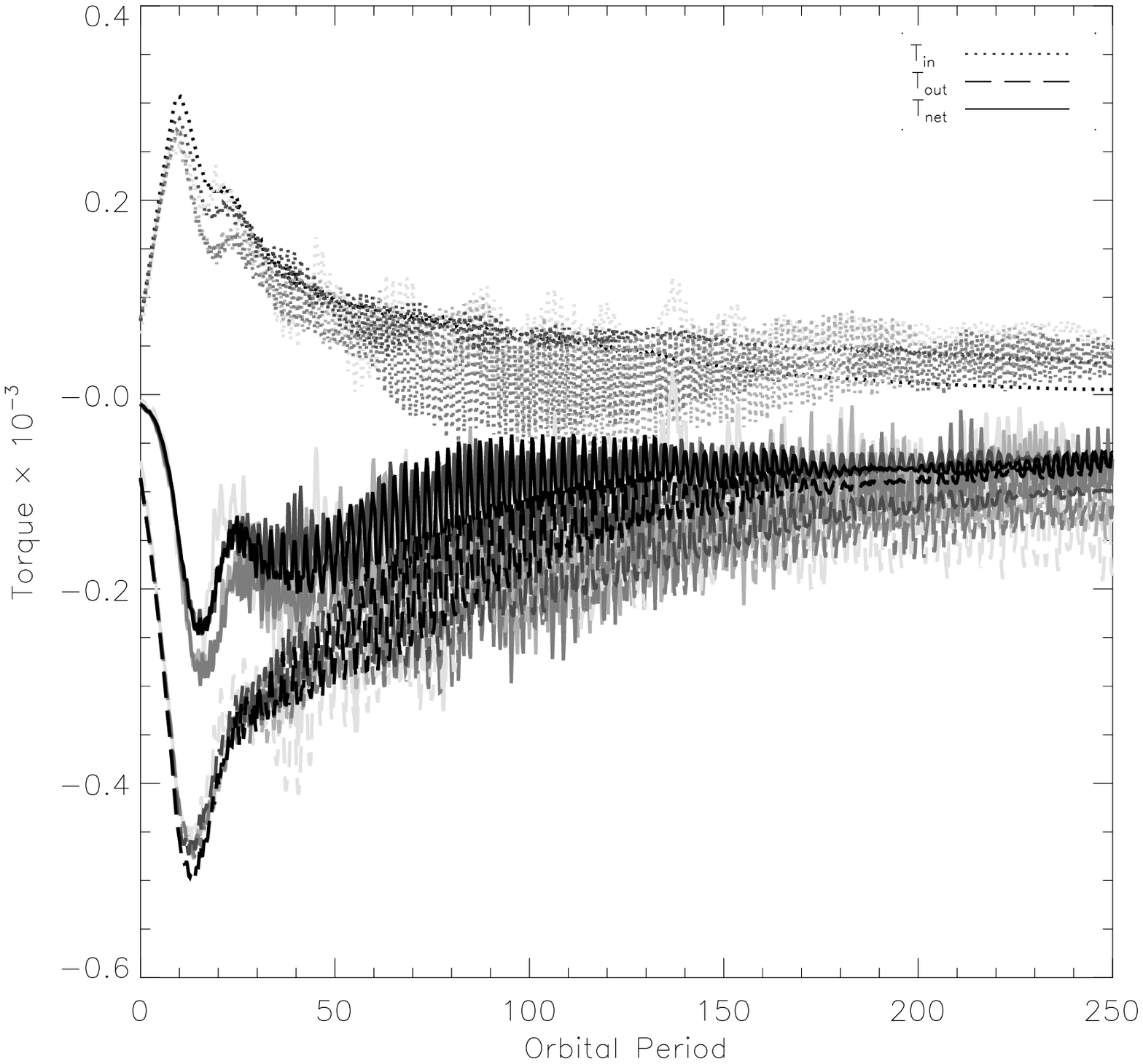}}
\end{center}
\caption {\label{fig:SJlim.avg} Left to right: azimuthally averaged
  density, radially averaged density in the gap region, and time
  evolution of the torque for different limiters.  All runs have
  resolution $N_r \times N_\q = 384 \times 384$ and density averages
  are taken at $100$ orbits.  The torque is broken up into components
  from the disk material inside (dotted) and outside (dashed) the
  planet's orbit.  Also plotted is the total net torque (solid).  The
  lines from darkest to palest correspond to the Minmod, Van Leer, MC,
  mixed (MB), and Superbee limiters, respectively.}  
\end{figure*} 

Figure \ref{fig:SJ0Hhu.Mcon} shows the total angular momentum in the
simulation as a function of time.  The standard solution set loses
angular momentum at a rate of 1\%/100 orbits, using the inertial
angular velocity causes a loss rate of 4\%/100 orbits and using the
local angular velocity causes a loss rate of 8\%/100 orbits.

\subsection{Effects of resolution and evidence of numerical convergence} \label{subsec:resolution}

The effects of numerical viscosity become more pronounced at lower
resolutions.  Figure \ref{fig:SJres.avg} shows the azimuthally and
radially averaged densities for several different resolutions above
and below that of the standard run.  The effects of increasing the
resolution are most apparent inside the planet's orbit.  Low radial
resolution appears to make the slope of the gap shallower on the
inside edge of the planet's orbit.  For the two runs with the lowest
radial resolution $N_r=128$, the libration islands of fluid do not
exist---likely the numerical viscosity at these resolutions is too
large for them to be maintained.  Also, as the radial resolution
increases, two distinct vortex lines and corresponding overdensities
(at approximately $r=0.55$ and $r=0.70$) become apparent inside the
planet's orbit, rather than just a single line, or none.

Note that increasing the azimuthal resolution relative to the radial
resolution widens the gap profile.  At resolutions where the results
have not yet converged, it also affects some of the details of the
structures present within the disk---the vortex lines and the the
libration islands---in a more complicated manner because the number of
iterations required to reach $100$ orbits differs amongst differing
resolutions by as much as a factor of three.  Thus, the effects of
numerical resolution are the result of a competition between an
increased amount of diffusion from an increased number of iterations
required, and a decreased amount of numerical viscosity due to the
increased grid resolution.

Figure \ref{fig:SJres.avg} also shows the torques as the resolution is
varied.  The runs with the three or four highest resolutions are
consistent with one another.  Note that while the magnitudes of the
torques from the inner and outer disk still increase slightly with 
higher resolution, the total torque remains the same, except for the
two lowest resolution runs.

\subsection{Effects of the limiter scheme} \label{subsec:limiters}

The results from the Sod shock tube test and the Kelvin-Helm\-holtz
instability in Section \ref{sec:basictests} have already illustrated
that the different limiters exhibit different levels of numerical
vis\-cos\-ity---both differing diffusion and dispersion.  Figure
\ref{fig:SJlim.avg} shows the azimuthally and radially averaged
densities for the Minmod, Van Leer, MC, mixed (MB), and Superbee schemes
used in the standard run.  As in Section \ref{sec:basictests}, the
Superbee scheme shows the least diffusion, but the most dispersion.
This conclusion is evident from the increased mass of the fluid in the
gap region, as well as the oscillatory density structure exhibited
there. The mixed and MC schemes also show a fair amount of dispersion.
The Van Leer limiter seems to show diffusion comparable to that of the
MC and mixed limiters but substantially less dispersion.  As before, the
Minmod limiter shows the most diffusion.  Simulations run with $\nu =
10^{-5}$ (not presented) show analogous results.  Figure 
\ref{fig:SJlim.avg} also shows the calculated torques for each choice
of limiter.  Again, these results are consistent with those previous. 

\section{Conclusions} \label{sec:conclusions}

We have developed a new, efficient, parallelized hydrodynamic code for
studying accretion disk processes. The current incarnation is
optimized to study planetary disk-planet interactions.  A FARGO-type
algorithm is implemented to help alleviate CFL time step restrictions
imposed by the rapidly rotating inner disk region.  Open-MP directives
are also implemented to obtain faster computations on shared memory
machines. Parallelization on distributed memory machines (such as with
Message Passing Interface (MPI) protocols) requires further development.   

We have shown that the RAPID code performs comparably to the well-established
piece-wise para\-bolic method (PPM) on standard hydrodynamic tests.  The
largest difference observed between the two algorithms 
is the level of pre- and postshock oscillations that are allowed.  PPM codes
flatten such oscillations quite stringently.  We note that this procedure is
not necessarily physically motivated and may lead to a decrease in the amount
of structure present in some simulations.  We have also compared how results
from our code differ depending on the choice of flux limiter. The amounts of
diffusion and dispersion vary quite substantially, but the relative amounts
amongst limiters are  observed to be qualitatively consistent on all tests.   

In addition, we presented a large series of comparisons on the
standard protoplanet problem, showing results that are consistent with
ensemble results from a wide variety of other codes documented in the 
protoplanet comparison project \citep{valborro06}.  In particular we
confirmed the existence of libration islands at the $L_4$ and $L_5$
points, an asymmetry in the density of those islands, the sign and
magnitude of the torque exerted on the planet by the disk and the
growth and merging of vortices outside the planet's orbit at low
viscosity. 

We found that the large vortex which forms outside the plan\-et's orbit
causes substantial torque oscillations on the planet.  These
oscillations correspond to repeated passes of the vortex by the
planet.  Increasing the viscosity beyond $\nu \ge 10^{-5}$ damps the 
formation of the vortex thereby removing the oscillatory signature
from the torque.  The existence of additional \textit{vortex lines}
to the inside of the planet's orbit are demonstrated.

We have shown that the \textit{FARGO}-like fast-advection algorithm
reduces the required simulation time by a factor of $6.5$ in standard
planet-disk setups.  We also illustrated that using the inertial
angular momentum rather than angular velocity as a solution variable
decreases the numerical viscosity present in the simulations by an
order or magnitude or more.  This finding supplements previous work by 
\citet{kley98}.  In addition, the choice of inertial angular momentum
as a solution variable conserves the total angular momentum on the
grid to higher precision. 

We determined the level of numerical viscosity present with\-in the code 
to be $\nu < 10^{-6}$ ($\alpha < 10^{-3.5}$), enabling the simulation 
of scenarios with Reynolds numbers on the order of $Re = UL \times
10^6 $. 

\section*{Acknowledgements}
We thank L.J. Dursi for helpful discussions and a careful read of
this manuscript.

\appendix 

\section{Appendix}
\label{sec:appendix}

Written out in component form for $\bu = (u_r,u_\q, u_z)$ and $\bx=(r,\q,z)$ , the equations solved for are 
  \begin{align}
& \frac{\textstyle \prho}{ \pt} + \frac{1}{r}\frac{\p}{\pr}[\rho r u_r]
            + \frac{1}{r}\frac{\p}{\p \q}[\rho \uq] 
            + \frac{\p}{\pz}[\rho u_z] =  0 \label{eq:cyl.d} \\
\begin{split}
& \frac{\prho u_r}{ \pt} + \frac{1}{r}\frac{\p}{\pr}[\rho ru_r^2+ pr]  
            + \frac{1}{r}\frac{\p}{\p\q}[\rho u_\q u_r] + \frac{\p}{\pz}[\rho u_z u_r]  \\
& \hspace{1pc} = -\rho\frac{\p \phi}{\pr} \!+\! \frac{1}{r}\frac{\p
            r\sigma_{rr}}{\pr} \!+\! \frac{1}{r}\frac{\p \sigma_{\q r}}{\p\q} \!+\! \frac{\p \sigma_{zr}}{\pz}  
            \!-\! \frac{\sigma_{\q\q}}{r} \!+\! \frac{p}{r} \!+\! \frac{\rho
            \uq^2}{r} 
\end{split} \label{eq:cyl.ur} \\
\begin{split} 
& \frac{\p {\cal H}}{ \pt} + \frac{1}{r}\frac{\p}{\pr}[r u_r{\cal H}]   +
            \frac{1}{r}\frac{\p}{\p\q}[u_\q {\cal H} + pr] + \frac{\p}{\pz}[u_z{\cal H}] \\
& \hspace{1pc} = -\rho \frac{\p \phi}{\p\q} + \frac{\p r\sigma_{r\q}}{\pr} + \frac{\p \sigma_{\q\q}}{\p\q} 
            + r\frac{\p \sigma_{z\q}}{\pz} + \sigma_{\q r} 
\end{split} \label{eq:cyl.momq} \\
\begin{split}
& \frac{\prho u_z}{ \pt} + \frac{1}{r}\frac{\p}{\pr}[\rho r u_r  u_z]  
            + \frac{1}{r}\frac{\p}{\p\q}[\rho \uq u_z]   + \frac{\p}{\pz}[\rho u_z^2+p]  \\
& \hspace{1pc} = -\rho \frac{\p \phi}{\pz} + \frac{1}{r}\frac{\p r\sigma_{rz}}{\pr} + \frac{1}{r}\frac{\p \sigma_{\q z}}{\p\q} 
            + \frac{\p \sigma_{zz}}{\pz}
\end{split} \label{eq:cyl.uz}  \\
\begin{split}
& \frac{\p \rho e}{\pt} \!+\! \frac{1}{r}\frac{\p}{\pr}[r u_r(\rho e \!+\! p)] 
  \!+\! \frac{1}{r}\frac{\p}{\p\q}[\uq(\rho e \!+\! p)] \!+\! \frac{\p}{\pz}[u_z(\rho e \!+\! p)] \\
& \hspace{1pc}  = -\rho \left[ u_r\frac{\p \phi}{\pr} + \frac{\uq}{r}\frac{\p \phi}{\p\q} + u_z \frac{\p \phi}{\pz} \right ] \\
& \hspace{1pc} + \frac{1}{r}\frac{\p}{\pr}[ru_i \sigma_{ir}] + \frac{1}{r}\frac{\p}{\p\q}[u_i \sigma_{i\q}] + \frac{\p}{\pz}[u_i \sigma_{iz}],
\end{split} \label{eq:cyl.e}
\end{align}
where ${\cal H} = \rho r(u_\q +r\Omega)$ is the fluid's inertial angular
momentum, and we use the notation $u_i \sigma_{ij} = u_r \sigma_{rj} + u_\q \sigma_{\q j} + u_z \sigma_{zj}$. 

The stress tensor for a compressible Newtonian fluid under the Stoke's assumption is
     \begin{equation*}
       \sigma_{ij} \!=\! \rho \nu 
         \!\left[\begin{split}
            &\sfrac{\p u_r}{\pr} {\scriptstyle -} \sfrac{1}{3}\mbox{\boldmath $\scriptstyle \nabla \cdot \bu$}& 
            &\!\!\!\sfrac{1}{2} \!\left (\sfrac{1}{r}\!\sfrac{\p u_r}{\p\q}{\scriptstyle +}\!\sfrac{\p\uq}{\pr}{\scriptstyle -}\!\sfrac{\uq}{r} \right)& 
            &\!\!\!\!\sfrac{1}{2} \!\left (\sfrac{\p u_r}{\pz} \scriptstyle{+}
             \sfrac{\p u_z}{\pr} \right) \\ 
            &\hspace{-0.2em}\sfrac{1}{2} \!\left (\sfrac{1}{r}\!\sfrac{\p u_r}{\p\q}\scriptstyle{+}\!\sfrac{\p\uq}{\pr}\scriptstyle{-}\!\sfrac{\uq}{r} \right)& 
            &\hspace{-0.2em}\!\!\!\sfrac{1}{r} \!\sfrac{\p\uq}{\p\q}{\scriptstyle+}\sfrac{u_r}{r} {\scriptstyle -} \sfrac{1}{3}\mbox{\boldmath $\scriptstyle \nabla \cdot \bu$}&
            &\hspace{-0.2em}\!\!\!\!\sfrac{1}{2} \!\left(\sfrac{\p\uq}{\pz} {\scriptstyle+}
             \sfrac{1}{r}\sfrac{\p u_z}{\p\q}\right) \\ 
            &\sfrac{1}{2} \!\left(\sfrac{\p u_r}{\pz} {\scriptstyle+} \sfrac{\p u_z}{\pr}\right)& 
            &\!\!\!\sfrac{1}{2} \!\left(\sfrac{\p\uq}{\pz} {\scriptstyle+} \sfrac{1}{r}\sfrac{\p u_z}{\p\q}\right)&  
            &\!\!\!\!\sfrac{\p u_z}{\pz} {\scriptstyle-}\sfrac{1}{3}\mbox{\boldmath $\scriptstyle \nabla \cdot \bu$}
         \end{split} \!\right]\!\!,
     \end{equation*}
     where 
     \begin{displaymath}
       \divm \bu = \frac{1}{r}\frac{\p r u_r}{\pr} + \frac{1}{r}\frac{\p \uq}{\p\q} + \frac{\p u_z}{\pz}.
     \end{displaymath}

In order to close the above equations, one requires an equation of
state relating the internal energy and pressure of the fluid.  For
adiabatic fluids, 
     \begin{subequations}
         \begin{align}
           &p = (\gamma-1) \rho \varepsilon, \\
           \intertext{in terms of the adiabatic index $\gamma$.  For isothermal fluids, the
             pressure is defined independently of the internal energy as}
           &p = \rho c_s^2,
         \end{align}
     \end{subequations}
and a set prescription is used to define the sound speed.

\end{document}